\documentclass[12pt,a4paper,final]{iopart}

\usepackage{iopams}  
\usepackage{graphicx}
\usepackage[breaklinks=true,colorlinks=true,linkcolor=blue,urlcolor=blue,citecolor=blue]{hyperref}
\usepackage{accents}
\newcommand{\be}{\begin{equation}}
\newcommand{\ee}{\end{equation}}

\def\stackrel#1#2{\mathrel{\mathop{#2}\limits^{#1}}}

\begin{document}

\title{Lifshitz holography}

\author{Marika Taylor}
\address{Mathematical Sciences and STAG, University of Southampton, Highfield, Southampton SO17 1BJ, UK.}

\ead{m.m.taylor@soton.ac.uk}

\begin{abstract}
{In this article we review recent progress on the holographic modelling of field theories with Lifshitz symmetry.
We focus in particular on the holographic dictionary for Lifshitz backgrounds - the relationship between bulk fields and boundary operators, operator correlation functions and underlying geometrical structure. The holographic dictionary is essential in identifying the universality class of strongly coupled Lifshitz theories described by gravitational models. }

\end{abstract}

\vspace{2pc}
\noindent
\submitto{\CQG}

\section{Introduction}

Gauge/gravity duality is a relationship between a (relativistic) field theory and a gravity theory in one higher spatial dimension. The best understood example is the anti-de Sitter/conformal field theory (AdS/CFT) correspondence relating string theory in negatively curved backgrounds to conformal field theories in one less non-compact spatial dimension. The strong coupling limit of the boundary field theory corresponds to the supergravity limit of the bulk string theory and hence holography is frequently used as a prototype to illustrate features of strongly coupled relativistic field theories. Such modelling has been applied to the physics of the quark/gluon plasma and to a wide variety of condensed matter systems, ranging from superconductors to (non) Fermi liquids. 

In condensed matter physics there exists a variety of systems which are believed to be described by strongly interacting non-relativistic physics and it is natural to ask whether one can use holography to gain intuition about such systems also. One example is the case of fermions at unitarity: these are fermions whose interactions are fine tuned to produce a scale invariant but non-relativistic system, one which can be experimentally realized using cold atoms. Another potential example is high temperature superconductors: it has been proposed that their unconventional properties are controlled by an underlying non-relativistic quantum critical point. 

Leaving aside applications to condensed matter physics, non-relativistic holography is certainly interesting in its own right. As an example of gauge/gravity duality without anti-de Sitter asymptotics, non-relativistic holography could provide deeper insights into the general principles of holography, and thus  perhaps lead to progress in longstanding questions such as holography for asymptotically flat and de Sitter spacetimes.  

In this review we will discuss gravity duals for non-relativistic field theories, focussing particularly on those admitting Lifshitz symmetry. Along the way we will also discuss aspects of theories with Schr\"{o}dinger symmetry, since their holographic realisations turn out to be intimately related to Lifshitz theories. 
We will focus here primarily on formal aspects of Lifshitz holography rather than applications of non-relativistic holography. 

Applications of holography require a detailed understanding of the correspondence between bulk fields and boundary operators, i.e. a detailed holographic dictionary.
The generalisation of the AdS/CFT dictionary to non-relativistic cases has turned out to be extremely subtle and is the main topic of this review. Despite considerable recent progress, the dictionary is still not completely understood for Lifshitz theories. Holographic models describe specific universality classes of strongly coupled Lifshitz theories and we will use the holographic dictionary to deduce universal features of holographic Lifshitz models. 

\bigskip

The structure of this review is as follows. In section \ref{two} we give a short introduction to non-relativistic symmetry groups. In section \ref{three} we discuss how non-relativistic symmetry can be realised as the isometry group of a higher dimensional geometry and discuss matter fields which can support such a geometry. We discuss the stability of such gravity backgrounds and features of the correlation functions for the corresponding dual non-relativistic field theories. In section \ref{four} we discuss hydrodynamics in Lifshitz theories from both field theory and holographic perspectives. We turn to the holographic dictionary for Lifshitz theories, i.e. the mapping between bulk fields and operators in the dual field theory, in section \ref{hol-dict}. In section \ref{newton} we discuss the role of Newton-Cartan structure in coupling non-relativistic field theories to background gravity and we explore the emergence of Newton-Cartan structure in certain holographic duals. In section \ref{seven} we summarise other developments in Lifshitz holography and we conclude by discussing key open questions in section \ref{eight}.

\section{Non-relativistic symmetry groups} \label{two}

Consider a relativistic field theory in $d$ spacetime dimensions. Such a theory is invariant under the Poincar\'{e} group consisting of translations $P^{\mu}$ and Lorentz transformations $M^{\mu \nu}$:
\begin{eqnarray}
P^{\mu} &:& \qquad x^{\mu} \rightarrow x'^{\mu} = x^{\mu} + a^{\mu}; \\
M^{\mu \nu} &:& \qquad x^{\mu} \rightarrow x'^{\mu} = M^{\mu}_{\; \nu} x^{\nu}, \nonumber
\end{eqnarray}
with $M^{\mu}_{\; \nu} \in SO(d,1)$. A theory is conformally invariant if in addition it is invariant under dilatations ${\cal D}$ and special conformal transformations $K^{\mu}$:
\begin{eqnarray}
{\cal D} &:& \qquad x^{\mu} \rightarrow x'^{\mu} = \lambda x^{\mu}; \\
K^{\mu} &:& \qquad x^{\mu} \rightarrow x'^{\mu} = \frac{x^{\mu} + k^{\mu} x \cdot x}{(1 + 2 k^{\mu} x_{\mu} + k^2 x^2 )}. \nonumber
\end{eqnarray}
Here $x \cdot x = \eta_{\mu \nu} x^{\mu} x^{\nu}$. 
It is widely believed, but not proven, that any scale invariant theory is also conformally invariant. The conformal group in $d$ spacetime dimensions is $SO(d,2)$ and it is enhanced in two dimensions to an infinite dimensional group associated with analytic coordinate transformations. 

Now consider a non-relativistic field theory in $D$ spatial dimensions, with coordinates $x^i$, and time coordinate $t$. A spatially isotropic and homogeneous theory is invariant under translations and spatial rotations, i.e.
\begin{eqnarray}
H &:& \qquad t \rightarrow t' = t + a; \label{sym-nr} \\
P^{i} &:& \qquad x^{i} \rightarrow x'^{i} = x^{i} + a^{i}; \nonumber \\
L^{ij} &:& \qquad x^{i} \rightarrow x'^{i} = L^{i}_{\; j} x^{j}, \nonumber
\end{eqnarray}
with $L^{i}_{\; j} \in SO(D)$. These transformations can be augmented by Galilean boosts $C^i$ which act as
\be
C^i : \qquad x^i \rightarrow x'^i = x^i - v^i t.
\ee
The resulting Galilean group is a group contraction of the Poincar\'{e} group. The Galilean algebra admits a central extension such that
\be
\left [ C^i, P^j \right ] = M \delta^{ij}.
\ee
$M$ is viewed as the particle number or non-relativistic mass. 

A Lifshitz invariant theory is spatially isotropic and homogeneous and admits the non-relativistic scaling symmetry
\begin{equation}
{\cal D}_z : \qquad t \rightarrow t' = \lambda^z t; \qquad x^{i} \rightarrow x'^{i} = \lambda x^i. \label{sym-lif}
\end{equation}
The parameter $z$ is the dynamical exponent. The symmetry group consisting of $(H,P^i,L^{ij}, {\cal D}_z)$ will be denoted as ${\bf Lif_D(z)}$; it is not 
a subgroup or contraction of a conformal group and Lifshitz theories do not admit Galilean boosts. 

An immediate generalisation of Lifshitz is to spatially anisotropic but homogeneous theories, i.e. those with translational symmetries $(H,P^i)$, in which
each direction scales differently
\be
{\cal D}_{z_i} : \qquad t \rightarrow t' = \lambda t; \qquad x^{i} \rightarrow x'^i = \lambda^{z_i} x^i,
 \ee
where $z_i$ are the dynamical exponents; one exponent may always be chosen to be one, by rescaling the parameter $\lambda$. We will denote this group as ${\bf Lif_D( \{ z_i \})}$. 

\bigskip

There is another non-relativistic and scale invariant symmetry group, the Schr\"{o}dinger group. This can be realised in $(D+2)$ dimensions, with $D$ spatial coordinates $x^i$ and two light-cone coordinates $x^{\pm}$. The Schr\"{o}dinger group ${\bf Sch_D(z)}$ consists of spatial translations $P^i$ and spatial rotations $L^{ij}$, as well translations of the light-cone coordinates
\be
H : \qquad x^+ \rightarrow x'^+ = x^+ + a; \qquad M : \qquad x^- \rightarrow x'^- = x^- + a,
\ee
and Galilean boosts $C^i$ and dilatation ${\cal D}_z$ acting as 
\begin{eqnarray}
C^i &:& \qquad x^i \rightarrow x'^i = x^i - v^i x^+, \qquad x^- \rightarrow x'^- = x^- - v^i x^i; \\ 
{\cal D}_{z} &:& \qquad x^{i} \rightarrow x'^i = \lambda x^i, \qquad x^{+} \rightarrow x'^+ = \lambda^z x^+, \qquad x^- \rightarrow  x'^- = \lambda^{2-z} x^-.  \nonumber 
\end{eqnarray}
For any value of $z$ this symmetry group is a subgroup of the conformal group in $(D+2)$ spacetime dimensions, $SO(D+2,2)$. We should note that in the literature the terminology Schr\"{o}dinger is sometimes reserved for only the $z=2$ case, i.e. $Sch_D(2)$. 

In the case of $z=2$ the symmetry group can be extended to include one special conformal symmetry $K$
\begin{eqnarray}
K &:& \qquad   x^i \rightarrow x'^i = \frac{x^i}{1 + k x^+}, \qquad x'^+ = \frac{x^+}{1 + k x^+}; \\
&& \qquad x^- \rightarrow x'^- = \frac{
x^- + \frac{1}{2} k x \cdot x}{1 + k x^+}, \nonumber
\end{eqnarray}
with $x \cdot x = 2 x^+ x^- + x^i x^i$. This is the $K^-$ special conformal symmetry of the $(D+2)$ dimensional conformal group. 

The name of the Schr\"odinger group originates from the fact that it is the maximal kinematical symmetry group of the free Schr\"odinger equation \cite{Niederer:1972zz}. 
The Schr\"{o}dinger group is more conventionally realised in $(D+1)$ dimensions with $D$ spatial coordinates $x^i$ and a time coordinate $t$; it consists of the centrally extended Galilean group $(H,P^i,L^{ij},C^i,M)$ together with dilatations ${\cal D}_z$ and the special conformal symmetry $K$ when $z=2$. The realisations in $(D+2)$ dimensions and $(D+1)$ dimensions can be connected as follows: the light-cone momentum $M$ plays the role of the mass parameter in the Schr\"{o}dinger algebra and the light-cone coordinate $x^+$ plays the role of time. In realistic non-relativistic theories the mass spectrum is discrete. Such a discrete spectrum for the light cone momentum can be obtained by compactifying the light-cone coordinate $x^-$, but such a discrete light-cone quantisation (DLCQ) is problematic.

\section{Gravity duals of non-relativistic theories} \label{three}

In this section we discuss the construction of geometries realising non-relativistic symmetry groups holographically. 

\subsection{Lifshitz geometry}

We begin with the Lifshitz symmetry group, ${\bf Lif_D(z)}$. The Lifshitz geometry \cite{Kachru:2008yh}
\be
ds^2 = \frac{dr^2}{r^2} + \frac{dx^i dx_i}{r^2} - \frac{dt^2}{r^{2z}} \label{geom-lif}
\ee
manifestly realises the desired symmetry group on a spacetime with $(D+2)$ dimensions. The metric admits the 
translations and spatial rotations (\ref{sym-nr}) as isometries and the scaling symmetry (\ref{sym-lif}) is also realised as a metric isometry
\be
{\cal D}_z: \qquad r \rightarrow r' = \lambda r, \qquad t' = \lambda^z t; \qquad x^{i} \rightarrow x'^{i} = \lambda x^i.
\ee
When $z=1$ the metric is anti-de Sitter and has full relativistic symmetry. 

The Ricci tensor for (\ref{geom-lif}) is
\begin{eqnarray}
R_{rr} &=& - \frac{1}{r^2} (z^2 + d - 2); \qquad
R_{tt} = \frac{1}{r^{2z}}  z (z+d-2); \\
R_{ij} &=&  -\frac{1}{r^2} (z+d-2) \delta_{ij}. \nonumber
\end{eqnarray}
The strong energy condition states that for any future directed timelike vector $v^{m}$ $R_{mn} v^{m} v^{n} \ge 0$ and this condition is clearly satisfied for Lifshitz for $z \ge 1$. 
One can similarly show that the null energy condition, $G_{mn} k^{m} k^{n} \ge 0$ for any future pointing null vector field, is also satisfied for $z \ge 1$. Therefore there is no obstruction to supporting the Lifshitz geometry with physically reasonable matter for $z \ge 1$. Pathologies in realising Lifshitz with $z < 1$ holographically caused by the violation of the null energy condition were discussed in \cite{Hoyos:2010at}.

\subsection{Bottom up models for Lifshitz}

The Lifshitz geometry (\ref{geom-lif}) has an anisotropic curvature tensor and solves the Einstein equations with a non-trivial stress energy tensor. A simple way to obtain a Lifshitz solution is by coupling Einstein gravity to a massive vector $A_{\mu}$ \cite{Taylor:2008tg}. We consider an action
\be
S = \frac{1}{16 \pi G_{d+1}} \int d^{d+1} x \sqrt{-g} \left ( R + d (d-1) - \frac{1}{4} F^2 - \frac{1}{2} M^2 A^2 \right ), \label{mas-vc}
\ee
where $d = D+1$ and $F_{\mu \nu}$ is the vector field strength. This action admits an AdS solution with unit AdS radius. When the mass of the vector field is equal to 
\be
M^2 = \frac{ z d (d-1)^2}{z^2 + z (d-2) + (d-1)^2}
\ee
the field equations also admit a solution with Lifshitz scaling symmetry given by 
\begin{eqnarray}
ds^2 &=& dr^2 - e^{2 z r/l} dt^2 + e^{2 r /l } dx^i dx_i; \label{domain-wall} \\
A &=& \sqrt{\frac{2(z-1)}{z}} e^{z r /l} dt, \nonumber
\end{eqnarray}
with 
\be
l^2 = \frac{z^2 + z (d-2) + (d-1)^2}{d (d-1)}. 
\ee
Note that the gauge field is real only for $z \ge 1$. Throughout this review we will switch back and forth between 
Poincar\'{e} type coordinates (\ref{geom-lif}) and domain wall coordinates (\ref{domain-wall}); we will in both cases denote the radial coordinate by $r$ but we will indicate which of the two coordinate choices is being used. 

The massive vector action is completely equivalent, at least classically, to the Chern-Simons system coupling a gauge field to a massless $(d-1)$ form used to engineer Lifshitz geometries in \cite{Kachru:2008yh}. For example, in four bulk dimensions, consider the coupled system
\be
- \int d^{d+1} x \sqrt{-g} \left ( \frac{1}{4} F^2 + \frac{1}{12} H^2 + c \epsilon^{mnpq} B_{mn} F_{pq} \right )
\ee
with $H_{mnp}$ the field strength of the two form $B_{mn}$. The coupled equations of motion are
\begin{eqnarray}
\nabla_{m} F^{mn} &=& \frac{c}{3} \epsilon^{nrst} H_{rst}; \\
\nabla_{p} H^{pmn} &=& c \epsilon^{mnpq} F_{pq}. \nonumber
\end{eqnarray}
This system of equations has the degrees of freedom of a massive vector field. To show this, first define
\be
C_{m} = \frac{1}{3!} \epsilon_{mnpq} H^{npq}.
\ee
Closure of the three form $H$ implies that $C$ is divergenceless. Now denoting the curvature of $C$ as ${\cal F}$ the coupled equations of motion reduce to 
\be
\nabla_{m} {\cal F}^{mn} = 2c  C^{n}
\ee
The massive vector is related to the gauge field strength via ${\cal F}_{mn} = 2 c F_{mn}$ and the effective mass of the vector is given by $M^2 = 2 c$. 

\subsection{Other realisations of Lifshitz}

Lifshitz geometries also arise as solutions to higher derivative gravity theories. For example, one can find Lifshitz solutions to three-dimensional higher derivative 
gravity theories, see \cite{AyonBeato:2009nh}, following the work of \cite{Bergshoeff:2009hq}. Four-dimensional Lifshitz can be found as a solution to $R^2$ gravity, and corresponding black hole solutions can also be found analytically, see \cite{Cai:2009ac}. Analytic Lifshitz black hole solutions to higher derivative theories in various dimensions were given in 
\cite{AyonBeato:2010tm,Brenna:2011gp,Gonzalez:2011nz,Dehghani:2011hf,Liu:2012yd,Lu:2012xu,Ghanaatian:2014bpa}. Note that all these higher derivative theories are non-unitary, but asymptotically Lifshitz solutions also arise in Lovelock theories \cite{Matulich:2011ct}, which have second order field equations. 

Lifshitz can also be engineered using the bulk stress energy tensor associated with a gas of fermions \cite{Hung:2010te}; as a solution to Brans-Dicke theory \cite{Maeda:2011jj}
and using higher spin gravity \cite{Gutperle:2013oxa,Beccaria:2015iwa}.

Such realisations have the advantage that analytic black hole solutions can be found, which is not possible in most massive vector realizations of Lifshitz. However, many bottom up models seem far removed from top down models; for example, non-unitary higher derivative theories do not arise from reductions of unitary string models. 

\subsection{Spatially anisotropic Lifshitz geometries}

Spatially anisotropic Lifshitz geometries can be realised using $(d-1)$ massive vectors \cite{Taylor:2008tg}:
\begin{eqnarray}
S &=& \frac{1}{16 \pi G_{d+1}} \int d^{d+1} x \sqrt{-g} \left ( R + \Lambda - \frac{1}{4} (F^2 + 2 M^2 A^2)  \right .  \\
&&  \hspace{60mm} \left . - \frac{1}{4} \sum_a (F_a^2 + 2 M_a^2 A_a^2)   \right ) \nonumber
\end{eqnarray}
where $a = 2, \cdots (d-1)$. The anisotropic geometry
\be
ds^2 = dr^2 + e^{2 \alpha_\mu r} \eta_{\mu \nu} dx^{\mu} dx^{\nu}
\ee
solves the equations of motion with fluxes
\be
A = {\cal A} e^{\alpha_0 r} dt; \qquad 
A_{a} = \delta_{i a} {\cal A}_{i} e^{\alpha_a r} dx^i, 
\ee
provided that
\be
\alpha_a (\alpha_0 + \sum_{b \neq a} \alpha_b) = M_a^2; \qquad \alpha_0 \sum_{a=1}^{d-1} \alpha_a = M^2, 
\ee
and $a_{0} \ge \alpha_{a} \ge \alpha_{a+1} \ge 0$. The fluxes are given by
\begin{eqnarray}
{\cal A}^2 &=& \frac{2}{\alpha_0} (\alpha_0 - \alpha_1); \\
{\cal A}_i^2 &=& \frac{2}{\alpha_i} (\alpha_{i-1} - \alpha_i) \qquad i \ge 2,
\end{eqnarray}
and
\be
\Lambda = \alpha_0^2 + \sum_{a=1}^{(d-1)} \alpha_a^2 + (d-2) \left  (\alpha_0 \alpha_1 + \alpha_1^2 + \sum_{a =1}^{(d-2)} \alpha_a \alpha_{a+1} \right )
\ee
Note that there is no flux along the $x^1$ direction.

\subsection{Lifshitz with running couplings and hyperscaling violating geometries}

Solutions of Einstein-Maxwell-Dilaton systems can realise Lifshitz geometries with running scalars. A simple example is provided by the action \cite{Taylor:2008tg}:
\be
S = \frac{1}{16 \pi G_{d+1}} \int d^{d+1} x \sqrt{-g} \left ( R + \Lambda - \frac{1}{2} (\partial \phi)^2 - \frac{1}{4} e^{\lambda \phi} F^2 \right )
\ee
where $\phi$ is a scalar field and $F_{mn}$ is a gauge field strength. The equations of motion admit a Lifshitz solution for the metric which in domain wall coordinates is 
\be
ds^2 = dr^2 - e^{2 zr } dt^2 + e^{2 r } dx^i dx_i
\ee
provided that we choose the coupling and cosmological constant as
\be
\lambda^2 = 2 \frac{d-1}{z-1}; \qquad \Lambda = (d - 1 + z)(d - 2 + z)
\ee
and 
\begin{eqnarray}
F_{rt} &=& f e^{(z+ d- 1) r}; \qquad \mu f^2 = 2 (z-1) (z + d-1); \\
e^{\lambda \phi} &=& \mu e^{2 (1 -d) r}. \nonumber
\end{eqnarray}
Note that the limit $z \rightarrow 1$ is not smooth as the coupling $\lambda$ diverges in this limit. The running scalar coupling breaks the Lifshitz symmetry in the dual field theory.

\bigskip

A further generalisation is the case of hyperscaling violating ({\bf HV}) geometries, introduced in \cite{Huijse:2011ef}, for which the Einstein frame metric can be expressed as 
\be
ds^2_E = u^{- \frac{2 (D-\theta)}{D}} \left ( u^{-2 (z-1)} dt^2 + du^2 + dx^i dx_i \right ), \label{hv-geom}
\ee
where $D$ is the number of spatial directions, $z$ is the Lifshitz dynamical scaling exponent and $\theta$ is the hyperscaling violation exponent. This metric is spatially homogeneous and scale covariant, with the scaling behaviour
\begin{eqnarray}
&& x^{i} \rightarrow x'^{i} = \lambda x^i, \qquad t \rightarrow t' = \lambda^z t, \qquad
u \rightarrow u' = \lambda u, \\
&& ds^2   \rightarrow  (ds^2)' = \lambda^{\frac{2 \theta}{D}}  ds^2. \nonumber
\end{eqnarray}
The metric can only be supported by a stress energy tensor satisfying the null energy condition provided that $z > 1$. The restriction $\theta \le 0$ is required for the dual theory to be well-defined in the UV. 

HV geometries arise as solutions to general Einstein-Maxwell-Dilaton theories, see \cite{Gath:2012pg,Gouteraux:2012yr}.
Following the approach and notation of 
\cite{Chemissany:2014xpa,Chemissany:2014xsa} it is useful to express the latter in the so-called dual frame, rather than in the Einstein frame, so that
\be
\hspace{-5mm} S = \frac{1}{2 \kappa^2} \int d^{d+1} x \sqrt{-g} e^{D \xi \phi} \left ( R - \alpha (\partial \phi)^2 - Z(\phi) F^2 
 - W(\phi) B^2 - V(\phi) \right ). \label{cp}
\ee
This action admits scaling solutions supported by a scalar and a vector:
\begin{eqnarray}
ds^2 &=& dr^2 - e^{2 z r } dt^2 + e^{2 r } dx^i dx_i; \\
B &=& \frac{Q}{\epsilon Z_0} e^{\epsilon r}dt; \nonumber \\
\phi &=&  \mu r, \nonumber
\end{eqnarray}
provided that we impose the following relations on the Lagrangian parameters and parameters of the solution:
\begin{eqnarray}
V(\phi) &=& - (D - \theta) (D + z - \theta) - (z -1) \epsilon; \\
Z(\phi) &=& Z_0 e^{\frac{2 (z - \epsilon)}{\mu} \phi}; \nonumber \\
W(\phi) &=& 2 Z_0 \epsilon ( D + z - \theta - \epsilon) e^{\frac{2 (z - \epsilon)}{\mu} \phi}; \nonumber \\
Q^2 &=& \frac{1}{2} Z_0 (z-1) \epsilon; \nonumber \\
\epsilon&=& \frac{ (\alpha \mu^2 + \theta(1 + \theta) + z (z-1))}{(z-1)}. \nonumber
\end{eqnarray}
where the hyperscaling exponent in the Einstein frame is 
\be
\theta = - D \xi \mu
\ee
The independent parameters in the action are thus $(\alpha, Z_0)$ and the independent parameters in the solutions are $(z,\xi,\mu)$. The presence of three independent parameters in the solutions indicates that the general solutions are characterised not just by the dynamical exponent $z$ and the (metric) hyperscaling violation $\theta$ but by an additional third parameter $\mu$ (or, equivalently, $\epsilon$). A physical interpretation of $z$ can be given in terms of the scaling of temperature; similarly the energy density scales as $(D + z - \theta)$. The interpretation of the third parameter is more subtle. On the bulk side $\epsilon$ characterises the hyperscaling violation of the vector field. From the field theory perspective, however, we will show later that the parameter $\mu$ is most naturally understood in terms of a dimensionally running coupling.

We can recover cases discussed earlier as follows. The Einstein-Proca model is obtained by setting $\alpha = 0$, $\theta = 0$ and hence $z = \epsilon$. The case of Lifshitz with a running coupling discussed above is obtained by setting $\xi=0$ (so the hyperscaling violation vanishes, $\theta = 0$) and choosing $\alpha$ and $\mu$ such that $\epsilon = (D+z)$.
More generally, the vector field has an unbroken $U(1)$ gauge invariance whenever $W(\phi) = 0$ (with $Q \neq 0$).

Non-conformal $Dp$-branes have $z=1$, $\theta = (p-3)^2/(p-5)$, $D=p$ together with \cite{}
\be
\alpha = \frac{4 (p-1)(4-p)}{(7-p)^2}, \qquad
\xi = \frac{2 (p-3)}{p (7-p)}, \qquad \mu = \frac{(7-p)(p-3)}{2 (5-p)}. 
\ee
Note that $\theta \le 0 $ for $p < 5$; there is no decoupling limit for $Dp$-branes with $p > 5$. 

\subsection{Schr\"{o}dinger geometries}

In realising the Schr\"{o}dinger group geometrically, it is useful to recall that the Schr\"{o}dinger group is a subgroup of the conformal group. Therefore one can look for deformations of Anti-de Sitter space which reduce the symmetry to the Schr\"{o}dinger group. ${\bf Sch_D(z)}$ is realized as the isometry group of the following $(D+3)$-dimensional geometry \cite{Balasubramanian:2008dm,Son:2008ye}:
\be
ds^2 = - \frac{b^2 (dx^+)^2}{r^{2 z}} + \frac{1}{r^2} \left ( dr^2 + dx^i dx_i + 2 d x^+ dx^- \right ). \label{schg}
\ee
The dilatation symmetry is realized as
\be
{\cal D} :  \qquad r \rightarrow \lambda r, \qquad x^i \rightarrow \lambda x^i, \qquad x^+ \rightarrow \lambda^{z} x^+, \qquad
x^- \rightarrow \lambda^{2-z} x^-. 
\ee
The parameter $b$ is arbitrary and can be removed by the rescalings
\be
x^+ \rightarrow x'^+ = b x^+, \qquad
x^- \rightarrow x'^- = x^-/b.
\ee
We will find it useful to retain the parameter $b$ explicitly, with $b=0$ being the anti-de Sitter metric and $b \neq 0$ corresponding to a deformation of the conformal theory. 

\bigskip

Schr\"{o}dinger geometries arise as solutions to models consisting of Einstein gravity coupled to massive vector fields \cite{Son:2008ye}:
\be
S = \frac{1}{16 \pi G_{d+1}} \int d^{d+1} x \sqrt{-g} \left ( R + d (d-1)  - \frac{1}{4} F^2 - \frac{1}{2} M^2 A^2 \right ) \label{sch}
\ee
where $F$ is the field strength of $A$ and 
\be
M^2 = z (z + d -2).
\ee
The vector field supporting the geometry is 
\be
B = \sqrt{ \frac{2 (z-1)}{z}} \frac{b}{r^z} dx^+ \label{vector}
\ee
For $z < 1$ we obtain solutions by the analytic continuations $x^+ \rightarrow i x^{+}$ together with $x^- \rightarrow - i x^-$, or equivalently $b \rightarrow i b$. Embeddings of Schr\"{o}dinger solutions into string theory were discussed in \cite{Maldacena:2008wh,Herzog:2008wg,Kraus}. Note that
Schr\"{o}dinger geometries also arise as solutions to topologically massive gravity theories in three bulk dimensions; holography for such solutions was analysed in detail in \cite{Guica:2010sw}.

\bigskip

The $z=0$ Schr\"odinger group cannot be realised using a massive vector since (\ref{vector}) is singular in this limit but can instead be realised using a massless scalar field. Consider a Lagrangian
\be
S = \frac{1}{16 \pi G_{d+1}} \int d^{d+1} x \sqrt{-g} \left ( R + d ( d-1) - \frac{1}{2} (\partial \chi)^2 \right ). \label{chi}
\ee
The equations of motion admit a $z=0$ Schr\"{o}dinger solution supported by the massless scalar field profile
\be
\chi = \sqrt{(d-2)} b x^+ \label{phi0}
\ee
${\bf Sch_D(0)}$ is related to a solution with Lifshitz symmetry in one less dimension \cite{Balasubramanian:2010uk,Costa:2010cn,Narayan:2011az}. This can be shown by rewriting the metric as
\be
ds^2 = \frac{1}{r^2} \left ( dr^2 + dx_i dx^i - \frac{(dx^-)^2}{b^2 r^2} \right ) + b^2 ( dx^+ + \frac{dx^-}{b^2 r^2})^2. \label{sch0}
\ee
Dimensionally reducing along the $x^+$ direction results in a $(D+2)$ dimensional metric with vector field $A$
\begin{eqnarray}
ds^2 &=& \frac{1}{r^2} \left ( dr^2 + dx_i dx^i - \frac{(dx^-)^2}{b^2 r^2} \right ); \\ 
A &=& \frac{1}{b^2 r^2} d x^-. \nonumber
\end{eqnarray}
The reduced system has Lifshitz symmetry with dynamical exponent two and corresponds to a massive vector model. Note however that strictly speaking the scalar field cannot be dimensionally reduced along the $x^+$ direction as the scalar field is linearly proportional to $x^+$. Furthermore, $x^+$ is asymptotically null, and thus from the field theory perspective the reduction is over a lightlike circle.  

The reduced system of the metric and massive vector equations is not sufficient to solve the higher dimensional equations unless an additional constraint is imposed \cite{Costa:2010cn}; the reduction is not a consistent truncation. Moreover, the action (\ref{chi}) does not seem to follow from a consistent truncation of a ten or eleven dimensional supergravity theory. 
However an extended system of dilaton and axion can be obtained as a consistent truncation of supergravity and this action admits solutions of the type (\ref{sch0}) which reduce to 
Lifshitz solutions \cite{Donos:2010tu,Cassani:2011sv,Halmagyi:2011xh,Petrini:2012bh}:
\be
S = \frac{1}{16 \pi G_{d+1}} \int d^{d+1}x \sqrt{-g} \left ( R + d (d-1) - \frac{1}{2} (\partial \chi)^2 - \frac{1}{2} e^{2 \chi} (\partial \phi)^2 \right ) \label{axion-dilaton}
\ee
The equations of motion for this model has solutions of the type (\ref{sch0}) and (\ref{phi0}), with $\phi$ constant. We will discuss the holographic dictionary and the reduction of this model in section \ref{newton}. 

\subsection{IR instabilities}

The Lifshitz metric (\ref{geom-lif}) has no curvature singularities as all local invariants constructed from the Riemann tensor are finite and constant everywhere. However, the metric is not geodesically complete and infalling probes generically experience large tidal forces as $r \rightarrow \infty$; it is said to have a null curvature singularity.  
The implications of the null curvature singularity in the Lifshitz geometry have been explored in a number of papers.

In \cite{Copsey:2010ya} it was pointed out that the initial value problem is not well posed for Lifshitz and that generic normalisable states would seem to evolve in such a way as to violate Lifshitz asymptotics in finite time. If one enforces the desired asymptotics, then the dynamics of the bulk fields deep in the interior seems to be over-constrained. Related issues are discussed in \cite{Keeler:2013msa,Keeler:2014lia,Keeler:2015afa}. Note however that to set up the initial value problem and dynamics in holography requires a real-time dictionary. This has been developed in detail for anti-de Sitter \cite{Skenderis:2008dh,Skenderis:2008dg} and the analogue setup for Lifshitz may well resolve the apparent pathologies. 

It was argued in \cite{Horowitz:2011gh} that strings will become infinitely excited as they propagate through the singularity, and thus the interior of the Lifshitz geometry would be expected to receive large corrections in string theory. However, it was later shown in \cite{Bao:2012yt} that scattering slows the string and prevents such divergent mode production. In \cite{Harrison:2012vy} it was argued that higher derivative corrections to the effective action may drive the deep interior region to be replaced by a relativistic fixed geometry such as $AdS_2 \times R^{d-1}$. These ideas have been explored further in \cite{Knodel:2013fua}. While the fate of the Lifshitz singularity is not completely understood, the singularity fortunately does not seem to play in many holographic computations of interest. 

\subsection{Top down models for Lifshitz}

Top down models in holography are desirable in order to have an explicit description of the dual field theory in terms of e.g. decoupling limits of branes. As a first step one would like to find explicit solutions of ten and eleven dimensional supergravity which are (warped) products of Lifshitz with compact geometries. As a second step it would be desirable to find associated consistent truncations, i.e. reductions of ten and eleven dimensional supergravity over compact manifolds consisting of gravity coupled to a small number of matter fields. Such consistent truncations would generally be expected to contain the fields of the bottom up models, together with additional fields needed to ensure that solutions of the lower dimensional theory are always solutions of the top down theory.  

Unfortunately very few top down realisations of Lifshitz geometries have been found.
Early attempts at finding string theory embeddings of Lifshitz and some no-go theorems for Lifshitz embeddings can be found in \cite{Azeyanagi:2009pr,Li:2009pf,Blaback:2010pp}. In \cite{Hartnoll:2009ns} string theory constructions of Lifshitz  using F-theory, polarised branes and charged Fermi gases were outlined; making such constructions explicit and quantitative is however challenging.   Other brane and string theory constructions of Lifshitz can be found in \cite{Singh:2010zs,Singh:2012un,Dey:2012tg,Narayan:2012hk}. 

 A systematic search for top down Lifshitz supergravity solutions was carried out in \cite{Donos:2010tu}. Families of supersymmetric $z=2$ embeddings were found, corresponding to reductions of $z=0$ Schr\"{o}dinger solutions over a null circle. Such reductions were further explored in a number of works \cite{Cassani:2011sv,Halmagyi:2011xh,Petrini:2012bh}.  Recently a $z=2$ supersymmetric Lifshitz solution of the STU model coupled to one hypermultiplet was constructed \cite{Cardoso:2015wcf}; related non-supersymmetric solutions interpolated between Lifshitz and a near-horizon Nernst brane were also found in this work. As isolated embedding  with $z \sim 39$ was found in \cite{Donos:2010ax}; it remains unclear why this specific value of $z$ was realised. 
 
An alternative approach to embedding Lifshitz into top down models was initiated in \cite{Gregory:2010gx}. This work constructed solutions of the six-dimensional Romans massive theory consisting of four-dimensional Lifshitz, with any value of $z$, cross a compact two-dimensional hyperbolic manifold and solutions of five-dimensional gauged supergravity consisting of three-dimensional Lifshitz, with any value of $z$, cross a compact two-dimensional hyperbolic manifold. The former can be uplifted to solutions of the massive IIA theory in ten dimensions while the latter can be uplifted to solutions of type IIB supergravity. All the $z \neq 1$ solutions break supersymmetry and therefore would be expected to be unstable. Stability, renormalization group flows and black holes in these models were explored in \cite{Braviner:2011kz,Barclay:2012he,Burda:2014jca}.

In \cite{Balasubramanian:2011ua} candidate string theory realisations were proposed for $z=2$ Lifshitz Chern-Simons gauge theories studied in \cite{Mulligan:2010wj}. It was argued that the field theories could be realised as light cone reductions of deformations of ${\cal N} = 4$ SYM. Using the standard holographic dictionary for the latter, one could then identify the bulk fields implementing the required deformations. In \cite{Copsey:2011ek} it was however noted that the Lifshitz duals have not only the usual null singularity but also regions of closed timelike curves. 
 
\subsection{Correlation functions}

It is well-known that in a conformal field theory (Euclidean) two point functions of operators are determined up to an overall normalisation by the operator dimension. Three point functions are also determined, up to overall coefficients, and the first non-trivial functional dependence in correlation functions arises for four point functions. 

In a Lifshitz invariant theory the symmetry is not sufficient to determine completely even two point functions of scalar operators: for a scalar operator ${\cal O}(t, \vec{x})$ of Lifshitz scaling dimension $\Delta_L$ the Euclidean two point function at separated points is 
\be
\langle {\cal O}(t, \vec{x}) {\cal O} (0,0) \rangle = \frac{1}{ | \vec{x}|^{2\Delta_L}} {\cal F} \left (  \frac{ |\vec{x}|^z} {t} \right ), 
\ee
where ${\cal F}(w)$ is an arbitrary function and $|\vec{x}|^z/t$ is invariant under scale transformations. Equivalently, Fourier transforming the operator to momentum space we find that
\be
\langle {\cal O}(\omega, \vec{k}) {\cal O} (0,0) \rangle = k^{2\Delta_L - (d+z -1)} \tilde{\cal F} \left (  \frac{\omega}{k^z} \right ), \label{mom-sca}
\ee

To obtain insight into the interpretation of Lifshitz two point functions it is useful to consider the following free field theory model in three dimensions:
\be
S =  \frac{1}{2} \int dt d^2 x \left [  (\partial_t \varphi)^2 - \kappa^2 (\nabla^2 \varphi)^2 \right ],
\ee 
where $\varphi(t,x)$ is a massless scalar field. The action manifestly breaks Lorentz symmetry but has $z=2$ Lifshitz symmetry. From the classical field equation 
\be
\left ( \partial_t^2 + \kappa^2 \nabla^4 \right ) \varphi = 0,
\ee
one can infer that the Fourier transform of the Euclidean propagator for $\varphi$ is
\be
\tilde{G} (\omega,k) = \frac{1}{(\omega^2 + \kappa^2 k^4)} 
\ee
Transforming back into position space, the propagator has a short distance logarithmic divergence which is often regularised as (see \cite{Ardonne:2003wa})
\be
G_R(t,\vec{x}) \equiv G(t,\vec{x}) - G(0,a) = - \frac{1}{8 \pi \kappa} \left [ \ln \left ( \frac{|\vec{x|}^2}{a^2} \right ) + \Gamma \left ( 0,\frac{|\vec{x}|^2}{4 \kappa t} \right ) \right ],
\ee
with $\Gamma(0,z)$ the incomplete Gamma function. The regularised propagator behaves as
\be
G_{R}(0,\vec{x}) = - \frac{1}{4 \pi \kappa} \ln \left  ( \frac{|\vec{x}|}{a} \right  ); \qquad
G_R(t,a) = - \frac{1}{8 \pi \kappa} \ln \left ( \frac{4 \kappa |t|}{a^2 \gamma}  \right ).
\ee
Charge operators in the theory can be defined as 
$
{\cal O}_n = e^{-in \varphi}
$
and their correlation functions are given by
\be
\langle {\cal O}_{n}^{\dagger} (t,\vec{x}) {\cal O}_{n}(t',x') \rangle = e^{n^2 G_R(t-t',\vec{x} - \vec{x}')}.
\ee
The equal time correlation functions behave as 
\be
\langle {\cal O}_{n}^{\dagger} (0, \vec{x}) {\cal O}_{n}(0,0) \rangle = \left ( \frac{a}{|\vec{x}|} \right )^{\frac{n^2}{4 \pi \kappa}}
\ee
which implies that the Lifshitz scaling dimension is $\Delta_L = n^2/8 \pi \kappa$. For $|x -x'| \rightarrow a$ 
\be
\langle {\cal O}_{n}^{\dagger} (t, 0) {\cal O}_{n}(0,a) \rangle = \left ( \frac{a^2 \gamma}{4 \kappa |t|} \right )^{\frac{n^2}{8 \pi \kappa}}
\ee
which is consistent with the scaling exponent $z=2$. 

Moving to real time physics, the analytic continuation of the Green function gives 
\be
\tilde{G}(\omega,k) = - \frac{1}{(\omega^2 - \kappa^2 k^4)},
\ee
i.e. the propagator has two simple poles on the real axis at $\omega = \pm \kappa k^2$. By appropriate choices of integration contour
one can thence define the usual Feynman, retarded and advanced propagators. A conceptual interpretation of these poles follows from realising that the field equation is the square of the Schr\"{o}dinger equation for a non-relativistic scalar in three dimensions, and the Green function for the latter describes diffusion in imaginary time. 

\bigskip

Now let us consider correlation functions in the holographic models. As in AdS/CFT a scalar operator is dual to a bulk scalar field. In the vacuum of the dual theory 
correlation functions for these scalar operators may be computed working perturbatively around the Lifshitz background. Let us compute the Euclidean two point function for an operator ${\cal O}_{\phi}$ dual to a bulk field $\phi$ which has an action
\be
S_{\phi} = - \frac{1}{2} \int d^{d+1} x \sqrt{g} \left (  (\partial \phi)^2 + m^2 \phi^2 + \cdots \right ); \label{bulk-act}
\ee
higher order interactions are denoted by ellipses but only quadratic terms are required to compute the two point function. 
The linearized field equation for the scalar field in the Euclidean Lifshitz background in the coordinates of (\ref{geom-lif})  is then
\be
\left (r^2 \partial_r^2 - (d+z-2) r \partial_r + r^{2z} \partial_t^2 + r^2 \partial_i \partial^i  - m^2 \right )\phi(r,t,x^i) =0.
\ee
Fourier transforming with respect to $t$ and $x^i$ the equation becomes an ordinary differential equation
\be
\left ( r^2 \partial_r^2 - (d+z-2) r \partial_r - \omega^2 r^{2z} - k^2 r^2 - m^2 \right )\phi(r,\omega,k) =0. \label{eom}
\ee
It is difficult to find analytic solutions for this equation for general values of $z$ but analytic solutions in terms of hypergeometric functions 
exist for integral $z$. Asymptotic analysis of the equation of motion near the conformal boundary $r \rightarrow 0$ implies that the two independent solutions behave
as
\be
\phi \sim r^{\Delta_{+}} \phi_{+}  + r^{\Delta_-} \phi_- 
\ee
where 
\be
\Delta_{\pm} = \frac{1}{2} (d + z-1) \pm \frac{1}{2}  \sqrt{ (d+z-1)^2 + 4 m^2} \equiv \frac{1}{2} (d + z-1) \pm N_z.
\ee
The requirement that $\Delta_{\pm}$ is real enforces
\be
m^2 \ge - \frac{1}{4} (d + z -1)^2,
\ee
which is the extension of the Breitenlohner-Freedman bound to Lifshitz. Following the same arguments as for AdS/CFT in \cite{Witten:1998qj}, one can interpret $\phi_-$ as the non-normalisable mode and $\phi_+$ the normalisable mode for a dual operator of Lifshitz dimension $\Delta_{+}$. 

For the relativistic case of $z=1$ the general solution to (\ref{eom}) is given in terms of Bessel functions:
\be
\phi(r, q) = A (q) r^{\frac{d}{2}} K_{N_1}(q r) + B (q) r^{\frac{d}{2}} I_{N_1}(q r)
\ee
with $A$ and $B$ arbitrary coefficients and $q^2 = \omega^2 + k^2$. The first solution is regular in the interior. For general $z$ the zero frequency limit is
\be
\phi(r, 0, k) = A (k)  r^{\frac{1}{2}(d+z -1)} K_{N_z}(k r) + B (k) r^{\frac{1}{2}(d+z-1)} I_{N_z}(k r).
\ee
At zero frequency the system therefore only feels the dynamical exponent through a modified effective dimension which sends $d \rightarrow (d+z-1)$. 

For $z=2$ the equation of motion is solvable analytically in terms of confluent hypergeometric functions
\begin{eqnarray}
\phi(r, \omega,k) &=& \ A(\omega, k) r^{\Delta_{-}} e^{- \frac{1}{2} \omega r^2}  U \left ( \frac{k^2}{4 \omega} + \frac{1}{2} (1 - N_2); (1 - N_2); \omega r^2 \right ) \label{reg}  \\
&& + B(\omega,k)  r^{\Delta_{-}} e^{- \frac{1}{2} \omega r^2}  M \left ( \frac{k^2}{4 \omega} + \frac{1}{2} (1 + N_2); (1 + N_2); \omega r^2 \right )  \nonumber
\end{eqnarray}
Here $M(a,b,z)$ and $U(a,b,z)$ are confluent hypergeometric functions of the first and second kind respectively. The U solution is regular in the interior, i.e. for $r \rightarrow \infty$, 
while the M solution is not. 

To work out the two point function one needs to renormalise the action (\ref{bulk-act}) using the asymptotic solutions of the scalar fields to compute the counterterms. The details of this analysis depend on $z$, the mass $m^2$ and the dimension. Expressing the asymptotic expansion in the usual way as 
\be
\phi(r,x) = r^{\Delta_-} \left ( \phi_0(x) + \cdots \right ) + r^{\Delta_+} \left ( \phi_{\Delta_+}(x) + \cdots \right )
\ee
then the renormalized action always contains the following counterterm
\be
S_R = - \frac{1}{2} \int d^{d+1} x \sqrt{g} \left (  (\partial \phi)^2 + m^2 \phi^2  \right ) + \int d^{d} x \sqrt{h} \left ( \frac{1}{2} \Delta_{- } \phi^2 + \cdots \right ),
\ee
since this counterterm is needed to remove the leading divergence. Now using the definition of the dual operator one point function as 
\be
\langle {\cal O}_{\phi} \rangle = - \frac{\delta S_{R}^{ \rm onshell}} {\delta \phi_0}
\ee
we can show that 
\be
\langle {\cal O}_{\phi} \rangle = \frac{1}{2} (\Delta_{+} - \Delta_{-}) \phi_{\Delta_+} + \cdots
\ee
where the ellipses denote possible terms involving local derivatives of the source $\phi_0$. The latter are only computable by working out, case by case, the full renormalized action. 
The two point function is then computed by functionally differentiating once more with respect to the source, i.e. 
\be
\langle {\cal O}_{\phi} (x) {\cal O}_{\phi} (y) \rangle = - \frac{1}{2} (\Delta_{+} - \Delta_{-}) \frac{\delta \phi_{\Delta_+} (x)}{\delta \phi_0(y)} + \cdots  \label{cf1}
\ee
The asymptotic expansion of the regular solution for $z=2$ given in (\ref{reg}) then allows us to read off
\be
\langle {\cal O}_{\phi} (\omega,k) {\cal O}_{\phi} (-\omega,-k) \rangle = - N_2 \omega^{N_2} \frac{\Gamma (-N_2)}{\Gamma(N_2)} \frac{ \Gamma \left ( \frac{1}{2}(1 + N_2) 
+ \frac{k^2}{4 \omega} \right )}{\Gamma  \left ( \frac{1}{2}(1 - N_2) + \frac{k^2}{4 \omega } \right ) } + \cdots \label{cff1}
\ee
where the ellipses denote local contributions. This expression is consistent with $z=2$ Lifshitz scaling, see (\ref{mom-sca}), but the momentum and frequency dependence is complicated. 

Our analysis applies for generic values of the scalar mass but the result is singular whenever $N_2$ is an integer, reflecting the need for more careful treatment of the regular solution (\ref{reg}) and of holographic renormalization in such cases.  For example, for a massless scalar field in four bulk dimensions, one finds that 
\begin{eqnarray}
\langle {\cal O}_{\phi} (\omega,k) {\cal O}_{\phi} (-\omega,-k) \rangle &=& - \frac{1}{16} (4 \omega^{2} - k^4) \psi \left ( \frac{3}{2} + \frac{k^2}{4 \omega} \right ) \label{cf2} \\
&& \qquad - \frac{1}{4} k^2 \omega 
- \frac{1}{16} (4 \omega^{2} - k^4) \log(\omega) + \cdots \nonumber
\end{eqnarray}
where the ellipses denote local terms. 

The correlators (\ref{cff1}) and (\ref{cf2}) are universal for any operator dual to a minimally coupled scalar field in the Lifshitz background and therefore constitute non-trivial predictions of holographic Lifshitz models. It was shown in \cite{Kachru:2008yh} that the Fourier transform of (\ref{cf2}) at large spatial separation (for fixed temporal separation) behaves as
\be
\langle {\cal O}_{\phi} (t,x) {\cal O}_{\phi} (0,0) \rangle \sim \frac{1}{|x|^8} \label{x8}
\ee
which is in agreement with the behaviour in the free field model above. Note however that the full analytic structure of the holographic correlators is much more complicated than in the free field model. For example, analytically continuing to real time we note that (\ref{cff1}) has poles at
\be
\frac{1}{2} (1+ N_2) + i \frac{k^2}{4 \omega} = - n,
\ee
with $n$ an integer, i.e. along the negative imaginary axis for the frequency
\be
\omega = - i \frac{k^2}{(4n +2 (1+ N_2))}.
\ee
Each such pole is associated with dissipative behaviour in the correlation function c.f. the quasiparticle poles in the free field model. Analogous behaviour is found for the pole structure of fermionic correlators, see \cite{Korovin:2011kw,Alishahiha:2012nm}. 
Moreover, an analysis of the linearised fluctuations of the metric and vector field of the Einstein-Proca action around the $z=2$ Lifshitz solution shows similar behaviour: quasi-normal modes are situated in the lower half-plane of complex frequency \cite{Zingg:2013xla}. It was also proposed in \cite{Zingg:2013xla} that the correlation functions had the structure of a logarithmic field theory with anisotropic scaling symmetry. Further analysis of Lifshitz quasi normal modes can be found in \cite{Sybesma:2015oha}.

It was argued in \cite{Keeler:2015afa} that, despite the apparently complicated pole structure, the imaginary part of the retarded scalar Green's function is exponentially suppressed at low frequencies. It would be interesting to analyse this further by explicitly computing retarded correlators using the real-time methods of \cite{Skenderis:2008dh,Skenderis:2008dg}.

Other aspects of correlation functions in Lifshitz have been discussed in the literature. Alternative (non-Dirichlet) boundary conditions for scalars were explored in \cite{Andrade:2012xy,Keeler:2012mb}, and related bounds on the scalar operator dimensions were proposed. Analogous analysis for metric and vector field fluctuations was carried out in \cite{Andrade:2013wsa}. 

\bigskip

There is a simple geometric way to understand why the fixed time correlators must take the form (\ref{x8}) for large spatial separation. In AdS/CFT in the limit of large mass $m$ (and hence large scaling dimension) it has been proposed that one can compete the two point function geometrically as
\be
\langle {\cal O}_{\phi} (t,x) {\cal O}_{\phi} (0,0) \rangle = e^{- m {\cal L} (t,x;0,0)}
\ee
where ${\cal L}(t,x;0,0)$ is the renormalized length of a bulk geodesic connecting the two boundary points $(t,x)$ and $(0,0)$. For a fixed time correlator the geodesic lies entirely within a constant time surface, given the symmetry of the bulk geometry. Using Poincar\'{e} coordinates, 
\be
ds^2 = \frac{dr^2}{r^2} + \frac{1}{r^2} \left ( dx^i dx_i - dt^2 \right )
\ee
this implies that the regularised length of the geodesic is
\be
{\cal L}_{\rm reg} = 2 \ln \left ( \frac{x}{\epsilon} \right ) 
\ee
where we impose the usual cutoff $r = \epsilon$. We now renormalise the geodesic length by subtracting the divergent boundary contributions i.e. 
\be
{\cal L} = {\cal L}_{\rm reg} + 2 \ln \epsilon = 2 \ln x
\ee
and hence 
\be
\langle {\cal O}_{\phi} (t,x) {\cal O}_{\phi} (t,0) \rangle = |x|^{-2 m}
\ee
Recalling that the relativistic conformal dimension behaves as $\Delta \sim m$ for large $m$ we reproduce the standard expression for the conformal two point function. Returning to Lifshitz, for fixed time correlators the geodesic still lies entirely within a constant time surface. However, such constant time surfaces are hyperbolic and insensitive to the dynamical exponent $z$, and therefore the form of the two point function is unchanged from the relativistic case. 

\bigskip

To end this section, let us note that imposing Galilean invariance constrains correlation functions further. For example, in the case of Schr\"{o}dinger invariance with $z=2$ the symmetry group suffices to fix the two point functions entirely. For an operator ${\cal O}$ of scaling dimension $\Delta$ and mass $m$ the two point function is 
\be
\langle {\cal O} (t,x) {\cal O} (0,0) \rangle = \frac{C}{t^{\Delta}} \exp [ - \frac{m |x|^2}{2t} ], 
\ee
while restrictions on higher correlation functions can be found in \cite{Henkel}. 

\section{Hydrodynamics} \label{four}

\subsection{Implications of Lifshitz invariance}

A non-relativistic field theory in $(D+1)$ dimensions admits an energy density ${\cal E}$, an energy flux ${\cal E}_i$, a momentum density ${\cal P}_i$ and a symmetric spatial 
stress tensor density $t_{ij}$. These quantities can be expressed as a non-symmetric rank two tensor ${\cal T}_{\mu \nu}$, with $ {\cal T}_{00} = {\cal E}$; $ {\cal T}_{i0} = {\cal E}_i$; 
$ {\cal T}_{0i} = {\cal P}_i$ 
and $ {\cal T}_{ij}  = t_{ij}$. The conservation equations (diffeomorphism Ward identities) in flat space can then be expressed as 
\be
\partial_t {\cal E} + \partial_i {\cal E}^i = \partial_t {\cal P}_{i} + \partial^j t_{ji} = 0,  \label{dff1}
\ee
or equivalently as 
\be
\partial^{\mu}  {\cal T}_{\mu \nu} = 0.
\ee
The dilatation Ward identity associated with the Lifshitz symmetry is expressed as 
\be
(z {\cal E} + t^{i}_{i} ) = 0 \label{dw1} 
\ee
with $z$ the dynamical exponent. 

An operator is marginal with respect to the Lifshitz symmetry if its dimension is $(D+z)$, irrelevant for dimensions greater than $(D+z)$ and relevant for dimensions smaller than $(D+z)$. 
The Lifshitz scaling dimensions of the energy momentum complex operators are $[ {\cal E} ] = [t^{i}_{i} ] = (D+z)$; $[ {\cal P}^{i} ] = (d+1)$ and $[ {\cal E}^{i} ] = (d+2z -1)$. Hence the energy flux is irrelevant for $z > 1$; the energy density and the spatial stress tensor are always marginal and the momentum density is relevant for $z > 1$. Note that these scaling dimensions for the energy momentum complex are consistent with the Ward identities given above.

\subsection{Lifshitz thermodynamics}

Consider a static equilibrium state in a Lifshitz invariant theory, in which the energy density is ${\cal E}$ and the homogeneous pressure is $p$, i.e. $t_{ij} = p \delta_{ij}$. 
The fundamental thermodynamic relation implies that
\be
{\cal E} + p = T s
\ee
where $T$ is the temperature and $s$ is the entropy density. Invariance under Lifshitz scaling requires
\be
z {\cal E} = (d-1) p, \label{eos}
\ee
which is the equation of state.
One can rewrite these relations as 
\be
M = \frac{d-1}{d+z-1} T S, \label{above}
\ee
where $M$ is the mass (with ${\cal E} = dM/dV$ where $V$ is the volume) and $S$ is the entropy (with $s = dS/dV$). The scaling dimension of the temperature is $[T] = z$
while the scaling dimensions of the energy density and pressure are both (as given in the previous section) $[{\cal E}] = [p] = (z + d -1)$.

The scaling relation between the temperature and the entropy is
\cite{Bertoldi:2009vn,Bertoldi:2009dt}
\be
S \sim T^{\frac{d-1}{z}},
\ee
which together with (\ref{above}) implies the first law of thermodynamics
\be
dM = T d S.
\ee
In a theory with scaling exponent $z$ and hyperscaling exponent $\theta$ it remains true that the scaling dimension of the temperature $[T]$ is $z$. However, the energy density scales as $[ {\cal E} ] = (D + z - \theta)$, as does the pressure. The fundamental thermodynamic relation then implies that the entropy density scales as 
\be
s \sim T^{\frac{D- \theta}{z}}.
\ee
Note that the hyperscaling exponent does not appear explicitly in the Ward identities (\ref{dff1}) and (\ref{dw1}), although when $\mu \neq 0$ additional terms arise in (\ref{dw1}) due to the running scalar coupling, see section \ref{hol-dict}.

\bigskip

In general mass is a difficult concept to define in general relativity and in particular the appropriate notion of mass for general asymptotics is poorly understood. The above definition of mass agrees with the mass obtained from holographic renormalization, i.e. from renormalising the onshell action with appropriately covariant counterterms and then defining the dual field theory energy momentum complex. For asymptotically locally anti-de Sitter spacetimes, it was shown in \cite{Papadimitriou:2005ii} that the mass obtained by holographic renormalization is always in agreement with that defined by Wald's approach \cite{Wald:1993nt}. The proof given in \cite{Papadimitriou:2005ii} should in principle be extendable to asymptotically locally Lifshitz spacetimes, although this analysis has not yet been done. 

In the context of Lifshitz holography, the mass defined by the holographic renormalised dual stress energy tensor is the natural mass to use but other definitions of mass for Lifshitz have appeared in the literature. A construction of a gauge invariant quasilocal mass was given in \cite{Wang:2008jy}; this construction is applicable to general asymptotics and would therefore define  a mass for Lifshitz. Another alternative proposal for the mass of Lifshitz black holes was given in \cite{Devecioglu:2011yi} although it was later shown in \cite{Gim:2014nba} that this mass was not consistent with the first law of thermodynamics. A notion of thermodynamic mass, based on the consistency between generalised Smarr relations and the first law of thermodynamics, was put forward and analysed recently in \cite{Brenna:2015pqa}. The analysis of \cite{Brenna:2015pqa} is based on the conjecture that the mass scales as $L^{d-2}$, where $L$ is a fiducial length scale for the system, in contrast to the scaling as $L^{d+z-1}$ given above. 

\subsection{Hydrodynamics}

The equation of state (\ref{eos}) can be understood as deriving from the dilatation Ward identity $z {\cal T}^{0}_{0} +  {\cal T}^{i}_{i}  = 0$, when we identify ${\cal T}_{00}  = {\cal E}$ and ${\cal T}_{ij}  = p \delta_{ij}$. One can generalise this identity to constant velocity $u^{\mu}$ as
\be
z  \langle {\cal T}^{\mu}_{\nu}  \rangle u_{\mu} u^{\nu} - \langle {\cal T}^{\mu}_{\nu} \rangle P^{\nu}_{\mu} = 0
\ee
where
\be
P_{\mu}^{\nu} = \delta^{\nu}_{\mu} + u_{\mu} u^{\nu} 
\ee
and $u^{\mu} u_{\mu} = -1$. Note that the stress energy tensor ${\cal T}_{\mu \nu}$ is not symmetric but it is conserved, i.e. $\partial_{\mu}  {\cal T}^{\mu \nu} =0$. 
Imposing a Landau frame condition ${\cal T}_{\mu \nu} u^{\nu}  = - {\cal E} u_{\mu}$ the conservation equation implies that
\be
{\cal T}_{\mu \nu}  = ({\cal E} + p) u_{\mu} u_{\nu} + p \eta_{\mu \nu} + \pi^S_{(\mu \nu)} + \pi^{A}_{[\mu \nu]} + (u_{\mu} \pi^A_{\nu \sigma} + u_{\nu} \pi^A_{\mu \sigma}) u^{\sigma} + \cdots
\ee
to first order in gradients. In a theory with spatial rotational invariance the dissipative terms can be expressed as \cite{Hoyos:2013eza}
\be
\pi^{S (\mu \nu)} = - \eta P^{\mu \sigma} P^{\nu \tau} \Delta_{\sigma \tau} - \frac{\zeta}{(d-1)} P^{\mu \nu} \partial_{\sigma} u^{\sigma},
\ee
where the shear tensor is $\Delta_{\sigma \tau} = 2 \partial_{(\sigma} u_{\tau )} - 2 P_{\sigma \tau} \partial_{\sigma} u^{\sigma}/(d-1)$, and
\be
\pi^{A [ \mu \nu ]} = - \alpha_D u^{[\mu} u^{\sigma} \partial_{\sigma} u^{\nu ]}. 
\ee
Hence the theory is characterised by three transport coefficients at this order: the shear viscosity $\eta$, the bulk viscosity $\zeta$ and a third dissipative coefficient $\alpha_D$. All three coefficients scale in the same way as the entropy with temperature. In the case of broken rotational invariance or additional conserved charges there are further transport coefficients, see \cite{Hoyos:2013eza,Hoyos:2013qna}. Computation of these transport coefficients in holographic realisations therefore characterises the dual field theories. 
Further discussion of Lifshitz (and Galilean) hydrodynamics from the field theory perspective may be found in 
\cite{Chapman:2014hja,Arav:2014goa,Jensen:2014ama,Hoyos:2015lra,Banerjee:2015uta}. 

\subsection{Lifshitz black holes}

A static Lifshitz black hole can be described by a metric 
\be
ds^2 = \frac{1}{r^2} \left ( \frac{dr^2}{f(r)} - r^{2 (1-z)} f(r) dt^2 + dx^i dx_i \right ), \label{lif-bh}
\ee
where $f(r)$ is the blackening factor. In the Lifshitz background (\ref{geom-lif}) $f(r) = 1$ and hence in a finite temperature black hole $f(r)$ should asymptotically approach one but have a simple zero at a finite value of $r$. 

Unfortunately only a handful of analytic black hole solutions have been found; in general the equations of motion do not reduce to a nested set of linear equations as they do in the relativistic case \cite{Pinzani-Fokeeva:2014cka}. Numerical solutions for Lifshitz black holes and branes have been found in \cite{Danielsson:2009gi,Bertoldi:2009vn,Bertoldi:2009dt,Mann:2009yx,Brynjolfsson:2009ct,Bertoldi:2011zr,Amado:2011nd,Tarrio:2011de}. Analytic solutions for black holes in massive vector theories were found in \cite{Balasubramanian:2009rx,Korovin:2013nha,Bertoldi:2009vn,Tarrio:2011de}. Such realisations have slightly undesirable features, however: the model used in  \cite{Balasubramanian:2009rx} has a scalar with no kinetic term; the analysis of \cite{Korovin:2013nha} is applicable only for dynamical exponents close to one and the analytic solutions in \cite{Bertoldi:2009vn} have spherical horizons. Analytic black hole solutions can also be found in three spacetime dimensions, see \cite{Shu:2014eza}, and analytic solutions for solitons were constructed in \cite{Mann:2011bt}.

Analaytic asymptotically Lifshitz analytic black hole solutions can be found in Einstein-Maxwell-Dilaton models in which the Lifshitz symmetry is broken by a running scalar, see \cite{Taylor:2008tg,Pang:2009ad}. Generalisations to charged black holes are also possible \cite{Pang:2009pd,Danielsson:2009gi}.
Analytic black hole solutions in higher derivative theories were found in \cite{AyonBeato:2009nh,Cai:2009ac,AyonBeato:2010tm,Brenna:2011gp,Gonzalez:2011nz,Dehghani:2011hf,Matulich:2011ct,Liu:2012yd,Lu:2012xu,Ghanaatian:2014bpa} and in Brans-Dicke and higher spin models in \cite{Maeda:2011jj,Gutperle:2013oxa}. As mentioned already, (most) higher derivative theories are not unitary and furthermore the holographic dictionary for such models has not been developed: the operator content of the dual Lifshitz theories is unclear. Note however that \cite{Gonzalez:2011nza} proposed an explanation of the microscopic entropy of three-dimensional asymptotically Lifshitz black holes using an analogue of the Cardy formula.

\bigskip

To analyse Lifshitz hydrodynamics holographically one follows the standard fluid/gravity approach \cite{Bhattacharyya:2008jc}: one first boosts the black hole solution (\ref{lif-bh}) to obtain
\be
ds_0^2 = - 2 \frac{f(r)}{r^z} u_{\mu} dx^{\mu} dr +  \frac{1}{r^2} \left ( P_{\mu \nu} - r^{2(1-z)} f(r) u_{\mu} u_{\nu} \right ) dx^{\mu} dx^{\nu}
\ee
where $u_{\mu} u^{\mu} = -1$ and $P_{\mu \nu} = \eta_{\mu \nu} + u_{\mu} u_{\nu}$. One then promotes the parameters in the metric to be spacetime dependent, i.e. one lets the function $f(r)$ become a function $f(r,x^{\mu})$ and works in a gradient expansion for the latter. Correspondingly the supporting matter is also boosted; for example, in the case of the massive vector one starts from the boosted equilibrium solution
\be
A _0 = {a} (r) u_{\mu} dx^{\mu} + b(r) dr
\ee
and then allows the functions to become spacetime dependent. In order to satisfy the field equations one must then add terms involving the derivatives of these functions to the metric and vector field i.e. one looks for a solution 
\be
g = g_0 + g_1 + \cdots; \qquad
A = A_0 + A_1 + \cdots 
\ee
where the metric $g_1$ and the vector $A_1$ contain first derivatives of the defining functions $(f(r,x^{\mu}), a(r,x^{\mu}),b(r,x^{\mu}))$.

In practice this approach is hard to implement in generality given that very few analytic solutions for Lifshitz black holes exist. One can however extract results for certain transport coefficients without solving for the fully back-reacted metric in generality. 
The conductivity and diffusion constant are discussed in \cite{Pang:2009wa}.
For all solutions of Einstein gravity with matter, the shear viscosity to entropy density takes the usual value of $1/4 \pi$, as expected \cite{Policastro:2002se}. General arguments can be used to predict how the shear viscosity to entropy density bound is affected by higher derivative corrections \cite{Adams:2008zk}. 

One can also obtain the bulk viscosity $\zeta$ from the null horizon focussing equation \cite{Hoyos:2013cba}. In all cases involving massive vector fields this bulk viscosity is found to be zero, which implies that the trace dilatation identity continues to hold to first order in derivatives. 
These results can be generalised to models with hyperscaling violation, in which one finds a relation between the bulk viscosity and the shear viscosity
\be
\frac{\zeta}{\eta} = - 2 \frac{\theta}{(d-1) (d-1 - \theta)}  + 2 \frac{ (z-1)}{(d -1 - \theta)}
\ee
The case of $z=1$ corresponds to non-conformal branes; the relation between the shear and bulk viscosity is understood in this case to originate from the dilatation Ward identity in a parent conformal field theory, from which the non-conformal theory is obtained by generalised dimensional reduction \cite{Gouteraux:2011qh}. It would be interesting to understand whether the general $z \neq 1$ result can be understood in a similar way, via generalised dimensional reduction of a parent Lifshitz theory. Along these lines, classes of Einstein-Maxwell-Dilaton solutions were embedded in higher-dimensional AdS and Lifshitz backgrounds in \cite{Gouteraux:2011ce}, explaining their hydrodynamic properties. 
Hydrodynamics for hyperscaling violating Lifshitz was recently analysed further in \cite{Kiritsis:2015doa}. 
Note that it is possible to obtain temperature dependent ratios of bulk to shear viscosity in certain Einstein-Maxwell-dilaton models \cite{Gouteraux:2011ce}. 

Further analysis of hydrodynamics and transport in Lifshitz models can be found in \cite{Sun:2013wpa,Sun:2013zga,Roychowdhury:2015jha,Roychowdhury:2015cva}.
Finally let us comment that hydrodynamics for Schr\"{o}dinger holographic models has also been explored in the literature. Schr\"{o}dinger hydrodynamics is facilitated by the existence of exact finite temperature string theory backgrounds, \cite{Herzog:2008wg}, and was first analysed in \cite{Rangamani:2008gi}. 

\subsection{Anomalies} \label{anomalie}

Consider a non-relativistic field theory in a background with coordinates $(t,x^{i})$. It is natural to couple the theory to a gravity multiplet consisting of a scalar $N$, a vector $N_{i}$ and a symmetric spatial tensor $\gamma_{ij}$; in the relativistic context this is simply the ADM decomposition into the lapse, shift and spatial metric. Consider now diffeomorphism invariant actions involving this gravity multiplet. As discussed in \cite{Griffin:2011xs}
one possibility is the so-called kinetic term
\be
S \sim \int dt d^D x \sqrt{\gamma} N \left ( K^{ij} K_{ij} - \lambda K^2 \right ) \label{kin}
\ee
with $\lambda$ a constant, $K_{ij}$ the extrinsic curvature
\be
K_{ij}= \frac{1}{2N} \left ( \partial_t \gamma_{ij} - \partial_i N_j - \partial_j N_i \right ) 
\ee
and $K = \gamma^{ij} K_{ij}$. The other possibility is potential terms of the form
\be
S \sim \int dt d^{D} x \sqrt{\gamma} N {\cal V}
\ee
where ${\cal V}$ is a scalar function built from the Riemann tensor of $\gamma_{ij}$ and its covariant derivatives. 

Now let us consider anisotropic Weyl invariance, which acts as
\be
N \rightarrow e^{z \omega} N; \qquad N_{i} \rightarrow e^{2 \omega} N_{i}; \qquad \gamma_{ij} \rightarrow e^{2 \omega} \gamma_{ij}
\ee
for an arbitrary function $\omega(t,x^i)$. For general $z$ and $D$ one cannot find diffeomorphism and Weyl invariant actions. Whenever $z=D$, however, such actions do 
exist: the constant $\lambda$ in (\ref{kin}) must take the value $1/D$. Focussing on the case of $z= D = 2$, the only possible potential term which is Weyl invariant is
\be
S  \sim \int dt d^2 x \sqrt{\gamma} N \left ( R +  \frac{\nabla^2 N}{N} - \left ( \frac{\nabla N}{N} \right )^2 \right )^2, \label{potw}
\ee
where $\nabla_i$ is the spatial covariant derivative. Therefore for $z=D=2$ the anomalous dilatation Ward identity is 
\be
2 {\cal E} + t^{i}_{i} = A_1  \left ( K^{ij} K_{ij} - \frac{1}{2} K^2 \right ) + A_2 N \left ( R +  \frac{\nabla^2 N}{N} - \left ( \frac{\nabla N}{N} \right )^2 \right )^2
\ee
where the anomaly coefficients $(A_1,A_2)$ characterize the theory. We will explore in section \ref{hol-dict} which anomalies are found in holographic realisations of $z=D=2$ Lifshitz. 
Further discussion of anomalies in Lifshitz invariant field theories can be found in \cite{Adam:2009gq,Baggio:2011ha,Gomes:2011di,Arav:2014goa,Jensen:2014hqa} and we will return to the coupling of non-relativistic theories to background gravity in section \ref{newton}. Note also that the trace anomaly was recently analysed for non-relativistic 
Schr\"{o}dinger theories in three dimensions \cite{Auzzi:2015fgg}; the general anomaly structure is very different from the Lifshitz case but there is a natural candidate for an $a$-theorem in this case.

\section{Holographic dictionary} \label{hol-dict}

In this section we consider the holographic dictionary for Lifshitz models, i.e. how to extract dual field theory data from a given asymptotically Lifshitz geometry. We will focus on massive vector and Einstein-Maxwell-dilaton models, deferring axion/dilaton models to section \ref{newton}. Holographic renormalization for the gravity sector is considerably more complicated than the renormalization used above to compute correlation functions in a (fixed) Lifshitz background. To set the stage for holographic renormalization in Lifshitz it is convenient first to recall relevant features of asymptotically AdS gravity.

\subsection{AdS holographic renormalization}

The bulk (Euclidean) action for gravity with a negative cosmological constant is taken to be
\be
S = - \frac{1}{16 \pi G_{d+1}} \int d^{d+1} x \sqrt{g} \left (R + d (d-1) \right ) 
\ee
so the Einstein equation is $R_{mn} = -d g_{mn}$. In this section we work primarily in Euclidean signature, and hence the overall sign of the action differs from the Lorentzian actions in previous sections.

Any asymptotically locally anti-de Sitter geometry can be expressed in the neighbourhood of the conformal boundary in Fefferman-Graham form as
\be
ds^2 = \frac{dr^2}{r^2} + \frac{1}{r^2} \left ( g^{(0)}_{\mu \nu} + r^2 g^{(2)}_{\mu \nu} + \cdots r^d (g^{(d)}_{ \mu \nu} + h^{(d)}_{ \mu \nu} ) + \cdots  \right ) dx^{\mu} dx^{\nu} \label{fgmet}
\ee
where all terms in the expansion are determined by $g^{(0)}_{ \mu \nu}$ and $g^{(d)}_{ \mu \nu}$. In particular, $g^{(2)}_{ \mu \nu}$ and $h^{(d)}_{ \mu \nu}$ are expressed locally in terms of derivatives of $g^{(0)}_{ \mu \nu}$ \cite{Henningson:1998gx,deHaro:2000xn}. 
The independent data in the Fefferman-Graham expansion corresponds to the conjugate pair of the source for the energy-momentum tensor in the dual conformal field theory, $g^{(0)} _{\mu \nu}$, and the expectation value of the energy momentum tensor, which can be expressed in terms of the normalizable mode $g^{(d)}_{ \mu \nu}$ and derivatives of $g^{(0)}_{ \mu \nu}$. 

 In the original approach to holographic renormalization  \cite{Henningson:1998gx,deHaro:2000xn} the asymptotic expansions were inserted into the bulk Einstein equations. Differentiating these equations recursively with respect to the radial coordinate $r$ and then setting $r \rightarrow 0$, one obtains a set of algebraic equations for the coefficients $g^{(n)}_{\mu \nu}$. As claimed above these equations determine recursively all coefficients as local functionals of $g^{(0)}_{\mu \nu}$, except for the traceless transverse part of $g^{(d)}_{\mu \nu}$. The exact form of the coefficients depends on the dimension $d$; for example
\be
g^{(2)}_{\mu \nu} = \frac{1}{(d-2)} \left ( R_{\mu \nu} [ g^{(0)} ] - \frac{1}{2(d-1)} R [g^{(0)}] g^{(0)}_{\mu \nu} \right ).
\ee
Note that this expression applies to pure gravity; in general all coefficients are modified by the addition of matter. 
Using the asymptotic solutions one can evaluate the onshell action regulating the volume divergences using a cutoff $r \ge \epsilon$. The general form of the regulated divergences in the onshell action is then
\be
S^{\rm div} = \frac{1}{16 \pi G_{d+1}} \int_{\epsilon} d^dx \sqrt{g^{(0)}} \left ( \frac{a^{(0)}}{\epsilon^d} + \cdots + {a}^{(d)} \log \epsilon + {\cal O} (\epsilon^0) \right ),
\ee
where the coefficients $a^{(n)}$ depend locally only on the non-normalisable data $g^{(0)}$ and $a^{(d)}$ is the conformal anomaly of the dual field theory. 

To renormalise the action one proceeds by expressing the divergent terms in terms of the induced fields on the hypersurface $r = \epsilon$, which entails inverting the asymptotic series to express the sources in terms of the induced metric on the hypersurface $h_{\mu\nu} = g_{\mu\nu}/\epsilon^2$. One can then rewrite the divergences of the regulated action in covariant form, and subtract covariant local counterterms. For example, in $d=4$ one finds that the required counterterm action for pure gravity is
\be
S^{\rm ct} = \frac{3}{8 \pi G_{5}} \int_{\epsilon} d^4 x \sqrt{h} \left (1 + \frac{1}{12} R[h] - \frac{1}{48} (R^{\mu \nu} [h] R_{\mu \nu} [h]- \frac{1}{3} R[h]^2) \log \epsilon^2 \right ).
\ee
The renormalised action is defined as 
\be
S^{\rm ren} = {\cal L}_{\epsilon \rightarrow 0} \left ( S^{\rm div} + S^{\rm ct} \right ),
\ee
and the renormalised one point function for the energy momentum tensor is defined as 
\be
\langle T_{\mu \nu} \rangle = \frac{2}{\sqrt{g^{(0)}}} \frac{\delta S^{\rm ren}}{\delta g^{(0) \mu \nu}}
\ee
which can be evaluated to give  
\be
\langle T_{\mu \nu} \rangle = \frac{d}{16 \pi G_{d+1}} g^{(d)}_{\mu \nu} + X_{\mu \nu}[g^{(0)}], \label{t-vev}
\ee
where 
explicit expressions for the functionals $X_{\mu \nu}[g^{(0)}]$  can be found in \cite{deHaro:2000xn}. Note that these functionals vanish
for pure gravity with $d$ odd; this follows on dimensional grounds, as no covariant functionals of the required dimension exist. 

\bigskip

The original method of holographic renormalization requires the computationally inefficient step of inverting asymptotic expansions so as to express the divergences in a covariant form. The Hamiltonian method of holographic renormalization developed in \cite{Papadimitriou:2004ap} is more elegant and efficient, and it is this approach which has been generalised to asymptotically Lifshitz spacetimes. 

Any asymptotically locally anti-de Sitter manifold admits a radial function normal to the boundary which can be used to foliate the space in radial slices, at least in a neighbourhood of the boundary. This fact implies that one can set up a radial Hamiltonian formalism, analogous to the usual ADM formalism which relies on the existence of a time function foliating spacetime into hypersurfaces of constant time. In the neighbourhood of the conformal boundary one can decompose the metric as 
\be
ds^2 = {\cal N}^2 dr^2 +h_{\mu \nu} (dx^{\mu} + {\cal N}^{\mu} dr) (dx^{\nu} + {\cal N}^{\nu} dr),
\ee
where ${\cal N}$ is the shift and ${\cal N}^{\mu}$ is the lapse. The curvature of the manifold can be expressed in terms of the intrinsic and extrinsic curvatures of the hypersurfaces via the Gauss-Codazzi equations. After gauge fixing to normal coordinates such that the lapse and shift are ${\cal N} =1$ and ${\cal N}^{\mu} = 0$ and the induced metric is $h_{\mu \nu}$ (with $\mu$ running over $d$ indices) the Gauss-Codazzi equations become
\begin{eqnarray}
{\cal K}^2 - {\cal K}_{\mu \nu} {\cal K}^{\mu \nu} &=& R + 16 \pi G_{d+1} T^B_{rr};  \label{g-c} \\
D_{\mu} {\cal K}^{\mu}_{\nu} - D_{\nu} {\cal K} &=& 8 \pi G_{d+1} T^B_{r \nu}; \nonumber \\
\dot{\cal K}^{\mu}_{\nu} + {\cal K} {\cal K}^{\mu}_{\nu} &=& R^{\mu}_{\nu} - 8 \pi G_{d+1} \left ( T^{\mu}_{\nu} - \frac{1}{d-1} T^{B m}_{m} \delta^{\mu}_{\nu} \right ) \nonumber 
\end{eqnarray}
where ${\cal K}_{\mu \nu}$ is the extrinsic curvature\footnote{Note that ${\cal K}_{\mu \nu}$ denotes the extrinsic curvature of radial slices whereas we use $K_{ij}$ for the extrinsic curvature of constant time surfaces of the boundary metric.}, which in normal coordinates is given by
\be
{\cal K}_{\mu \nu} = \frac{1}{2} \partial_{r} h_{\mu \nu}
\ee
while $\dot{\cal K}^{\mu}_{\nu} = \partial_r \left ( h^{\mu\rho} {\cal K}_{\rho \nu} \right )$ and ${\cal K}= h^{\mu \nu} {\cal K}_{\mu \nu}$.  
$T^B_{mn}$ is the bulk stress energy tensor, i.e. the Einstein equations are 
\be
G_{mn} = 8 \pi G_{d+1} T^B_{mn}.
\ee
For pure cosmological constant $8 \pi G_{d+1} T^{B}_{mn} = \frac{1}{2} d (d-1) g_{mn}$. The Gauss-Codazzi equations in the form (\ref{g-c}) are however applicable to general bulk stress energy tensors. 

One can define a canonical momentum conjugate to the induced metric $h_{\mu \nu}$ as 
\be
\pi^{\mu \nu} = \frac{\delta L}{\delta \dot{h}_{\mu \nu}} =  \frac{1}{16 \pi G_{d+1}} \sqrt{h} ({\cal K}^{\mu \nu} - {\cal K} h^{\mu \nu} )
\ee
where the action is $S = \int dr L$. The Hamilton equation for the induced metric is equivalent to the third equation in (\ref{g-c}), upon applying Einstein's equations; viewed as an equation for the momentum it is first order in radial derivatives.  The momenta conjugate to the shift and lapse vanish identically, giving the momentum and Hamiltonian constraints, respectively. The latter are equivalent to the first two equations in (\ref{g-c}) (after applying Einstein's equations); expressed in terms of the momentum conjugate to $h_{\mu \nu}$ these are constraints, rather than differential equations involving radial derivatives. 

The next step is use the equations of motion to determine the asymptotic form for the momentum in terms of the induced metric. Both the momentum and the induced metric transform covariantly under diffeomorphisms on the radial slice and thus the Hamiltonian formulation is manifestly covariant. The momentum admits an asymptotic expansion in terms of the eigenfunctions of the dilatation operator $\delta_{D}$, i.e. 
\be
\pi_{\mu \nu} = \sqrt{h} \left ( \pi^{(0)}_{\mu \nu} + \cdots + \pi^{(d)}_{\mu \nu} + \tilde{\pi}^{(d)}_{\mu \nu} \log e^{- 2 r} + \cdots \right  ) \label{pi-exp}
\ee
where 
\begin{eqnarray}
\delta_{D} \pi^{(n) \mu}_{\nu} &=& - n \pi^{(n) \mu}_{\nu} \qquad n \neq d \\
\delta_{D} \pi^{(d) \mu}_{ \nu} &=& - d \pi^{(d) \mu }_{\nu} - 2 \tilde{\pi}^{(d) \mu}_{\nu} \nonumber \\
\delta_{D} \tilde{\pi}^{(d) \mu}_{\nu} &=& - d \tilde{\pi}^{(d) \mu}_{\nu} \nonumber
\end{eqnarray}
i.e. the terms transform homogeneously, except for $\pi^{(d) \mu}_{\nu}$. The eigenfunctions which are arise in the expansion are self-consistently determined by recursively solving the field equations. The dilatation operator is identified asymptotically with the radial derivative, i.e. 
\be
\delta_{D} + {\cal O}(e^{-r}) = \partial_{r} = \int d^d x \left ( 2 h_{\mu \nu} \frac{\delta}{\delta h_{\mu \nu}} \right ) + {\cal O}(e^{-r}).
\ee
More precisely, one can express the radial derivative in the form 
\be
\delta_r = \delta_D + \delta^{(1)} + \cdots + \tilde{\delta}^{(d)} \log e^{-2r} + \cdots \label{d-exp}
\ee
where $\delta^{(n)}$ are covariant functional operators of defined dilatation weight. 

The regulated onshell action can also be expressed as a sum of eigenfunctions of the dilatation operator
\be
S^{\rm div} =  \frac{1}{8 \pi G_{d+1}} \int_{\Sigma_r} d^d x \sqrt{h} \lambda
\ee
where the integration is over an arbitrary radial hypersurface $\Sigma_r$ and 
\be
\lambda = \lambda^{(0)} + \cdots + \lambda^{(d)} + \tilde{\lambda}^{(d)} \log e^{-2r} + \cdots \label{l-exp}
\ee
where again all terms except $\lambda^{(d)}$ transform homogeneously. Note that $\lambda$ satisfies a differential equation 
\be
\partial_{r} (\lambda + {\cal K}) + {\cal K}(\lambda + {\cal K}) = \frac{8 \pi G_{d+1}}{(d-1)} T^{B m}_{m}, \label{doe}
\ee
which is again first order in radial derivatives. 

One can then insert the covariant expansions (\ref{pi-exp}), (\ref{d-exp}) and (\ref{l-exp}) into the equations (\ref{g-c}) and (\ref{doe}) and solve the equations iteratively for the coefficients. 
As in the original method, non-local terms in the expansions are not determined by the recursion relations: the traceless and transverse part of $\pi^{(d)}_{\mu \nu}$ is undetermined, as is 
$\lambda^{(d)}$. The divergent terms in the onshell action are automatically expressed in covariant form, since the coefficients $\lambda^{(n)}$ are covariant, and hence the required counterterm action is simply
\be
S^{\rm ct} = - \frac{1}{8 \pi G_{d+1}} \int_{\Sigma_{r_{0}}} d^d x \sqrt{h} \left ( \sum_{n < d} \lambda^{(n)} + \tilde{\lambda}^{(d)} \log e^{-2 r_0} \right ).
\ee
The renormalised action is then given by 
\be
S^{\rm ren} = {\cal L}_{r_0 \rightarrow \infty} \left ( S^{\rm div} + S^{\rm ct} \right ) =  \frac{1}{8 \pi G_{d+1}} \int d^d x \sqrt{h} \lambda^{(d)},
\ee
i.e. in terms of the non-local coefficient which is undetermined by the asymptotic analysis. The renormalised one point functions are also expressed compactly in terms of the non-local terms in the expansion (\ref{pi-exp}):
\be
\langle T_{\mu \nu} \rangle = 2 \frac{\pi^{(d)}_{\mu \nu}}{\sqrt{h}} = - \frac{1}{8 \pi G_{d+1}} \left ( {\cal K}^{(d)}_{\mu \nu} - {\cal K}^{(d)} h_{\mu \nu} \right ). \label{t-vev2}
\ee 
To compare this expression with (\ref{t-vev}) one needs to express the coefficients of the extrinsic curvature in terms of the coefficients of the asymptotic expansion of the metric. Note that the expression (\ref{t-vev2}) makes manifest the geometric origin of the expression for the one point function. 

An advantage of the Hamiltonian method is that the Ward identities can be obtained immediately from the Hamiltonian and momentum constraints. These constraints must hold at each order in the expansion in eigenfunctions. The momentum constraint at order $d$ becomes 
\be
D_{\mu} \pi^{(d) \mu \nu} = 0,
\ee 
where $D_{\mu}$ is the covariant derivative in the boundary metric,
and hence from (\ref{t-vev2}) we obtain the diffeomorphism Ward identity
\be
D_{\mu} \langle T^{\mu \nu} \rangle = 0. 
\ee
The dilatation Ward identity arises from a Weyl transformation of the renormalized action
\be
\delta_{\sigma} S^{\rm ren} = - \frac{1}{8 \pi G_{d+1}} \int_{r_{(0)}}  d^d x \sqrt{h} (\tilde{\lambda}^{(d)}) \delta \sigma = 2 \int_{r_{(0)}} d^d x \sqrt{h} \pi^{(d) \mu}_{\mu} \delta \sigma
\ee
and thus 
\be
\langle T^{\mu}_{\mu} \rangle = - \frac{1}{8 \pi G_{d+1}} \tilde{\lambda}^{(d)},
\ee
so as anticipated $\tilde{\lambda}^{(d)}$ is the conformal anomaly. Note that the conformal anomaly vanishes for pure gravity with $d$ odd, again on dimensional grounds, as no covariant scalar functional of the required dimension exists. 

\bigskip

Now let us turn to Lifshitz gravity, with the supporting matter being the massive vector (\ref{mas-vc}). The basic questions to be answered are the 
following: what is the appropriate definition of asymptotically locally Lifshitz? Put differently, what are the independent non-normalisable modes of the bulk fields and their corresponding conjugate normalisable modes? To what operators in the dual Lifshitz field theory do these modes correspond? If we change the bulk matter supporting the Lifshitz geometry, how does the operator content of the dual field theory change? 

\subsection{Deformations of relativistic conformal theories}

We begin with the case where Lifshitz can be viewed as a deformation of AdS, see \cite{Korovin:2013bua}.  
Consider the action (\ref{mas-vc}) and its Lifshitz solution in the limit that $z = 1 + \delta^2$ with $\delta^2 \ll 1$. In this limit the Lifshitz metric (\ref{geom-lif}) becomes
\be
ds^2 = \frac{dr^2}{r^2} + \frac{1}{r^2} \left (dx^i dx_i -  (1 - 2 \delta^2 \log r )dt^2 \right ) + \cdots
\ee
with the massive vector being
\be
A = \sqrt{2} \frac{\delta}{r} dt + \cdots; \qquad M^2 = (d-1) + \cdots \label{source}
\ee
Here ellipses denote terms higher order in $\delta$. Clearly when $\delta = 0$ the background reduces to $AdS_{d+1}$ with no vector field. 
Using the standard AdS/CFT dictionary we can interpret the vector field as a source for the time component of a vector operator $J^{\mu}$ of dimension $d$ in the dual field theory: 
\be
S_{CFT} \rightarrow S_{CFT} +  \sqrt{2} \delta \int d^{d} x J^t + \cdots
\ee
This source breaks relativistic invariance and conformal symmetry but preserves the non-relativistic Lifshitz symmetry with dynamical exponent $z = 1 + \delta^2$. 
In this case the role of the bulk matter used to support the Lifshitz geometry is therefore clear.

A similar interpretation can be given to the Schr\"{o}dinger geometries (\ref{schg}) and (\ref{vector}) for any value of $z$. Working perturbatively in the parameter $b$, the background with $b=0$ is again $AdS_{d+1}$. The vector field corresponds to a source for a null component of a vector operator $V^{\mu}$ of dimension $(d+z-1)$ \cite{Guica:2010sw,Costa:2010cn}:
\be
S_{CFT} \rightarrow S_{CFT} +  b \int d^{D+2} x V^+ + \cdots
\ee
Again this deformation breaks relativistic and conformal invariance but for all values of $(b,z)$ it preserves the Schr\"{o}dinger symmetry group. Note that the deformation is irrelevant from the perspective of the original conformal field theory whenever $z \ge 1$ and relevant for $z \le 1$.

Returning to the Lifshitz case, 
at $z=1$ the system is anti-de Sitter, with the metric $g^{(0)}_{\mu \nu}$ in (\ref{fgmet}) being the source for a symmetric relativistic stress energy tensor $T_{\mu \nu}$. For the vector field
\be
A = \frac{A^{(0)}_{ \mu }}{r}  dx^{\mu}+ \cdots 
\ee
$A^{(0)}_{ \mu}$ is the source for the dual vector operator $J_{\mu}$. The corresponding diffeomorphism Ward identity in the conformal field theory is then
\be
\nabla^{\mu}  \langle T_{\mu \nu}  \rangle = A^{(0)}_{ \nu} \nabla_{\mu} \langle J^{\mu} \rangle - F^{(0)}_{ \nu \mu} \langle J^{\mu} \rangle,   
\ee 
with $F^{(0)}_{\mu \nu}$ the curvature of $A^{(0)}_{ \mu}$. Now switching on the specific source (\ref{source}) and working to order $\epsilon^2$ this identity becomes
\be
\nabla^{\mu} \left (  \langle T_{\mu \nu} \rangle -  \sqrt{2} \delta  \langle J_{\mu} \rangle \delta_{\nu t} \right ) = 0,
\ee
i.e. the non-symmetric stress tensor ${\cal T}_{\mu \nu} =  (T_{\mu \nu} -  \sqrt{2} \delta  J_{\mu}  \delta_{\nu t})$ is conserved. The source for the conserved stress tensor ${\cal T}_{\mu \nu}$ is not just $g^{(0)}_{ \mu \nu}$, but rather a combination of the non-normalizable modes of the metric and the vector field. We will show in subsequent sections that similar behaviour is found for general $z$, i.e. the conserved energy momentum complex of the non-relativistic theory is sourced by the boundary values of the bulk metric and of the (spatial part of) the massive vector. 

The dilatation Ward identity depends on the spacetime dimension. In the case of $D=2$ the terms appearing in the Ward identity  up to order $\delta^2$ are
\be
\langle T^{\mu}_{\mu} \rangle - \frac{1}{2} A^{(0) \mu} \langle T_{\mu \nu} \rangle A^{(0) \nu} = A^{(0)}_{\mu} \langle J^{\mu} \rangle 
\ee
Switching on the specific source (\ref{source}) we obtain
\be
z \langle {\cal T}^{t}_{t} \rangle + \langle {\cal T}^{i}_{i} \rangle  = \sqrt{2} \delta \langle {J}^{t} \rangle.
\ee
In the pure Lifshitz geometry $\langle J^{\mu} \rangle = 0$ and thus the expected Lifshitz trace identity is obtained.

\subsection{Vielbein approach}

In this section we consider holographic renormalisation for general asymptotically Lifshitz solutions of massive vector models.
We will give a definition of an asymptotically locally Lifshitz spacetime in terms of boundary conditions for vielbeins. This boundary data is identified with
sources for dual field theory operators and renormalised one point functions for these operators are computed using the radial Hamiltonian approach to holographic
renormalisation. This section follows the analysis given in \cite{Ross:2009ar,Ross:2011gu,Griffin:2011xs}; related work can be found in \cite{Baggio:2011cp,Mann:2011hg,Baggio:2011ha,Holsheimer:2013ula}.  Earlier work on holographic renormalisation can be found in \cite{Hohm:2010jc}. 

One can express a general asymptotically Lifshitz metric in the form
\begin{eqnarray}
ds^2 &=& \frac{dr^2}{r^2} + h_{\mu \nu} (r, x^{\mu}) dx^{\mu} dx^{\nu} \label{Gauss-metric} \\
&\equiv & \frac{dr^2}{r^2} - N^2 dt^2 + \gamma_{ij} (dx^i + N^i dt) (dx^j + N^j dt) \nonumber
\end{eqnarray}
where we have fixed a radial gauge (Gaussian normal coordinates) for the bulk metric. Since the dual theory is non-relativistic it is natural to work with vielbeins, which we define as
\be
ds^2 = \eta_{MN} E^{M}_{m} E^{N}_{n} dx^m dx^n \equiv \eta_{U V} e^{U}_{\mu} e^{V}_{\nu} dx^{\mu}dx^{\nu} + \frac{dr^2}{r^2},
\ee
i.e. $(M,N)$ denote $(d+1)$-dimensional frame indices and $(U,V)$ denote $d$-dimensional frame indices. We furthermore denote
spatial frame indices as $(I,J)$, i.e. $U = (0,I)$. The boundary as $r \rightarrow 0$ of an asymptotically Lifshitz background has vielbeins given by
\be
e^{0} = N dt; \qquad
e^{I} = e^{I}_{i} (N^i dt + dx^i) \label{viel}
\ee
We focus on the case where the bulk action is the massive vector action (\ref{mas-vc}). Then the vector field can be expressed as
\be
A_{\mu} = e^{U}_{\mu} A_{U} \qquad A_{U} = (C + \psi) \delta^0_U
\ee
where $C^2 = 2(z-1)/z$ is a constant and $\psi$ represents the perturbation relative to the exact Lifshitz background. Note that the equation of motion for the massive vector field induces a radial component for the vector field but the latter is not independent. 

The boundary conditions may then be specified by characterising the asymptotic behaviour of both $e^{U}_{\mu}$ and $\psi$. Analysis of the field equations shows that 
\begin{eqnarray}
e^{0}_{\mu} &=& \frac{1}{r^z} e^{(0) 0}_{\mu} + \cdots; \\ 
e^{I}_{\mu} &=& \frac{1}{r} e^{(0) I}_{\mu} + \cdots; \nonumber \\ 
\psi &=& \frac{1}{r^{\Delta_{-}}} \psi^{(0)} + \cdots \nonumber
\end{eqnarray}
where
\be
\Delta_{-} = \frac{1}{2} \left ( z + D - \sqrt{ (z+D)^2 + 8(z-1)(z-D)} \right ). \label{deltam}
\ee
Here each leading order term in the asymptotic expansion corresponds to an independent source for a dual operator. Implicitly in (\ref{viel}) we have set to zero $e^{(0) 0}_{i}$; we will explain why below. Note that the dimension of the operator dual to $\psi^{(0)}$ is  $\Delta_{+} = (D + z - \Delta_{-})$. 

\bigskip

Having set up the boundary conditions,
let us now turn to holographic renormalization. As anticipated, the starting point is the radial Hamiltonian formalism for the action, where the radial coordinate plays the role of Hamiltonian time. Writing the bulk metric as
\be
ds^2 = {\cal N}^2 dr^2 + h_{\mu \nu} (dx^{\mu} + {\cal N}^{\mu} dr) (dx^{\nu} + {\cal N}^{\nu} dr)
\ee
the (radial) Hamiltonian takes the form 
\be
{ H} = \int d^{d}x ({\cal N} {\cal H} + {\cal N}_{\mu} {\cal H}^{\mu} ),
\ee
and thus the Hamiltonian and momentum constraints are, respectively, ${\cal H} = 0$ and ${\cal H}^{\mu} = 0$. These constraints give first order flow equations. 

Fixing Gaussian normal coordinates (\ref{Gauss-metric}), the fields on each constant radial slice are the induced metric and the vector field, or equivalently the vielbein and the scalar field $\psi$. We can define canonical momenta for the latter via the variation of the onshell action
\be
\delta S^{\rm onshell} =  \int d^{d}x  \left ( \pi^{\mu}_{\; U} \delta e^{U}_{\mu} + \pi_{\psi} \delta \psi \right )
\ee
where
\be
\pi^{U}_{\; V} =  e^{U}_{\mu} \frac{ \delta S^{\rm onshell}}{\delta e^{V}_{\mu}}
\ee
and 
\be
\pi_{\psi} =  \frac{\delta S^{\rm onshell}}{\delta \psi}. 
\ee
Using the explicit expression for the vielbein (\ref{viel}) one finds that 
\begin{eqnarray}
\pi^{0}_{\; 0} &=& N \frac{\delta S^{\rm onshell}}{\delta N}; \\
\pi^{0}_{\; I} &=& N \frac{\delta S^{\rm onshell}}{\delta N^{I}}; \nonumber \\
\pi^{I}_{\; J} &=&  \left ( N^I \frac{\delta S^{\rm onshell}}{\delta N^{J}} +  2 e^{I}_{i} e_{Jj} \frac{\delta S^{\rm onshell}}{\delta \gamma_{ij}} \right ). \nonumber
\end{eqnarray}
Note that these canonical momenta are defined in terms of the bare onshell action, which is divergent. The equations of motion can be conveniently expressed in terms of these canonical momenta. The Hamiltonian and momentum constraint equations are automatically satisfied when the equations of motion are satisfied; the approach of \cite{Ross:2009ar,Ross:2011gu,Griffin:2011xs} was to analyse the equations of motion rather than the constraints. 

For the relation to field theory, note that $T^{\mu}_{U}$ is the operator sourced by $e^{(0) U}_{\mu}$, and its expectation value is contained in $\pi^{\mu}_{U}$. The choice  $e^{(0) 0}_{i} =0$ implies that one cannot access the expectation value of the dual operator $T^{I}_{\; 0}$, i.e. the source for the energy flux has been switched off. The other components of the vielbein source the energy density ($e^{(0)}_{0}$) and the momentum density and stress tensor ($e^{(0) I}_{\mu}$). 
The scalar $\psi^{(0)}$ acts as a source for a dual scalar operator ${\cal O}_{\psi}$ whose expectation value is contained in $\pi_{\psi}$. From the asymptotic falloff behaviour (\ref{deltam}) and its dimension we can see that this operator is marginal when $z=D$, relevant for $z < D$ and irrelevant for $z > D$. 

\bigskip
 
The onshell action can be expressed as a function of the boundary fields as 
\be
S^{\rm onshell}  = \frac{1}{16 \pi G_{d+1}} \int_{\epsilon}  dt  d^D x {\cal L}.
\ee
We can organise the divergent terms in ${\cal L}$ according to their dilatation weight under the operator 
\be
\delta_{D} = \int dt d^{D} x \left ( z e^{0}_{\mu} \frac{\delta }{\delta e^{0}_{\mu}} + e^{I}_{\mu} \frac{\delta}{\delta e^{I}_{\mu}} - \Delta_- \psi \frac{\delta}{\delta \psi} \right ).
\ee
Then (analogously to the relativistic case) ${\cal L}$ can be expressed as a sum
\be
{\cal L} = \sum_{m \ge 0} {\cal L}^{(m)} + \tilde{\cal L}^{z+D} \log (r),
\ee
with 
\begin{eqnarray}
\delta_{D} {\cal L}^{(m)} &=& - m {\cal L}^{(m)} \qquad m \neq (z+D)  \\
\delta_{D} {\cal L}^{(z+D)} &=& - (z+D) {\cal L}^{(z+D)} + \tilde{\cal L}^{(z+D)} \nonumber \\
\delta_D \tilde{\cal L}^{(z+D)} &=& - (z+D) \tilde{\cal L}^{(z+D)}. \nonumber
\end{eqnarray}
The logarithmic term captures the Weyl anomaly while the terms with $m < (z+D)$ compute the other required counterterms; the counterterms are hence
\be
S_{ct} = - \frac{1}{16 \pi G_{d+1}} \int_{\epsilon} dt d^{D} x  \left ( \sum_{0 \le m < (z+D)} {\cal L}^{(m)} + \tilde{\cal L}^{(z+D)} \log \epsilon \right ),
\ee
where the cutoff is $r = \epsilon$. The essence of the Hamiltonian approach is hence to decompose the onshell action into terms of different dilatation weight which respect
the required covariance, so that one can immediately extract appropriately covariant counterterms.  

\bigskip

To explicitly compute the counterterms we expand the canonical momenta in terms of eigenstates of the dilatation operator, i.e. we let
\be
\pi^{U}_{V} = \sqrt{\gamma} N \sum_{m} \pi^{(m) U}_{V} + \cdots \qquad
\pi^{\psi} = \sqrt{\gamma} N \sum_{m} \pi^{(m) \psi} + \cdots 
\ee
where the weights which occur are determined by recursive analysis of the equations of motion and the ellipses denote anomalous terms, i.e. terms which include $\log(r)$ 
factors. Such terms will arise for specific values of $D$ and $z$, in order to satisfy the equations of motion. Note that we follow the conventions of \cite{Papadimitriou:2004ap}, i.e. we suppress the explicit representations of the dilatation eigenfunctions (in terms of the radial coordinate), so the momenta components depend only on the boundary coordinates. By substituting these expansions into the equations of motion and working recursively one can thence determine the explicit expansions in terms of the induced fields on each radial slice. 

As in the relativistic case the operator expectation values may then be expressed in terms of specific dilatation weight components of the momenta:
\begin{eqnarray}
\langle T^{0}_{\; 0} \rangle &=& \pi^{(z+D) 0}_0 \label{op-ex} \\
\langle T^{I}_{\; V} \rangle & =& \pi^{(z+D) I}_{V} \nonumber \\
\langle {\cal O}_{\psi} \rangle &=& \pi_{\psi}^{(\Delta_+)}. \nonumber
\end{eqnarray}
The variation of the onshell action under dilatations at order $(z+D)$ gives the dilatation Ward identity
\be
\tilde{\cal L}^{(z+D)} = - z \langle T^{0}_{\; 0} \rangle - \langle T^{I}_{\; I} \rangle + \Delta_- \psi^{(0)} \langle {\cal O}_{\psi} \rangle,
\ee
indicating that $\tilde{\cal L}^{(z+D)}$ is the Weyl anomaly. 

The computation of explicit expressions for the counterterms and anomalies is somewhat involved. Here we highlight the main results only; see \cite{Ross:2009ar,Ross:2011gu,Griffin:2011xs,Baggio:2011cp,Mann:2011hg,Baggio:2011ha,Holsheimer:2013ula} for details. 
For $D = 2$, $z < 2$ the required counterterms were given in \cite{Ross:2011gu}. In the case of $z=D=2$ the realisation with a massive vector gives a non-vanishing Weyl anomaly, but only one of the two possible terms is found: (\ref{kin}) arises but (\ref{potw}) does not \cite{Griffin:2011xs,Baggio:2011cp}. In the case of $D=2$ and $z > 2$ one needs to take into account that the scalar operator is irrelevant, and thus the source for this operator must be treated perturbatively. 

A number of other issues remain open in this analysis. The diffeomorphism Ward identity should follow from the momentum constraint, but this was not analysed in the vielbein approach in \cite{Ross:2009ar,Ross:2011gu}. Given any asymptotically Lifshitz solution of the massive vector theory the key computable is operator expectation values. The expressions (\ref{op-ex}) are expressed as terms in the canonical momenta of specific dilatation weight but the relation of the latter to terms in the asymptotic expansions of the metric and vector field, or equivalently the asymptotic expansions of the vielbein and scalar, have not been calculated in generality.  Comparing with the relativistic case, this implies that the analogue of (\ref{t-vev2}) is known but the analogue of (\ref{t-vev}) is not known for general $D$ and $z$. Finally, while the analysis of \cite{Ross:2011gu} determines the most general asymptotically locally Lifshitz boundary conditions, the analysis is more transparent in the metric formalism of the following section.

\subsection{Metric formalism}

In  \cite{Chemissany:2014xpa,Chemissany:2014xsa,Papadimitriou:2014lia} the authors work in the metric formalism, rather than the veilbein formalism, and introduce a St\"{u}ckelburg field $\omega$ so that the vector field is gauge invariant. In other words, they write the massive vector field as
\be
A_{m} = {\cal A}_{m} - \partial_{\mu} \omega.
\ee
The bulk fields are again decomposed using a radial Hamiltonian approach as
\be
ds^2 = ( {\cal N}^2 + {\cal N}_{\mu} {\cal N}^{\mu} )dr^2 + 2 {\cal N}_{\mu} dx^{\mu} dr + h_{\mu \nu} dx^{\mu} dx^{\nu}
\ee
and
\be
{\cal A} = {\cal A}_r dr + {\cal A}_{\mu} dx^{\mu}.
\ee
The radial Hamiltonian then takes the form 
\be
H = \int d^{d}x \left ( {\cal N} {\cal H} + {\cal N}_{\mu} {\cal H}^{\mu} + {\cal A}_{r} {\cal F} \right )
\ee
where we interpret the shift ${\cal N}$, the lapse ${\cal N}_{r}$ and the radial component of the vector ${\cal A}_r$ as Lagrange multipliers enforcing the Hamiltonian, momentum and gauge constraints, ${\cal H} = {\cal H}^{\mu} = {\cal F} = 0$. These constraints, the Hamilton-Jacobi equations, provide first order equations encapsulating all of the dynamics, expressed in terms of the canonical momenta
\be
\pi^{\mu \nu} = \frac{\delta S}{\delta h_{\mu \nu}} \qquad
\pi^{\mu} = \frac{\delta S}{\delta {\cal A}_{\mu}} \qquad
\pi_{\omega} = \frac{\delta S}{\delta \omega}.
\ee  
For example, the momentum and gauge constraints are
\begin{eqnarray}
{\cal H}^{\mu} &=& -2 D_{\nu} \pi^{\nu \mu} + F^{\mu}_{\; \; \nu} \pi^{\nu} - A^{\mu} \pi_\omega = 0 \\
{\cal F} &=& - D_{\mu} \pi^{\mu} + \pi_{\omega} = 0 \nonumber 
\end{eqnarray}
while the Hamiltonian constraint is more complicated, see \cite{Chemissany:2014xpa,Chemissany:2014xsa} for explicit expressions. 
Solving the Hamilton-Jacobi equations asymptotically allows one to regulate the onshell action, derive counterterms and hence obtain renormalized correlation functions for the dual operators. 

The second step in \cite{Chemissany:2014xpa,Chemissany:2014xsa} is to introduce a near boundary expansion in terms of two commuting operators. The first operator is the usual (relativistic) dilatation operator
\be
\hat{\delta} = \int d^{d}x \left (2 h_{\mu \nu} \frac{\delta}{\delta h_{\mu \nu}} + A_{\mu} \frac{\delta}{\delta A_{\mu}} \right ).
\ee
This operator counts the number of derivatives transverse to each radial slice. 
The second operator is expressed in terms of fields in the decomposition of the metric $h_{\mu \nu}$, namely 
\be
h_{\mu \nu} = \sigma_{\mu \nu} + Y^{-1} A_{\mu} A_{\nu}. \label{sigma-def}
\ee
The operator is 
\be
\delta_A = \int d^{d}x \left ( 2 Y^{-1} A_{\mu} A_{\nu} \frac{\delta}{\delta h_{\mu \nu}} + A_{\mu} \frac{\delta}{\delta A_{\mu}} \right ).
\ee
Lifshitz boundary conditions enforce that $A_{\mu}$ is asymptotically aligned with the unit normal to constant time slices, i.e. asymptotically
\be
A_{\mu} = \sqrt{ \frac{2(z-1)}{z}} \frac{n^{(0)}_{\mu}}{r^z} + \cdots \label{a-asymp}
\ee
where $n_{\mu}^{(0)}$ is the unit normal to constant time surfaces. Imposing such boundary conditions, 
the operator $\delta_A$ effectively counts the number of time derivatives. 

Since the two operators commute, we can expand every quantity in terms of simultaneous eigenfunctions of both operators. For example, the onshell action (in radial gauge in which ${\cal N} = 1$ and ${\cal N}_{\mu} = 0$) can be expressed as
\be
S^{\rm onshell} = \int d^{d} x  \left (  \sum_{k,l} {\cal L}^{(k,l)} \right )
\ee
with 
\be
\hat{\delta} {\cal L}^{(k,l)} = (d - k) {\cal L}^{(k,l)} \qquad
\delta_{A} {\cal L}^{(k,l)} = (1 - l) {\cal L}^{(k,l)}. 
\ee
Now one can proceed to solve the Hamilton-Jacobi equations in a graded expansion, imposing the Lifshitz boundary conditions. As before the divergent terms in the onshell action can be removed by counterterms, which by construction respect the required symmetries, and Weyl anomalies are found in the case of $z=D$. 

There is always an independent solution of the Hamilton-Jacobi equations with dilatation weight zero which is therefore UV finite:
\be
S^{\rm reg} = \int d^d x \left ( h_{\mu \nu} \hat{\pi}^{\mu \nu} + A_{\mu} \hat{\pi}^{\mu} \right ), \label{Reg}
\ee
where $\hat{\pi}^{\mu \nu}$ and $\hat{\pi}^{\mu}$ are not determined by the asymptotic analysis, although they are subject to constraints arising from the Hamilton-Jacobi equations (including the momentum constraint). The holographic dictionary identifies this  term in the regulated action with the generating functional of correlation functions in the dual quantum field theory and the functions $\hat{\pi}^{\mu \nu}$ and $\hat{\pi}^{\mu}$ are related to regularized one point functions of local operators. 

Let us parameterise the boundary induced metric as in the previous section as 
\be
ds^2 = - N^2 dt^2 + \gamma_{ij} (dx^{i} + N^i dt) (dx^j + N^j dt) \label{bmet1}
\ee
and the boundary gauge field ${\cal A}$ as
\be
{\cal A}_{\mu} dx^{\mu} = a dt + a_i dx^i.
\ee
The Hamilton-Jacobi equations determine the leading asymptotic behaviours of the metric components to be
\be
{N} \sim \frac{{N}^{(0)}(x)}{r^z} \qquad
N_{i} \sim \frac{N^{(0)}_i(x)}{r^2} \qquad
\gamma_{ij} \sim \frac{\gamma^{(0)}_{ij}(x)}{r^2} \label{bmet2}
\ee
with each of these source terms being arbitrary. The asymptotic form for the vector field is 
\begin{equation}
A_{\mu} \sim \sqrt{\frac{2(z-1)}{z}} \frac{N^{(0)}}{r^z} \delta_{\mu t} \left (1 + r^{\Delta_-} \psi^{(0)} (x) \right ) \label{aboun}
\end{equation}
and 
\begin{eqnarray}
\omega &=& \omega^{(0)} (x); \\
a & \sim & \sqrt{\frac{2(z-1)}{z}} \frac{N^{(0)}}{r^{z+\Delta_-}}  \psi^{(0)} (x) + \partial_t \omega^{(0)}(x) ;  \nonumber \\
a_i & \sim & \partial_i \omega^{(0)}(x), \nonumber 
\end{eqnarray}
where $\Delta_-$ is as given in (\ref{deltam}). 
Here $\psi^{(0)}(x)$ and $\omega^{(0)}(x)$ are arbitrary sources but the latter corresponds to a gauge transformation and therefore does not source an independent operator. Comparing with (\ref{a-asymp}) we note that $n^{(0)}_{\mu} = N^{(0)} \delta_{\mu t}$. 

The modes conjugate to the independent sources are the renormalised one-point functions and these can be expressed in terms of the functions appearing in (\ref{Reg}), namely
\begin{eqnarray}
\langle {T}^{\mu \nu} \rangle &=& - \frac{1}{\sqrt{-h}} \left ( 2 \hat{\pi}^{\mu \nu} + \sqrt{\frac{2(z-1)}{z}} N^{(0)} \delta^{\mu t} \delta^{\nu t} \hat{\pi}^{t} \right ); \\
\langle {\cal O}_{\psi} \rangle &=& \frac{1}{\sqrt{-h}} N_{(0)} \hat{\pi}^{t}; \nonumber \\
\langle {\cal E}^{\mu} \rangle &=&  \frac{1}{\sqrt{-h}} \sqrt{\frac{2(z-1)}{z}}  \sigma^{(0) \mu}_{\nu} \hat{\pi}^{\nu}, \nonumber
\end{eqnarray}
where $\sigma^{(0) \mu \nu}$ is the boundary value of the metric $\sigma^{\mu \nu}$ defined in (\ref{sigma-def}). We can now extract the energy density ${\cal E}$, the momentum density ${\cal P}^{\mu}$ and the spatial stress tensor $t^{\mu \nu}$ by the following projections:
\begin{eqnarray}
\cal{E} &=& - n_{\mu}^{(0)} n_{\nu}^{(0)} \langle {T}^{\mu \nu} \rangle; \\
{\cal P}^{\mu} &=& - \sigma^{(0) \mu}_{\; \; \; \; \; \; \nu} n_{\rho}^{(0)} \langle {T}^{\nu \rho} \rangle; \nonumber \\
t^{\mu \nu} &=& \sigma^{(0) \mu}_{\; \; \; \; \; \; \rho}\sigma^{(0) \nu}_{\; \; \; \; \; \; \tau} \langle {T}^{\rho \tau} \rangle. \nonumber
\end{eqnarray}
These expressions are given in covariant form but reduce to analogous expressions to those of the previous section when we fix $n^{(0)}_{\mu} = N^{(0)} \delta_{\mu t}$ as above. 

Using the momentum constraint at the required order in the recursive expansion one can show that (for a flat background) the diffeomorphism Ward identities are 
\begin{eqnarray}
&& D_{\mu} t^{\nu}_{\nu} + n^{(0) \nu} D_{\nu} P_{\mu} + D_{\mu} \psi^{(0)} \langle {\cal O}_{\psi} \rangle =0; \\
&& n^{(0) \mu} D_{\mu} {\cal E} + D_{\mu} {\cal E}^{\mu} =0; \nonumber \\
&& D_{\mu} {\cal P}^{\mu} = 0, \nonumber
\end{eqnarray}
where $D_{\mu}$ is the covariant derivative with respect to the boundary metric $\sigma^{(0)}_{\mu \nu}$. 
The dilatation Ward identity takes the expected form of 
\be
z {\cal E} + t^{i}_{i} + (z + D - \Delta_-)  \psi^{(0)}  \langle {\cal O}_{\psi} \rangle = {\cal A},
\ee
where ${\cal A}$ are anomalous terms, which can be present when $z = D = (d-1)$. 

The results in this formalism are in agreement with those obtained using the vielbein formalism. Note in particular the spatial vector source for the bulk vector field is set to zero by the asymptotically locally Lifshitz boundary conditions. This spatial vector sources the irrelevant operator, the energy flux, and correlation functions for the latter could as usual be obtained by switching on the vector source perturbatively.  Although the source was not switched on in the analysis of  \cite{Chemissany:2014xpa,Chemissany:2014xsa}, it was nonetheless possible to infer the expectation value of the energy flux. 

\bigskip

The analysis of \cite{Chemissany:2014xpa,Chemissany:2014xsa} is conceptually complete. However, it is computationally complicated to extract explicit expressions for the energy momentum complex in terms of coefficients in the asymptotic expansion of the bulk metric and vector field, i.e. the analogue of the relativistic expression (\ref{t-vev2}) is known but the analogue of (\ref{t-vev}) is hard to compute. Note that the latter would in practice be the most useful for computing the energy momentum complex of a state dual to a given geometry. 

The methods of \cite{Chemissany:2014xpa,Chemissany:2014xsa} could certainly be developed further to compute correlation functions in asymptotically locally Lifshitz backgrounds. To facilitate such computations it would be desirable to implement the renormalization procedure in a symbolic computation package.

\subsection{Lifshitz with running scalars and hyperscaling violating Lifshitz}

The formalism developed above can immediately be generalised to Lifshitz with running scalars and hyperscaling violating Lifshitz backgrounds \cite{Chemissany:2014xpa,Chemissany:2014xsa,Papadimitriou:2014lia}. Both cases can be analysed simultaneously using the action in the dual frame (\ref{cp}). 
In this more general case we can parameterise the metric in domain wall type coordinates (\ref{domain-wall}) as s
\be
ds^2 = dr^2  + h_{\mu\nu} dx^{\mu} dx^{\nu},
\ee
where as $r \rightarrow \infty$ the boundary conditions for $h_{\mu \nu}$ can be expressed as in (\ref{bmet1})  with
\be
N \sim e^{z r} N^{(0)}(x) \qquad
N_i \sim e^{2r} N_{i}^{(0)}(x) \qquad
\gamma_{ij} \sim e^{2r } \gamma_{ij}^{(0)}. \label{bmet3}
\ee
(These expressions are identical to those in (\ref{bmet2}), expressed in domain wall coordinates.) 
Relative to the massive vector model, we have one additional bulk field, the dilaton, 
which asymptotes to 
\be
\phi = \mu r + \phi_{(0)} (x)
\ee
where $\phi_{(0)} (x)$ is identified as the source for a dual operator ${\cal O}_\phi$. The asymptotic form for the vector field is analogous to (\ref{aboun}), namely
\be
B_{\mu} \sim \sqrt{\frac{(z-1)}{2 \epsilon  Z_0}} N^{(0)} e^{\frac{\epsilon -z}{\mu} \phi_{(0)}}  e^{\epsilon r} \delta_{\mu t} \left (1 + e^{- \Delta_- r} \psi^{(0)} (x) \right ) \label{bboun}
\ee
Here $\Delta_{-}$ is a complicated function of the parameters, see \cite{Chemissany:2014xsa}, and the dimension of the dual operator to $\psi^{(0)}(x)$ is
\be
\Delta_{+} = (d + z - \theta - \Delta_-).
\ee
The operator content of the dual field theory is summarised in table \ref{tabl1}.

\Table{\label{tabl1} Operators and their dual sources}
\br

Operator & Covariant Expression & Dimension &  Source  \\
\mr
${\cal E}$ &  ${\cal E} = - n_{\mu}^{(0)} n_{\nu}^{(0)} {T}^{\mu \nu}$ & $(D+z  - \theta)$ & $N^{(0)}$  \\
${\cal E}^{i}$ & & $(D + 2z-1 - \theta)$ & 0  \\
${\cal P}^i$ &  ${\cal P}^{\mu} = - \sigma^{(0) \mu}_{\; \; \; \; \; \; \nu} n_{\rho}^{(0)} {T}^{\nu \rho}$ & $ (D+ 1- \theta)$ & $N^{(0)}_i$  \\
$t^{ij}$ & $t^{\mu \nu} = \sigma^{(0) \mu}_{\; \; \; \; \; \; \rho}\sigma^{(0) \nu}_{\; \; \; \; \; \; \tau} {T}^{\rho \tau} $ & $(D+z - \theta)$ & $\gamma^{(0)}_{ij}$  \\
${\cal O}_{\phi}$ &  & $(D+z - \theta)$ & $\phi^{(0)}$  \\
${\cal O}_{\psi}$ &  & $\Delta_{+}$ & $\psi^{(0)}$  \\
\br
\end{tabular}

\end{indented}
\end{table} 

The diffeomorphism Ward identities take the covariant form 
\begin{eqnarray}
&& D_{\nu} t^{\nu}_{\mu} + n^{(0) \nu} D_{\nu} {\cal P}_{\mu} + D_{\mu} \phi^{(0)} {\cal O}_{\phi} + D_{\mu} \psi^{(0)} {\cal O}_{\psi} = 0 \\
&& 
n^{(0) \mu} D_{\mu} {\cal E} + D_{\mu} {\cal E}^{\mu} + n^{(0) \mu} D_{\mu} \phi^{(0)} = 0 \nonumber \\
&& 
D_{\mu} {\cal P}^{\mu} = 0, \nonumber 
\end{eqnarray}
where $D_{\mu}$ denotes the covariant derivative with respect to the boundary metric $\sigma^{(0)}_{\mu \nu}$. 

The dilatation Ward identity is 
\be
z {\cal E} + t^{\mu}_{\mu} + \Delta_- \psi^{(0)} {\cal O}_{\psi} = \mu {\cal O}_{\phi}. 
\ee
Clearly whenever $\mu \neq 0$ the Lifshitz symmetry is broken by the running scalar coupling: such theories have generalized Lifshitz structure
rather than Lifshitz invariance. It was argued in \cite{Chemissany:2014xpa,Chemissany:2014xsa} that such theories do not have Weyl anomalies, as one can absorb
any logarithmic divergences into the running dilaton, removing any explicit dependence on the cutoff. This makes sense from the dual perspective as the field theories have running couplings and therefore the concept of a Weyl anomaly is ill-defined. 

The arguments of \cite{Chemissany:2014xpa,Chemissany:2014xsa} assume that the dilaton continues to run and there is no true fixed point in the ultraviolet. The non-conformal branes are counterexamples to this assumption. For example, the D4-branes are dual to supersymmetric Yang-Mills in five dimensions, which runs to the six dimensional $(2,0)$ CFT in the ultraviolet. In this case one therefore chooses the renormalization scheme to respect the ultraviolet fixed point symmetries, and one can therefore not absorb the logarithmic divergences into the running dilation: the theory does have a Weyl anomaly, inherited from the $(2,0)$ theory \cite{Kanitscheider:2008kd}. 

It is an open question whether hyperscaling violating Lifshitz solutions can generally be viewed as related to generalized reductions of solutions with scaling symmetry, although examples which uplift to AdS and Lifshitz can be found in \cite{Gouteraux:2011ce}. In such cases it would natural to choose a renormalization scheme which respects the symmetries of the fixed point. In such a setup there could be Weyl anomalies inherited from the parent scale invariant theory, as in the case of non-conformal branes. 

It is interesting to note that when the massive vector theory is coupled to a scalar (with $\mu = 0$) the second type of Weyl anomaly (\ref{potw}) at $z= D=2$ (involving four spatial derivatives) is generated. Thus the holographic models provide examples of Lifshitz theories in which the detailed balance condition is violated. 

\subsection{Summary}

Holographic renormalization allows one to determine the operator content of the dual Lifshitz theory for massive vector and Einstein-Maxwell-Dilaton models. In a $(d+1)$-dimensional massive vector model, the boundary conditions are a $d$ dimensional metric $h_{\mu \nu}$ and vector $A_{\mu}$. This gives $\frac{1}{2} d(d+1) + d$ independent boundary conditions in total, or equivalently $\frac{1}{2} D^2 + \frac{5}{2} D + 2$ boundary conditions (expressed in terms of the spatial dimension $D$). 

The energy momentum complex for a non-relativistic theory consists of the energy density ${\cal E}$, the energy flux ${\cal E}_i$ ($D$ components), the momentum density ${\cal P}_{i}$ ($D$ components) and the spatial stress tensor density $t_{ij}$ ($\frac{1}{2} D(D+1)$ components). Thus the energy momentum complex has in total $\frac{1}{2} D^2 + \frac{5}{2} D + 1$ components. The bulk theory therefore has enough boundary conditions to source the energy momentum complex together with one additional scalar operator. Similarly an Einstein-Maxwell-Dilaton theory has one additional boundary condition (for the bulk scalar) and therefore the boundary conditions source the energy momentum complex together with two scalar operators. The boundary conditions for a bulk relativistic theory cannot provide sources only for the energy momentum complex; the additional boundary conditions source scalar operators. 
These counting arguments are in agreement with the detailed analysis, see table \ref{tabl1}.

\section{Newton-Cartan geometry} \label{newton}

In this section we explore the role of Newton-Cartan geometry for non-relativistic theories and discuss the holographic dictionary for Lifshitz theories which are obtained by null reductions of Schr\"{o}dinger theories. 

\subsection{Galilean invariance and Newton-Cartan structure} \label{newton1} 

We consider first Galilean invariant field theories and their coupling to background geometry. This section follows
in particular \cite{Jensen:2014aia,Jensen:2014hqa}; see also
\cite{Andringa:2010it,Christensen:2013rfa,Hartong:2014oma,Hartong:2014pma,Hartong:2015wxa}. 

We can define a Newton-Cartan structure on a $d$-dimensional spacetime in terms of a one-form $n_{\mu}$, a symmetric positive semi-definite rank $(d-1)$ tensor $\sigma_{\mu \nu}$ and a $U(1)$ connection $a_{\mu}$, with the requirement that 
\be
{h}_{\mu \nu} = n_{\mu} n_{\nu} + \sigma_{\mu \nu} \label{new-met1}
\ee
is positive definite. These tensors determine a velocity vector $v^{\mu}$ and an inverse metric $\sigma^{\mu \nu}$ via the algebraic constraints
\be
v^{\mu} n_{\mu} = 1 \qquad
\sigma_{\mu \nu} v^{\mu} = 0 \qquad
\sigma^{\mu \nu} n_{\nu} = 0 \qquad
\sigma^{\mu \rho} \sigma_{\rho \nu} = \delta^{\mu}_{\nu} - v^{\mu} n_{\nu}.
\ee
The velocity vector effectively defines a local time direction and $\sigma_{\mu \nu}$ gives a metric on spatial slices.
In torsionless Newton-Cartan geometry the one form $n_{\mu}$ is exact, i.e. $n_{\mu} = \partial_{\mu} t$ where $t$ is a global time function. In torsional Newton-Cartan (TNC) 
geometry there is no such condition imposed on the one form. In twistless torsion Newton-Cartan geometry (TTNC) one imposes that there is a preferred foliation on equal time slices; the terminology TTNC was introduced in \cite{Christensen:2013lma,Christensen:2013rfa}.
Torsional Newton-Cartan structure (TNC) is particularly relevant for holography, see \cite{Christensen:2013lma,Christensen:2013rfa,Hartong:2015wxa,Hartong:2015zia,Bergshoeff:2014uea}. Note that Newton-Cartan geometry can also be expressed in a vielbein formulation, as advocated in \cite{Andringa:2010it}.

One can define many possible covariant derivatives using the Newton-Cartan data; we will consider here a covariant derivative ${\cal D}_{\mu}$ which satisfies
\be
{\cal D}_{\mu} n_{\nu} = 0 \qquad
{\cal D}_{\mu} \sigma^{\nu \rho} = 0. 
\ee
The temporal part of the torsion is proportional to $d n$ and hence vanishes if $n$ is exact. The compatibility conditions do not entirely fix the connection: even when the temporal part of the torsion vanishes the spatial part of the torsion remains arbitrary. The most general expression for a torsion compatible connection can be found in 
\cite{Bekaert:2014bwa}.

A theory with Galilean symmetry couples to such a Newton-Cartan geometry; the $U(1)$ connection couples to the particle number. Such a Galilean theory is invariant under both coordinate reparameterisations and gauge transformations. One may in addition impose Milne boosts which leave $(n_{\mu},\sigma^{\mu \nu})$ invariant and act as
\begin{eqnarray}
v^{\mu} &\rightarrow & v^{\mu} + c^{\mu} \qquad
a_{\mu} \rightarrow a_{\mu} + c_{\mu} - \frac{1}{2} n_{\mu} c^2 \\
\sigma_{\mu \nu} & \rightarrow & \sigma_{\mu \nu} - (n_{\mu} c_{\nu} + n_{\nu} c_{\mu} ) + n_{\mu} n_{\nu} c^2 \nonumber
\end{eqnarray}
with $c^{\mu} = h^{\mu \nu} c_{\nu}$ and $c^2 = c^{\mu} c_{\mu} > 0$. Here $c_{\mu}$ is spatial so $v^{\mu} c_{\mu} = 0$ and implicitly we assume that $n$ is closed. 
One can conveniently define boost invariants as
\begin{eqnarray}
\hat{v}^{\mu} &=& v^{\mu} - \sigma^{\mu \nu} a_{\nu} \label{inv} \\
\hat{\sigma}_{\mu \nu} &=& \sigma_{\mu \nu} + n_{\mu} a_{\nu} + n_{\nu} a_{\mu} \nonumber \\
\Phi &=& - v^{\mu} a_{\mu} + \frac{1}{2} \sigma^{\mu \nu} a_{\mu} a_{ \nu}, \nonumber
\end{eqnarray}
so that a theory which is invariant under the boost symmetry couples to $(\hat{v}^{\mu},n_{\mu}, \sigma^{\mu \nu}, \hat{\sigma}_{\mu \nu}, \Phi)$. In the context of TTNC geometry one can characterise the torsion as
\be
\tau_{\mu} = {\cal L}_{\hat{v}} n_{\mu} \label{torsion}
\ee
and we will refer to $\tau_{\mu}$ as the torsion vector. 

Now consider the generating functional $W[n_{\mu},v^{\mu},\sigma^{\mu \nu},a_{\mu}]$ for such a theory. We can define a number current $J^{\mu}$ (also called the mass density) and stress energy tensor complex via
\be
\delta W = \int d^{d} x \sqrt{ {h} } \left [ \delta a_{\mu} J^{\mu} - \delta \bar{v}^{\mu} {\cal P}_{\mu} - \delta n_{\mu} {\cal E}^{\mu} - \frac{1}{2} \delta  \bar{\sigma}^{\mu \nu} t_{\mu \nu} \right]  \label{gen-nc}
\ee
where as in previous sections ${\cal P}_{\mu}$ is the momentum current, ${\cal E}^{\mu}$ is the energy current (the energy density and the energy flux) and $t_{\mu \nu}$ is the spatial stress tensor. Here the barred quantities arise from the fact that the variations of $\delta v^{\mu}$ and $\delta \sigma^{\mu\nu}$ contain terms involving $\delta {n}_{\mu}$ as well as arbitrary variations $\delta \bar{v}^{\mu}$ and $\delta \bar{\sigma}^{\mu}$, i.e.
\be
\delta v^{\mu} = - v^{\mu} v^{\nu} \delta n_{\nu} + P^{\mu}_{\nu} \delta \bar{v}^{\nu}. 
\ee
Diffeomorphism invariance leads to conservation equations for the energy current and the spatial stress tensor. Gauge invariance implies that $J^{\mu}$ is conserved and Milne invariance implies the relation 
\be
{\cal P}_{\mu} = \sigma_{\mu \nu} J^{\nu}.
\ee
A Galilean theory which is scale invariant with a critical exponent $z$ is invariant under 
\be
n_{\mu} \rightarrow e^{z \omega} n_{\mu} \qquad
\sigma_{\mu \nu} \rightarrow e^{2 \omega} \sigma_{\mu \nu} \qquad
a_{\mu} \rightarrow e^{(2-z) \omega} a_{\mu}.
\ee
The dilatation Ward identity is then 
\be
z n_{\mu} {\cal E}^{\mu} - \sigma^{\mu \nu} t_{\mu \nu} = {\cal A},
\ee
where ${\cal A}$ denotes the Weyl anomaly.

\bigskip

As an example of a boost invariant theory, one can consider free fields charged under the $U(1)$ symmetry 
\be
S = \int d^d x \sqrt{h} \left ( i v^{\mu} \Psi^{\ast} \stackrel{\leftrightarrow}{ {\cal D}_{\mu} } \Psi - \frac{1}{m} \sigma^{\mu \nu} {\cal D}_{\mu} \Psi^{\ast} D_{\nu} \Psi \right )
\ee
where $\Psi^{\ast} \stackrel{\leftrightarrow}{ {\cal D}_{\mu} } \Psi = \Psi^{\ast} { {\cal D}_{\mu} } \Psi - \Psi { {\cal D}_{\mu} } \Psi^{\ast}$. The equation of motion for the free field is the (covariant) Schr\"{o}dinger equation. This action can be written as \cite{Jensen:2014aia,Hartong:2014pma}
\be
S = \int d^d x \sqrt{h} \left (  i \hat{v}^{\mu} \Psi^{\ast} \stackrel{\leftrightarrow}{ \partial_{\mu} } \Psi - \frac{1}{m} \sigma^{\mu \nu} \partial_{\mu} \Psi^{\ast} \partial_{\nu} \Psi - m \Phi \Psi^{\ast} \Psi \right ),
\ee
using the boost invariant quantities defined in (\ref{inv}). Noting that the metric determinant is boost invariant this action is now manifestly invariant under boosts and expressed in a form which demonstrates the coupling to the background Newton-Cartan geometry. This free theory has a $z=2$ scaling symmetry. The action remains boost invariant under the addition of a potential depending only on $|\Psi |$ and the specific potential 
\be
V(\Psi) = V_0 | \Psi |^{\frac{2 (d+2)}{d}} 
\ee 
respects the $z=2$ scaling symmetry. 

\bigskip

From the holographic perspective, it is natural to expect asymptotically locally Schr\"{o}dinger spacetimes to be dual to field theories coupled to background Newton-Cartan geometries. Indeed, TNC is related to the warped geometry proposed in \cite{Hofman:2014loa} to couple to two-dimensional warped conformal field theories, which are in turn related to three-dimensional warped geometries (including Schr\"{o}dinger).  
A priori one would not think that Newton-Cartan geometry is relevant to Lifshitz invariant theories, as the latter do not possess full Galilean symmetry: no particle number symmetry exists in general and thence there is no Milne boost symmetry. In the rest of this section we will explore whether certain holographic realisations of Lifshitz admit an enhanced symmetry group, associated with Newton-Cartan structure.

Before moving to discuss holographic models, we should note that Newton-Cartan structure is interesting in its own right. For example, TTNC has been proposed to be relevant to condensed matter physics, in the context of an effective field theory for the fractional quantum Hall effect, see \cite{Son:2013rqa,Geracie:2014nka}. 

\subsection{Reduction of axion/dilaton models}

In this section we discuss the reduction of the axion/dilaton model (\ref{axion-dilaton}) to give asymptotically locally Lifshitz solutions in four bulk dimensions; the restriction to $d=4$ is unnecessary but only this case has so far been analysed in detail in the literature, see \cite{Chemissany:2011mb,Chemissany:2012du,Christensen:2013rfa}. 

The holographic dictionary for asymptotically locally AdS solutions of the five-dimensional theory was analysed in \cite{Papadimitriou:2011qb}. Any such asymptotically locally AdS solution can be expressed near the conformal boundary as 
\begin{eqnarray}
ds^2 &=& \frac{dr^2}{r^2} + \frac{1}{r^2} h_{mn} dx^{m} dx^{n}; \label{upstairs} \\
h_{mn} &=& h^{(0)}_{mn} + r^2 h^{(2)}_{mn}  + r^2 h^{(4)}_{mn} + r^2 \log r \tilde{h}^{(4)}_{mn} + \cdots \nonumber \\
\phi &=& \phi^{(0)} + r^2 \phi^{(2)} + r^4 \phi^{(4)} + r^4 \log r \tilde{\phi}^{(4)} + \cdots  \nonumber \\
\chi &=& \chi^{(0)} + r^2 \chi^{(2)} + r^4 \chi^{(4)} + r^4 \log r \tilde{\chi}^{(4)} + \cdots \nonumber 
\end{eqnarray}
where $(h^{(0)}_{mn}, \phi^{(0)}, \chi^{(0)})$ are non-normalisable source terms and $(h^{(4)}_{mn}, \phi^{(4)}, \chi^{(4)})$ are normalisable modes. All other coefficients in the asymptotic expansion are determined in terms of these and, in particular, the terms at order $r^2$ and $r^4 \ln r$ given above are determined in terms of the sources. Note that there are twelve independent unconstrained non-normalisable modes (ten from the metric and two from the scalars) which are sources for the dual stress energy momentum tensor and two scalar operators. 

The operators dual to the sources are the energy momentum tensor $T_{mn}$ and two scalar operators ${\cal O}_{\phi}$ and ${\cal O}_{\chi}$. Their expectation values are given by 
\begin{eqnarray}
\langle T_{mn} \rangle &=& \frac{1}{16 \pi G_{5}} \left ( 2 h^{(4)}_{mn} - 2 X_{mn}(h^{(0)}, \phi^{(0)}, \chi^{(0)}) \right ); \\
\langle {\cal O}_{\phi} \rangle &=& \frac{1}{16 \pi G_5} \left (  - 2 \phi^{(0)} + {\cal O}_{\phi}(h^{(0)},\phi^{(0)},\chi^{(0)}) \right ); \nonumber \\
\langle {\cal O}_{\chi} \rangle &=& \frac{1}{16 \pi G_5} \left ( - 2 e^{2 \phi^{(0)}} \chi^{(0)} + {\cal O}_{\chi}(h^{(0)},\phi^{(0)},\chi^{(0)} ) \right ) \nonumber
\end{eqnarray}
where the (scheme dependent) functions of the sources $(X_{mn},{\cal O}_{\phi},{\cal O}_{\chi})$ may be found in \cite{Papadimitriou:2011qb}. These one point functions satisfy the expected dilatation and diffeomorphism Ward identities 
\begin{eqnarray}
&& \langle T^{m}_m \rangle = {\cal A}(h^{(0)},\phi^{(0)},\chi^{(0)}); \\
&& \nabla^{(0)}_{m} \langle T^{mn} \rangle = - \langle {\cal O}_{\phi} \rangle \partial^{n} \phi^{(0)} - \langle {\cal O}_{\chi} \rangle \partial^{n} \chi^{(0)}, \nonumber
\end{eqnarray}
where the Weyl anomaly is given by 
\begin{eqnarray}
{\cal A} &=& \frac{1}{8} (R^{(0)}_{mn} - \frac{1}{2} \partial_{m} \phi^{(0)} \partial_n \phi^{(0)} - \frac{1}{2} e^{2 \phi^{(0)}} \partial_{m} \chi^{(0)} \partial_n \chi^{(0)})^2  \\
&& - \frac{1}{72} \left (R^{(0)} - \frac{1}{2} (\partial \phi^{(0)})^2 - \frac{1}{2} e^{2 \phi^{(0)}} (\partial \chi^{(0)})^2 \right ) \nonumber \\
&& + \frac{1}{16} \left ( \Box^{(0)} \phi^{(0)} - e^{2 \phi^{(0)}} (\partial \chi^{(0)})^2 \right )^2 \nonumber \\
&& + \frac{1}{16} e^{2 \phi^{(0)}} \left ( \Box^{(0)} \chi^{(0)} + 2 \partial_{a} \phi^{(0)} \partial^a \chi^{(0)} \right ) \nonumber 
\end{eqnarray}
where indices are contracted with the metric $h^{(0)}_{mn}$, whose curvature is $R^{(0)}_{mn}$.  

\bigskip

Now following  \cite{Chemissany:2011mb,Chemissany:2012du} we consider the following Scherk-Schwarz reduction 
\begin{eqnarray}
ds_5^2 &=& \frac{dr^2}{r^2} + \frac{h_{mn}}{r^2} dx^{m} dx^{n} = \frac{dr^2}{r^2} + \frac{h_{\mu \nu}}{r^2} dx^{\mu} dx^{\nu} + e^{2 \Phi} (du + A_{\mu} dx^{\mu})^2 \label{red-map} \\
&\equiv & ds_4^2 + e^{2 \Phi} (du + A_{\mu} dx^{\mu})^2 \nonumber \\
\chi_5 &=& \chi + k u \nonumber \\
\phi_5 &=& \phi, \nonumber 
\end{eqnarray} 
where we denote the five-dimensional fields by a subscript to distinguish them from their four-dimensional counterparts. Note that we are choosing to preserve the radial gauge under the dimensional reduction, which implies that the reduced metric is in the dual frame rather than the Einstein frame. Implicitly we assume that the five-dimensional metric admits a hypersurface orthogonal Killing vector $\partial_{u}$ in making this reduction; we assume that the fields $(h_{\mu \nu}, \chi,\phi, \Phi,A_{\mu})$ do not depend on $u$. 

Taking the periodicity of the circle to be $2\pi L$ the reduced bulk action is 
\begin{eqnarray}
S &=& \frac{1}{16 \pi G_{4}} \int d^4 x \sqrt{-g} \left ( e^{\Phi} R - \frac{1}{4} e^{3 \Phi} F^2 - \frac{1}{2} e^{\Phi} (\partial \phi)^2 - \frac{1}{2} k^2 e^{\Phi + 2 \phi} B^2 
\right . \nonumber \\
&& \qquad \qquad \qquad  \qquad \left . - e^{2 \Phi}  V(\phi,\Phi) \right ) \label{ax-dil}
\end{eqnarray} 
where the potential is
\be
V (\phi,\Phi) =  \left ( \frac{1}{2} k^2 e^{-3 \Phi + 2 \phi} - 12 e^{-\Phi} \right ) 
\ee
and 
\be
G_4 = \frac{G_5}{2 \pi L}. 
\ee
The massive vector $B$ is expressed in terms of the fields in the reduction as 
\be
- k B = d \chi - k A.
\ee
(Implicitly we assume $k \neq 0$.) A key difference between this action and that analysed in previous sections is the presence of two scalars, $\phi$ and $\Phi$. It is  straightforward to show that one cannot truncate the model (\ref{ax-dil}) to a pure massive vector model, i.e. the field equations do not allow the scalar fields $\phi$ and $\Phi$ to be set to constants for generic solutions. 

Without showing that a truncation is consistent (i.e. all solutions of the lower-dimensional equations of motion are also solutions of the higher-dimensional equations of motion), one cannot be sure that the dimensional reduction of the renormalised action gives the correct counterterms for the lower-dimensional action.  
It was not shown explicitly in \cite{Chemissany:2011mb,Chemissany:2012du} that the reduced action (\ref{ax-dil}) is a consistent truncation of the action (\ref{axion-dilaton}): the reduction was carried out at the level of the action, rather than the field equations. It was however shown in unpublished work by these authors \footnote{We thank Jelle Hartong for bringing this work to our attention.} that the five-dimensional equations of motion do reduce to the four-dimensional equations of motion. 

The equations of motion following from (\ref{ax-dil}) admit a $z=2$ Lifshitz solution 
\begin{eqnarray}
ds^2 &=& \frac{dr^2}{r^2} - e^{-2 \Phi^o} \frac{dt^2}{r^4} + \frac{1}{r^2} \left ( dx^2 + dy^2 \right ) \\
B &=& - e^{2 \Phi^o} \frac{dt}{r^2} \nonumber \\
\Phi &=& \Phi^o = \phi^o + \ln (\frac{k}{2}) \qquad \phi = \phi^o \nonumber
\end{eqnarray}
where $\phi^o$ is a constant. 

The main subtlety in relating the five-dimensional asymptotically locally AdS holographic dictionary to the four-dimensional asymptotically locally Lifshitz dictionary is that the appropriate definition of asymptotically locally Lifshitz in four dimensions is not a priori clear. The required definition could be found by systematically analysing the reduced equations of motion near the boundary as in section \ref{hol-dict}. The approach used in \cite{Chemissany:2012du} was however to infer the holographic dictionary directly from the dimensional reduction. In principle one can infer the counterterms and renormalised one point functions for the lower-dimensional asymptotically locally Lifshitz spacetime by dimensionally reducing those for asymptotically locally AdS using the reduction map (\ref{red-map}). 

This approach however has a number of subtleties. Since from the dual field theory perspective the reduction is on a null circle, the usual techniques of Kaluza-Klein reduction of the dual field theory spectrum are not directly applicable so one can only use the gravity realisation of the dimensional reduction. On the gravity side one can use the relation (\ref{red-map}) between five-dimensional and four-dimensional fields to relate the asymptotic expansions in five dimensions to those in four dimensions. The Scherk-Schwarz reduction (\ref{red-map}) does not however specify whether the Killing vector $\partial_{u}$ is timelike, spacelike or null; (\ref{red-map}) is not in itself sufficient to define the notion of asymptotically locally Lifshitz invariance in the reduced spacetime. The authors of \cite{Chemissany:2012du} therefore defined an asymptotically locally Lifshitz spacetime to be one for which the Killing vector is asymptotically hypersurface orthogonal and null and in addition the five-dimensional sources (\ref{upstairs}) satisfy
\begin{eqnarray}
\phi^{(0)} &=& {\rm constant}; \\
\chi^{(0)} &=& k u + \tilde{\chi}^{(0)}(x). \nonumber
\end{eqnarray}
In the four-dimensional language the corresponding boundary conditions for the fields (\ref{red-map}) are 
\begin{eqnarray}
h_{tt} & \sim & \frac{h^{(0)}_{tt}}{r^2} \qquad h_{ti} \sim h^{(0)}_{ti} \\
h_{ij} & \sim & h^{(0)}_{ij}  \qquad B_{t} \sim \frac{B^{(0)}_{t}}{r^2} \nonumber \\
B_{i} & \sim & B^{(0)}_{i} \qquad \Phi \sim \Phi^{(0)} \qquad \phi \sim \phi^{(0)}. \nonumber
\end{eqnarray}
It was found in \cite{Chemissany:2012du} that not all of these non-normalisable modes are independent:
\be
h^{(0)}_{tt} = - (B_t^{(0)})^2 e^{2 \Phi^{(0)}} \qquad 
\Phi^{(0)} = \phi^{(0)} + \log \frac{k}{2} \label{rel-sources}
\ee
and $B_{i}^{(0)} = 0$. Further relations between subleading terms in the asymptotic expansions in four dimensions can be found in \cite{Chemissany:2012du}. 

Thus, although the upstairs asymptotically locally AdS model has twelve independent boundary conditions, the reduced theory has fewer boundary conditions, corresponding to fewer sources for operators. From the field theory perspective, the original conformal field theory was first deformed by a Schr\"{o}dinger symmetry preserving deformation and then reduced. Neither the deformation nor the reduction would remove operators from the spectrum, but the non-relativistic dimension for an operator can be irrelevant so that the corresponding source for the operator must either be switched off (to preserve the Lifshitz symmetry in the UV) or treated perturbatively. In other words, the boundary conditions chosen above must implicitly be switching off certain operator sources; we will return to this issue in the next section. 

The relations (\ref{rel-sources}) are similar to the boundary conditions for the massive vector and Einstein-Maxwell-Dilaton models (\ref{bmet2}), (\ref{aboun}), (\ref{bmet3}) and (\ref{bboun}). Indeed one can directly compare with (\ref{bmet3}) and (\ref{bboun}) by setting $\Phi = \log \frac{k}{2} + \phi$; the parameters of the Einstein-Maxwell-Dilaton model are $D = \epsilon = z = 2$, $\xi = \frac{1}{2}$, $\alpha = 1$ and $Z_0 = \frac{1}{4}$. Comparison of the results of the dimensional reduction approach  \cite{Chemissany:2012du} to those of the metric Hamiltonian approach from section \ref{hol-dict} in this limit gives agreement for power law divergent counterterms but not for the logarithmic counter term (the Weyl anomaly); see
\cite{Chemissany:2014xsa} for more details. To complete the analysis of these models, it would be nice to show that the reduction used is consistent and explain the disagreement with the anomaly found in \cite{Chemissany:2014xsa}.

\subsection{Newton-Cartan structure in holographic models}

Following on from the work of  \cite{Chemissany:2012du}, further analysis of the reduced axion-dilaton model has indicated that the dual field theories couple to torsionful Newton-Cartan geometry \cite{Christensen:2013lma,Christensen:2013rfa}.  Given that the bulk models are obtained by the reduction of asymptotically locally 
Schr\"{o}dinger geometries, it is perhaps unsurprising that these holographic Lifshitz theories should admit such additional structure. 

In subsequent works it has been proposed that asymptotically locally Lifshitz solutions may be dual to field theories coupled to Newton-Cartan geometry more generically, i.e. for classes of bulk theories consisting of gravity coupled to a massive vector and scalar which are not obtained from reductions of Schr\"{o}dinger models \cite{Hartong:2014oma,Hartong:2014pma,Hartong:2015wxa}.

Detailed analysis of the holographic dictionary in such generic massive vector and scalar theories has not yet been carried out. In any theory with a $U(1)$ gauge symmetry the spectrum of operators contains not only the stress energy tensor complex but also a number current, as in (\ref{gen-nc}). Yet the holographic analysis of section \ref{hol-dict} identifies sources for the stress energy tensor complex (and additional scalar operators) but does not find sources for number currents. 

The works \cite{Christensen:2013lma,Christensen:2013rfa,Hartong:2014oma,Hartong:2014pma,Hartong:2015wxa} have proposed that there are nonetheless number currents in such theories, which are sourced by certain boundary conditions for bulk fields. Note however that these papers use a different definition of operator source than \cite{Witten:1998qj,Papadimitriou:2004ap,Ross:2009ar,Ross:2011gu,Baggio:2011cp}. In holography the sharpest definition of operator source and expectation values is via the symplectic form; whatever does not appear in the symplectic form is not a source/vev pair. Pure gauge parameters such as the shift and lapse functions do not appear in the symplectic form and are thus not usually viewed as sources. If a free function appears in the Fefferman-Graham expansion but does not appear in the variation of the renormalised onshell action then this function is interpreted as a pure gauge parameter (as it does not appear in the symplectic form) and not as a source. Such gauge degrees of freedom are dual to Ward identities which can be viewed as null operators in the dual theory. 

In contrast to earlier works on holography, the papers \cite{Christensen:2013lma,Christensen:2013rfa,Hartong:2014oma,Hartong:2014pma,Hartong:2015wxa} interpret free functions which appear in the asymptotic expansions of vielbeins etc but which do not appear in the onshell action in terms of sources for physical operators, rather than null operators. It would be interesting to reconcile this perspective with the analysis of section \ref{hol-dict}. 

Another important question is the relationship between the scalar operator discussed in section \ref{hol-dict} and the scalar in the Newton-Cartan approach:the latter has dimension $2 (z-1)$ (see \cite{Hartong:2015wxa}), c.f. the complicated expression for the dimension of the scalar operator given in section \ref{hol-dict}. This mismatch suggests that the operators are redefined in a highly non-trivial way in the Newton-Cartan approach, relative to the holographic renormalisation given in  section \ref{hol-dict}. 

\subsection{Dynamical Newton-Cartan geometry}

Throughout this review we have concentrated on relativistic gravity theories and their field theory duals but following the work of Ho\v{r}ava \cite{Horava:2009uw} there has been a resurgence of interest in non-relativistic gravity theories. In the context of Newton-Cartan geometry it is natural to ask what (non-relativistic) theory of gravity is obtained when one allows the background geometry to become dynamical. This question has been explored recently in a number of works, see \cite{Hartong:2015zia,Bergshoeff:2015uaa,Bergshoeff:2015ija}. 

In section \ref{newton1} we discussed boost invariants: the spatial metric $\hat{\sigma}_{\mu \nu}$; the inverse metric $\sigma^{\mu \nu}$, the velocity field $\hat{v}^{\mu}$, the one-form $n_{\mu}$ and the scalar $\Phi$, see (\ref{inv}). For TTNC geometry there is in addition a torsion vector $\tau_{\mu}$, see (\ref{torsion}). In section \ref{hol-dict} we coupled the non-relativistic theory to background gravity using an ADM parameterisation of the metric i.e.
\be
ds^2 = - N^2 dt^2 + \gamma_{ij} (dx^i + N^i dt) (dx^j + N^j dt),
\ee 
where $N$ is the lapse, $N^i$ is the shift and $\gamma_{ij}$ is the spatial metric. Comparing with (\ref{new-met1}) we see that the one-form $n_{\mu}$ plays the role of the lapse. Newton-Cartan geometry, for which $n$ is exact, corresponds to the case in which $N$ depends only on time, so-called projectable Ho\v{r}ava-Lifshitz gravity. In TTNC 
the lapse $N$ depends on both time and spatial coordinates; this is non-projectable Ho\v{r}ava-Lifshitz gravity. We see therefore that torsional Newton-Cartan geometry is a natural geometric framework underlying Ho\v{r}ava-Lifshitz gravity.

One can define covariant building blocks for an action of a dynamical gravity theory as follows. The extrinsic curvature of spatial slices can be written as 
\be
K_{\mu \nu} = - \hat{\sigma}_{\nu \rho} D_{\mu} \hat{v}^{\rho}
\ee
and one can define an intrinsic spatial curvature tensor $R^{\rho}_{\mu \nu \sigma}$. The invariant integration measure is the determinant of the complete vielbein defined by the one form $n_{\mu}$ and the vielbein for the spatial metric, which is equivalent to the determinant of the metric defined in (\ref{new-met1}). Focussing on three dimensions, the only terms which are relevant or marginal are 
\be
S = \int d^3 x \sqrt{h} \left [ A_1 (\sigma^{\mu \rho} \sigma^{\nu \sigma} K_{\mu \nu} K_{\rho \sigma} - \lambda (\sigma^{\mu \nu} K_{\mu \nu})^2 ) - {\cal V}  \right ]  \label{horlif}
\ee
where the potential ${\cal V}$ for $1 <  z  < 2$ is
\be
{\cal V} = -2 \Lambda + c_1 \tau^{\mu} \tau_{\mu} + c_2 {R}
\ee
while at $z=2$ the following additional terms are allowed
\be
c_3 ( \tau^{\mu} \tau_{\mu})^2 + c_4 (\tau_{\rho} \tau^{\rho}) D_{\nu} \tau^{\nu} + c_5 (D_{\mu} \tau_{\nu})(D^{\nu} \tau^{\mu}) + c_5 R^2 + c_6 R D_{\mu} \tau^{\mu} + c_7 R (\tau_{\mu} \tau^{\mu}), 
\ee
where $\tau^{\mu} = \sigma^{\mu \nu} \tau_{\nu}$. The kinetic terms in (\ref{horlif}) agree with those discussed earlier in the context of anomalies for $z=2$ Lifshitz theories, see section \ref{anomalie}, while the potential terms are generalised to include torsion. 
The case of dynamical TNC (as opposed to TTNC) has not yet been fully explored. Newton-Cartan supergravity has been developed in \cite{Bergshoeff:2015uaa,Bergshoeff:2015ija}. 

\bigskip

Dynamical Newton-Cartan structure has also arisen in the context of non-relativistic limits of holography. Using relativistic gravity in the bulk to realise a dual non-relativistic field theory is somewhat unnatural: as we saw in section \ref{hol-dict} the boundary degrees of freedom for a bulk relativistic theory need to be organised into non-relativistic operator multiplets. It is more natural to try to realise holographic duals for non-relativistic field theories using non-relativistic gravity. In \cite{Bagchi:2009my,Bagchi:2009pe} it was proposed that one could take non-relativistic limits of AdS/CFT to obtain holographic dualities between bulk Newton-Cartan theories and boundary Galilean conformal field theories. 

Proposals for non-relativistic holographic duals of field theories which have a number current may be found in  
\cite{Janiszewski:2012nb}, and their interpretations in terms of non-relativistic scaling limits of relativistic gauge/gravity dualities are discussed in \cite{Janiszewski:2012nf}. Further analysis of non-relativistic limits from the field theory perspective was given in \cite{Jensen:2014wha}.
Arguments that Ho\v{r}ava-Lifshitz gravity is the minimal holographic dual for Lifshitz field theories can be found in \cite{Griffin:2012qx}. Finally, we should mention that one can also consider ultra-relativistic limits in which one sends the speed of light to zero; such limits and the corresponding geometry of Carrollian spacetimes were explored in \cite{Bekaert:2015xua,Hartong:2015xda}.  

\subsection{Summary}

Newton-Caftan structure arises naturally in coupling Galilean invariant theories to background gravity. At least some holographic realisations of Lifshitz (those related to Schr\"{o}dinger models in one higher dimension) admit Newton-Cartan structure. There is currently an active research programme to understand whether and how Newton-Cartan structure is realised in generic holographic Lifshitz models, and to explore dynamical Newton-Cartan gravity. 

\section{Other developments in Lifshitz holography} \label{seven}

\subsection{Entanglement entropy}

Ryu and Takayanagi proposed in \cite{Ryu:2006bv} that entanglement entropy can be computed holographically in terms of the area of a codimension two (static) minimal surface. This proposal has undergone many checks in the context of AdS/CFT and has also been applied to other holographic backgrounds. For Lifshitz backgrounds static codimension two minimal surfaces are insensitive to the dynamical exponent $z$; all computations of the entanglement entropy for static cases therefore agree exactly with those for anti-de Sitter. While this result is trivial from the gravity side, it is a non-trivial statement about the universality class of the dual Lifshitz field theories. For example, the entanglement entropy for a spatial interval of length $l$ in the ground state of a two-dimensional CFT $S_{EE}$ is 
given by 
\be
S_{\rm EE} = \frac{c}{3} \log \left ( \frac{l}{\epsilon} \right ) 
\ee
where $\epsilon$ is the UV cutoff and $c$ is the central charge, which is related to the Newton constant in $AdS$ as $c  = 3/2G_{3}$. Precisely the same expression holds for holographic Lifshitz models; it would be interesting to relate the coefficient $c$ to correlation functions of the energy momentum complex. 

One way to probe the effect of the dynamical exponent is to consider the time evolution of the entanglement entropy for a boundary region, see \cite{Alishahiha:2014cwa}. 
Non-trivial behaviour is also found for the entanglement entropy in excited states: a first law of entanglement entropy is found relating the change in entanglement entropy to a change in energy, but the massive vector field in the bulk contributes to this relation \cite{Chakraborty:2014lfa}. Entanglement entropy in finite temperature Lifshitz was explored in \cite{Fischetti:2014zja}; again one would like to understand the implications of the holographic results for the nature of the dual Lifshitz theories. Entanglement in two-dimensional Lifshitz theories at zero and finite temperature was recently analysed in \cite{Hosseini:2015gua}.

\subsection{Modelling of CMT systems}

As mentioned in the introduction, non-relativistic holography was motivated originally by applications to condensed matter physics. Despite the subtleties in the holographic dictionary, Lifshitz and related models have been widely used to model condensed matter phenomena such as strange metal phases (relevant to high temperature superconductors). 
The reviews  \cite{Sachdev:2011wg,Adams:2012th} include various applications of non-relativistic holography to condensed matter physics.  

An interesting application of non-relativistic holography is to unconventional superconductors, in particular, to the strange metal phase above the critical temperature. An early attempt at model building for strange metal phases was initiated in \cite{Hartnoll:2009ns}: this work considered D-brane probes in a finite temperature Lifshitz background, to model charge carriers coupled to a Lifshitz invariant field theory. Transport coefficients such as the resistivity $\rho(T)$ and the ac conductivity $\sigma(\omega)$  were found to exhibit scaling behaviour as functions of the temperature $T$ and the frequency $\omega$:
\be
\rho(T) \sim T^{v_1} \qquad
\sigma(\omega) \sim \omega^{-v_2}
\ee
A range of values of $(v_1,v_2)$ can be realised in Einstein-Maxwell-Dilaton models and in particular one can obtain the linear scaling of the resistivity with temperature ($v_1  =1$) and $v_2 \sim 2/3$ observed in strange metal phases. (Both exponents cannot be simultaneously realised in such models, however.) Generalised Einstein-Maxwell-Dilaton holographic models for quantum critical theories  were proposed and explored in  \cite{Charmousis:2010zz,Gouteraux:2011ce}. 

Hyperscaling geometries (\ref{hv-geom}) were introduced in \cite{Huijse:2011ef} to describe compressible states with hidden Fermi surfaces. As discussed in section \ref{four} the scaling of the entropy density is fixed by the dynamical exponent $z$ and the hyperscaling violation $\theta$; this scaling agrees with what one would expect for a compressible state. One can also show that the dependence of the entanglement entropy on the shape of the entangling region, temperature and charge density agrees with what would expect for such compressible states. Following  \cite{Huijse:2011ef}, a number of works explored holography for hyperscaling geometries, see for example \cite{Dong:2012se,Gath:2012pg}.

We should note that hyperscaling models of the type analysed in section \ref{hol-dict} are unlikely to be sufficient to describe strange metal phases. To discuss charge transport, one needs to work at finite charge density and thus there must exist a gauge field in the bulk, whose boundary behavior captures the chemical potential and charge density in the field theory. While the models discussed in section \ref{hol-dict} do have a bulk vector field, this field is related to the non-relativistic energy momentum complex rather than to a charge current in the field theory. We will come back to this point below. 

To understand the scaling properties of generic scaling regimes arising in compressible matter, it has been proposed that (at least) three exponents are needed: the dynamical scaling exponent $z$, the hyperscaling violation exponent $\theta$ and a third exponent $\delta_{\Phi}$. As in earlier sections, the hyperscaling violation and dynamical scaling determine the scaling dimensions of the entropy density and of free energy density:
\be
[ s ] = D - \theta \qquad
[ {\cal E} ] = z  + D - \theta 
\ee 
where $D$ is the number of spatial dimensions. The third exponent $\delta_\Phi$  characterises the scaling of the charge density operator $n$:
\be
[ n] = D - \theta + \delta_{\Phi}
\ee
and this in turn determines the scaling of the chemical potential $\mu$, the electric field $E$ and the magnetic field $B$:
\be
[ \mu ] = z - \delta_{\Phi} \qquad
[ E] = 1 + z - \delta_{\Phi} \qquad
[ B] = 2 - \delta_{\Phi}. 
\ee
It was shown in \cite{Hartnoll:2015sea} that the experimentally observed scaling behaviour in cuprates of the electrical resistivity, the Hall angle, the Hall Lorenz ratio, the magnetoresistance and the thermopower can all be reproduced within a model with $z = 4/3$, $\theta = 0$ and $\delta_\Phi = -2/3$. A non-trivial $\delta_{\Phi}$ seems essential to reproduce the required scalings but no hyperscaling violation is required. Note that there seems to be some tension between achieving the required scalings of transport coefficients together with the required scalings of thermodynamic quantities; see \cite{Hartnoll:2015sea} for further details. Moreover the postulated scalings of the electromagnetic field are somewhat puzzling since they would not appear to be consistent with the dimensions of a conserved current in a non-relativistic theory.  

The exponent $\delta_{\Phi}$ has also been found to be present in generic holographic models in \cite{Gouteraux:2012yr,Gath:2012pg,Gouteraux:2013oca,Karch:2014mba}. For example, in  \cite{Gouteraux:2012yr}, generic actions of the type (\ref{cp}) were explored, and scale invariant solutions were classified. By allowing the potential to be exponential ($V = V_0 \exp(-\delta \phi)$), and tuning the other potentials and parameters appropriately, one can indeed find scale invariant solutions in which the bulk vector field scales independently to the metric. In Einstein frame these solutions are of the form 
\begin{eqnarray}
ds^2 &=& r^{\theta} \left [ \frac{L^2}{r^2 } dr^2 + \frac{1}{r^2} (dx^2 + dy^2) - \frac{dt^2}{r^{2z}} \right ]; \\
B &=& B_0 r^{\zeta -z} dt; \nonumber \\
\phi &=& \frac{\theta}{\delta (z,\theta,\zeta)} \ln(r), \nonumber
\end{eqnarray}
where $(L,B_0)$ are fixed in terms of the exponents and $V_0$ and $Z_0$, while $\delta$ depends on a specific combination of the three exponents $(z,\zeta,\theta)$. The exponent $\zeta$ is linearly related to $z$ and the exponent $\delta_{\Phi}$ above. 

The exponent $\zeta$ does not affect the scaling of the entropy and thermodynamic quantities would therefore seem to scale as in the hyperscaling violating models discussed earlier. In \cite{Gouteraux:2012yr} the exponent $\zeta$ was interpreted in terms of the impact of the charge density on the physics of the dual critical theory. The detailed analysis of the holographic dictionary in section \ref{hol-dict} however suggests that one should not interpret the bulk vector field as dual to a current. In the models with constant potential analysed in section \ref{hol-dict} the bulk vector field was always associated with the energy momentum complex of the dual field theory, rather than a dual vector operator/current, and 
analogous behaviour is found for general exponential potentials in \cite{Chemissany:2014xpa,Chemissany:2014xsa}. Note that the latter analysis parameterises the solutions as in section \ref{hol-dict}, in terms of two dynamical exponents
$(z,\theta)$ and a running coupling $\mu$, rather than in terms of a third dynamical exponent $\zeta$. 

In the AdS/CMT literature it is common to analyse solutions of EInstein-Maxwell-Dilaton models which are asymptotically anti-de Sitter (hence dual to UV conformal theories) but which approach non-relativistic fixed points in the interior (dual to IR fixed points). One can discuss properties of such solutions either in the language of the UV relativistic field theory or in the language of the IR non-relativistic field theory. The operator content of the former is a relativistic stress energy tensor, a current and a scalar operator and it is common to discuss hydrodynamics in terms of these relativistic operators. As we have emphasised above, the operator content of the non-relativistic field theory is the energy momentum complex and scalar operators (no current). Since hydrodynamics is by definition applicable at low energies, it would be interesting to express the hydrodynamics in such models in terms of the non-relativistic operators of the IR theory.

\section{Conclusions and key open questions} \label{eight}

The main focus of this review has been answering the questions: what is the nature of field theories dual to Lifshitz gravitational backgrounds?  what is the interpretation of the bulk matter supporting the Lifshitz background? 

The latter question was addressed in detail in sections \ref{hol-dict} and \ref{newton}. In the simplest model, Einstein-Proca theory, the profile for the massive vector breaks the relativistic symmetry and the vector combines with the metric to source the dual non-relativistic energy momentum complex, together with an additional scalar operator. More generally, part of the bulk matter is always associated with the dual energy momentum complex in holographic hyperscaling violating models, with the remaining degrees of freedom being associated with additional operators. 

We would expect analogous behaviour in realisations of Lifshitz using higher derivative gravity theories. The additional boundary conditions associated with higher derivative equations of motion are associated with sources for (new) operators in the dual field theory \cite{Skenderis:2009nt}. In the context of Lifshitz the complete set of boundary conditions for higher derivative gravity should provide the required sources for the non-relativistic energy momentum complex, together with certain other operators. 

Turning to the nature of the dual field theories, we showed in section \ref{three} that holographic correlation functions have qualitatively different analytic structure to that of correlation functions in perturbative Lifshitz models, perhaps unsurprisingly given the lack of quasi-particles in the holographic realisations. Lifshitz symmetry does not suffice to determine completely even two functions and thus two point functions are a prediction of holographic models. As we discussed in section \ref{four} holographic Lifshitz models can also be characterised by their hydrodynamic transport coefficients, with certain predictions such as shear viscosity to entropy ratio, and bulk viscosity, being very generic. It would be interesting to explore further whether there are any condensed matter systems whose properties are captured by holographic models. 

Despite considerable progress, a number of issues remain open in Lifshitz holography. For example, calculations in Lifshitz are often technically challenging. As we discussed in the context of black holes, numerics is often required for problems which could be solved analytically in anti-de Sitter. Holographic renormalization is considerably more complicated for Lifshitz than for relativistic cases and further progress would probably require implementing the procedure in a symbolic computation package. Many calculations have not yet been extended to Lifshitz because of technical challenges e.g. real time holography or because of the lack of top down embeddings e.g. three point and higher correlation functions. (The latter require knowledge of interactions and are therefore most sensibly computed in a top down model.)

Another outstanding issue is the role of Newton-Cartan geometry and enhanced symmetry of holographic Lifshitz theories. As discussed in section \ref{newton} it would be nice to understand further what classes of holographic models admit enhanced symmetry. A related question is whether bulk relativistic gravity is the most appropriate way to realise non-relativistic holography; it would seem much more natural to use non-relativistic gravity in the bulk and the work reviewed in section \ref{newton} provides a basis for such an approach.  

Finally, while the original motivation for Lifshitz holography was condensed matter physics, most works in AdS/CMT still use relativistic holography, in part because the holographic dictionary is better understood and easier to use. At the end of section \ref{seven} we mentioned the uses of hyperscaling violating geometries in modelling compressible states and possibly strange metal phases. However, the results of section \ref{hol-dict} suggest that the dual interpretation of hyperscaling geometries may differ from that given in the AdS/CMT literature and it would be interesting to revisit the CMT interpretation of these models. 
 
\section*{Acknowledgments}

We thank Kristan Jensen, Niels Obers, Ioannis Papadimitriou and Kostas Skenderis for discussions and correspondence. 
This work was supported by the Science and Technology Facilities Council (Consolidated Grant ``Exploring the Limits of the Standard Model and Beyond'') and by the Engineering and Physical Sciences Research Council. 
This work was also supported in part by National Science Foundation Grant No. PHYS-1066293 and the hospitality of the Aspen Center for Physics. We thank the Galileo Galilei Institute for Theoretical Physics for hospitality and the INFN for partial support during the completion of this work. We also thank the Simons Center for Geometry and Physics for hospitality during the 2015 summer workshop.


\section*{References}
\begin{thebibliography}{10}

\bibitem{Niederer:1972zz}
  U.~Niederer,
  ``The maximal kinematical invariance group of the free Schrodinger equation,''
  Helv.\ Phys.\ Acta {\bf 45} (1972) 802.


\bibitem{Kachru:2008yh}
S.~Kachru, X.~Liu, and M.~Mulligan, ``{Gravity Duals of Lifshitz-like Fixed
  Points},'' \href{http://dx.doi.org/10.1103/PhysRevD.78.106005}{{\em
  Phys.Rev.} {\bf D78} (2008)  106005},
\href{http://arxiv.org/abs/0808.1725}{{\tt [arXiv:0808.1725 [hep-th]]}}.

\bibitem{Hoyos:2010at}
  C.~Hoyos and P.~Koroteev,
  ``On the Null Energy Condition and Causality in Lifshitz Holography,''
{\href{ http://dx.doi.org/10.1103/PhysRevD.82.084002}
{ {\em Phys.\ Rev.\ D } {\bf 82} (2010) 084002,}}
 {\href{ http://dx.doi.org/10.1103/PhysRevD.82.109905}
  {  [ {\em Phys.\ Rev.\ D } {\bf 82} (2010) 109905]}}
\href{  http://arxiv.org/abs/arXiv:1007.1428}
{\tt  [arXiv:1007.1428 [hep-th]]}.



\bibitem{Taylor:2008tg}
M.~Taylor, ``{Non-relativistic holography},''
\href{http://arxiv.org/abs/0812.0530}{{\tt arXiv:0812.0530 [hep-th]}}.

\bibitem{AyonBeato:2009nh}
  E.~Ayon-Beato, A.~Garbarz, G.~Giribet and M.~Hassaine,
  ``Lifshitz Black Hole in Three Dimensions,''
 {\href{http://dx.doi.org/10.1103/PhysRevD.80.104029}
 { {\em Phys.\ Rev.\ D } {\bf 80} (2009) 104029 }}
{\href{ http://arxiv.org/abs/arXiv:0909.1347}
 {\tt [arXiv:0909.1347 [hep-th]]}}.

\bibitem{Bergshoeff:2009hq}
  E.~A.~Bergshoeff, O.~Hohm and P.~K.~Townsend,
  ``Massive Gravity in Three Dimensions,''
{\href{  http://dx.doi.org/10.1103/PhysRevLett.102.201301}
{ {\em  Phys.\ Rev.\ Lett.\  } {\bf 102} (2009) 201301,}}
{\href { http://arxiv.org/abs/arXiv:0901.1766}
{\tt  [arXiv:0901.1766 [hep-th]]}}.


\bibitem{Cai:2009ac}
  R.~G.~Cai, Y.~Liu and Y.~W.~Sun,
  ``A Lifshitz Black Hole in Four Dimensional R**2 Gravity,''
{\href{  http://dx.doi.org/10.1088/1126-6708/2009/10/080}
{ {\em  JHEP} {\bf 0910} (2009) 080,}}
{\href {http://arxiv.org/abs/arXiv:0909.2807}
{\tt   [arXiv:0909.2807 [hep-th]]}}.


\bibitem{AyonBeato:2010tm}
  E.~Ayon-Beato, A.~Garbarz, G.~Giribet and M.~Hassaine,
  ``Analytic Lifshitz black holes in higher dimensions,''
 {\href{http://dx.doi.org/10.1007/JHEP04(2010)030,}
 {{\em  JHEP} {\bf 1004} (2010) 030}}
{\href{ http://arxiv.org/abs/arXiv:1001.2361}{ \tt  [arXiv:1001.2361 [hep-th]]}}.

\bibitem{Brenna:2011gp}
  W.~G.~Brenna, M.~H.~Dehghani and R.~B.~Mann,
  ``Quasi-Topological Lifshitz Black Holes,''
 {\href{ http://dx.doi.org/10.1103/PhysRevD.84.024012}
 { {\em    Phys.\ Rev.\ D } {\bf 84} (2011) 024012}}
  {\href{http://arxiv.org/abs/arXiv:1101.3476}
{\tt  [arXiv:1101.3476 [hep-th]]}}.
 
   \bibitem{Gonzalez:2011nz}
  H.~A.~Gonzalez, D.~Tempo and R.~Troncoso,
  ``Field theories with anisotropic scaling in 2D, solitons and the microscopic entropy of asymptotically Lifshitz black holes,''
{ \href{ http://dx.doi.org/10.1007/JHEP11(2011)066}
{ {\em JHEP} {\bf 1111} (2011) 066}}
 \href{ http://arxiv.org/abs/arXiv:1107.3647}
{\tt  [arXiv:1107.3647 [hep-th]]}. 
  
  
\bibitem{Dehghani:2011hf}
  M.~H.~Dehghani and S.~Asnafi,
  ``Thermodynamics of Rotating Lovelock-Lifshitz Black Branes,''
{\href{ http://dx.doi.org/10.1103/PhysRevD.84.064038}
{ {\em  Phys.\ Rev.\ D } {\bf 84} (2011) 064038}}
{\href{http://arxiv.org/abs/arXiv:1107.3354 } 
{\tt   [arXiv:1107.3354 [hep-th]].}}
  


\bibitem{Liu:2012yd}
  Y.~Liu,
  ``Spatially homogeneous Lifshitz black holes in five dimensional higher derivative gravity,''
{\href{  http://dx.doi.org/10.1007/JHEP06(2012)024}
{{\em  JHEP} {\bf 1206} (2012) 024,}}
{\href{http://arxiv.org/abs/arXiv:1202.1748}{\tt  [arXiv:1202.1748 [hep-th]]}}.
  
  
  
\bibitem{Lu:2012xu}
  H.~Lu, Y.~Pang, C.~N.~Pope and J.~F.~Vazquez-Poritz,
  ``AdS and Lifshitz Black Holes in Conformal and Einstein-Weyl Gravities,''
 {\href{http://dx.doi.org/10.1103/PhysRevD.86.044011}
{ {\em  Phys.\ Rev.\ D } {\bf 86} (2012) 044011,}}
{\href{http://arxiv.org/abs/arXiv:1204.1062}
{\tt  [arXiv:1204.1062 [hep-th]]}}.

\bibitem{Ghanaatian:2014bpa}
  M.~Ghanaatian, A.~Bazrafshan and W.~G.~Brenna,
  ``Lifshitz Quartic Quasitopological Black Holes,''
{\href{  http://dx.doi.org/10.1103/PhysRevD.89.124012}
{ {\em  Phys.\ Rev.\ D } {\bf 89} (2014) 12,  124012}}
{\href{  http://arxiv.org/abs/arXiv:1402.0820}
{\tt  [arXiv:1402.0820 [hep-th]]}}.

\bibitem{Matulich:2011ct}
  J.~Matulich and R.~Troncoso,
  ``Asymptotically Lifshitz wormholes and black holes for Lovelock gravity in vacuum,''
{\href{http://dx.doi.org/10.1007/JHEP10(2011)118}
{ {\em  JHEP} {\bf 1110} (2011) 118}}
{\href{http://arxiv.org/abs/arXiv:1107.5568}
{\tt  [arXiv:1107.5568 [hep-th]]}}.


  
\bibitem{Hung:2010te}
  L.~Y.~Hung, D.~P.~Jatkar and A.~Sinha,
  ``Non-relativistic metrics from back-reacting fermions,''
{\href { http://dx.doi.org/10.1088/0264-9381/28/1/015013}
{ {\em  Class.\ Quant.\ Grav.\}  {\bf 28} (2011) 015013, }}}
{\href{http://arxiv.org/abs/arXiv:1006.3762}
{\tt  [arXiv:1006.3762 [hep-th]]}}.

\bibitem{Maeda:2011jj}
  H.~Maeda and G.~Giribet,
 ``Lifshitz black holes in Brans-Dicke theory,''
{\href{http://dx.doi.org/10.1007/JHEP11(2011)015} {\em   JHEP {\bf 1111} (2011) 015,}}
{\href{http://arxiv.org/abs/arXiv:1105.133}{ \tt arXiv:1105.1331 [gr-qc]}}.

\bibitem{Gutperle:2013oxa}
  M.~Gutperle, E.~Hijano and J.~Samani,
  ``Lifshitz black holes in higher spin gravity,''
 {\href{http://dx.doi.org/10.1007/JHEP04(2014)020}
 { {\em JHEP} {\bf 1404} (2014) 020,}}
 {\href {http://arxiv.org/abs/arXiv:1310.0837}
{\tt  [arXiv:1310.0837 [hep-th]]}}.

\bibitem{Beccaria:2015iwa}
  M.~Beccaria, M.~Gutperle, Y.~Li and G.~Macorini,
  ``Higher spin Lifshitz theories and the Korteweg-de Vries hierarchy,''
{\href{  http://dx.doi.org/10.1103/PhysRevD.92.085005}
{ {\em  Phys.\ Rev.\ D } {\bf 92} (2015) 8,  085005,}}
\href{  http://arxiv.org/abs/arXiv:1504.06555}
{\tt  [arXiv:1504.06555 [hep-th]]}.



\bibitem{Huijse:2011ef}
  L.~Huijse, S.~Sachdev and B.~Swingle,
  ``Hidden Fermi surfaces in compressible states of gauge-gravity duality,''
{\href{  http://dx.doi.org/10.1103/PhysRevB.85.035121}
{ {\em  Phys.\ Rev.\ B } {\bf 85} (2012) 035121,}}
 \href{ http://arxiv.org/abs/arXiv:1112.0573}
{\tt  [arXiv:1112.0573 [cond-mat.str-el]]}.
 
 
 
\bibitem{Gath:2012pg}
  J.~Gath, J.~Hartong, R.~Monteiro and N.~A.~Obers,
  ``Holographic Models for Theories with Hyperscaling Violation,''
 {\href{http://dx.doi.org/10.1007/JHEP04(2013)159}
{ {\em  JHEP} {\bf 1304} (2013) 159,}}
 \href{ http://arxiv.org/abs/arXiv:1212.3263}
{\tt   [arXiv:1212.3263 [hep-th]]}.
  
\bibitem{Gouteraux:2012yr}
  B.~Gouteraux and E.~Kiritsis,
  ``Quantum critical lines in holographic phases with (un)broken symmetry,''
{\href{ http://dx.doi.org/10.1007/JHEP04(2013)053}
{ {\em  JHEP} {\bf 1304} (2013) 053,}}
\href{  http://arxiv.org/abs/arXiv:1212.2625}
{\tt  [arXiv:1212.2625 [hep-th]]}.

 
\bibitem{Chemissany:2014xpa}
  W.~Chemissany and I.~Papadimitriou,
  ``Generalized dilatation operator method for non-relativistic holography,''
  \href{http://dx.doi.org/10.1016/j.physletb.2014.08.057}
  {{ \em Phys.\ Lett.\ B} {\bf 737} (2014) 272,}
  \href{ http://arxiv.org/abs/arXiv:1405.3965}
  {{ \tt [arXiv:1405.3965 [hep-th]]}}.

\bibitem{Chemissany:2014xsa}
  W.~Chemissany and I.~Papadimitriou,
  ``Lifshitz holography: The whole shebang,''
 {\href{ http://dx.doi.org/10.1007/JHEP01(2015)052}
  {{\em JHEP} {\bf 1501} (2015) 052,}}
  \href{http://arxiv.org/abs/arXiv:1408.0795}
  {{\tt [arXiv:1408.0795 [hep-th]]}}.


\bibitem{Balasubramanian:2008dm}
K.~Balasubramanian and J.~McGreevy, ``{Gravity duals for non-relativistic
  CFTs},'' \href{http://dx.doi.org/10.1103/PhysRevLett.101.061601}{{\em
  Phys.Rev.Lett.} {\bf 101} (2008)  061601},
\href{http://arxiv.org/abs/0804.4053}{{\tt [arXiv:0804.4053 [hep-th]]}}.

\bibitem{Son:2008ye}
D.~Son, ``{Toward an AdS/cold atoms correspondence: A Geometric realization of
  the Schrodinger symmetry},''
  \href{http://dx.doi.org/10.1103/PhysRevD.78.046003}{{\em Phys.Rev.} {\bf D78}
  (2008)  046003},
\href{http://arxiv.org/abs/0804.3972}{{\tt [arXiv:0804.3972 [hep-th]]}}.


\bibitem{Maldacena:2008wh}
J.~Maldacena, D.~Martelli, and Y.~Tachikawa, ``{Comments on string theory
  backgrounds with non-relativistic conformal symmetry},''
  \href{http://dx.doi.org/10.1088/1126-6708/2008/10/072}{{\em JHEP} {\bf 0810}
  (2008)  072},
\href{http://arxiv.org/abs/0807.1100}{{\tt [arXiv:0807.1100 [hep-th]]}}.

\bibitem{Herzog:2008wg}
  C.~P.~Herzog, M.~Rangamani and S.~F.~Ross,
  ``Heating up Galilean holography,''
 {\href{http://dx.doi.org/10.1088/1126-6708/2008/11/080}
{  {\em  JHEP} {\bf 0811} (2008) 080,}}
\href{  http://arxiv.org/abs/arXiv:0807.1099}
{\tt   [arXiv:0807.1099 [hep-th]]}.

\bibitem{Kraus}
P.~Kraus and E.~Perlmutter,
  ``Universality and exactness of Schrodinger geometries in string and M-theory,''
 \href{ http://dx.doi.org/10.1007/JHEP05(2011)04} {{\em JHEP} {\bf 1105}, 045 (2011)},
 \href{http://arxiv.org/abs/arXiv:1102.1727}{{\tt [arXiv:1102.1727]}}.



\bibitem{Guica:2010sw}
M.~Guica, K.~Skenderis, M.~Taylor, and B.~C. van Rees, ``{Holography for
  Schrodinger backgrounds},''
  \href{http://dx.doi.org/10.1007/JHEP02(2011)056}{{\em JHEP} {\bf 1102} (2011)
   056},
\href{http://arxiv.org/abs/1008.1991}{{\tt [arXiv:1008.1991 [hep-th]]}}.

\bibitem{Balasubramanian:2010uk}
K.~Balasubramanian and K.~Narayan, ``{Lifshitz spacetimes from AdS null and
  cosmological solutions},''
  \href{http://dx.doi.org/10.1007/JHEP08(2010)014}{{\em JHEP} {\bf 1008} (2010)
   014},
\href{http://arxiv.org/abs/1005.3291}{{\tt [arXiv:1005.3291 [hep-th]]}}.

\bibitem{Costa:2010cn}
R.~Caldeira~Costa and M.~Taylor, ``{Holography for chiral scale-invariant
  models},'' \href{http://dx.doi.org/10.1007/JHEP02(2011)082}{{\em JHEP} {\bf
  1102} (2011)  082},
\href{http://arxiv.org/abs/1010.4800}{{\tt [arXiv:1010.4800 [hep-th]]}}.

  \bibitem{Singh:2010zs}
  H.~Singh,
  ``Special limits and non-relativistic solutions,''
{\href{  http://dx.doi.org/10.1007/JHEP12(2010)061}
{ {\em  JHEP} {\bf 1012} (2010) 061,}}
 \href{http://dx.doi.org/10.1007/JHEP12(2010)061}
{\tt  [arXiv:1009.0651 [hep-th]]}.

\bibitem{Singh:2012un}
  H.~Singh,
  ``Lifshitz/Schr\'odinger Dp-branes and dynamical exponents,''
{\href{  http://dx.doi.org/10.1007/JHEP07(2012)082}
{ {\em  JHEP} {\bf 1207} (2012) 082,}}
 \href{ http://arxiv.org/abs/arXiv:1202.6533}
{\tt  [arXiv:1202.6533 [hep-th]]}.


\bibitem{Narayan:2011az}
K.~Narayan, ``{{Lifshitz-like systems and AdS null deformations}},"  
{\href{http://dx.doi.org/10.1103/PhysRevD.84.086001}
{ {\em Phys.Rev.} {\bf D84} (2011) 086001,}}
  \href{http://xxx.lanl.gov/abs/1103.1279}{{\tt [arXiv:1103.1279]}}.



\bibitem{Donos:2010tu}
A.~Donos and J.~P. Gauntlett, ``{Lifshitz Solutions of D=10 and D=11
  supergravity},'' \href{http://dx.doi.org/10.1007/JHEP12(2010)002}{{\em JHEP}
  {\bf 1012} (2010)  002},
\href{http://arxiv.org/abs/1008.2062}{{\tt [arXiv:1008.2062 [hep-th]]}}.

\bibitem{Cassani:2011sv}
D.~Cassani and A.~F. Faedo, ``{Constructing Lifshitz solutions from AdS},''
  \href{http://dx.doi.org/10.1007/JHEP05(2011)013}{{\em JHEP} {\bf 1105} (2011)
   013},
\href{http://arxiv.org/abs/1102.5344}{{\tt [arXiv:1102.5344 [hep-th]]}}.





\bibitem{Halmagyi:2011xh}
N.~Halmagyi, M.~Petrini, and A.~Zaffaroni, ``{Non-Relativistic Solutions of
  N=2 Gauged Supergravity}",  \href{http://dx.doi.org/10.1007/JHEP08(2011)041}{{\em JHEP} {\bf 1108} (2011) 041},
  \href{http://xxx.lanl.gov/abs/1102.5740}{{\tt [arXiv:1102.5740 [hep-th]]}}.



\bibitem{Petrini:2012bh}
M.~Petrini and A.~Zaffaroni, {A Note on Supersymmetric Type II Solutions
  of Lifshitz Type},  \href{http://dx.doi.org/10.1007/JHEP07(2012)051}{{\em JHEP} {\bf 1207} (2012) 051},
  \href{http://xxx.lanl.gov/abs/1202.5542}{{\tt [arXiv:1202.5542 [hep-th]] }}.

\bibitem{Cardoso:2015wcf}
  G.~L.~Cardoso, M.~Haack and S.~Nampuri,
  ``Nernst branes with Lifshitz asymptotics in N=2 gauged supergravity,''
{\href{  http://arxiv.org/abs/arXiv:1511.07676}
{\tt  arXiv:1511.07676 [hep-th]}}.




\bibitem{Copsey:2010ya}
  K.~Copsey and R.~Mann,
  ``Pathologies in Asymptotically Lifshitz Spacetimes,''
\href{http://dx.doi.org/10.1007/JHEP03(2011)039} {\em JHEP {\bf 1103} (2011) 039},
\href{http://arxiv.org/abs/arXiv:1011.3502}{{\tt  arXiv:1011.3502 [hep-th]}}.

\bibitem{Keeler:2013msa} 
  C.~Keeler, G.~Knodel and J.~T.~Liu,
  ``What do non-relativistic CFTs tell us about Lifshitz spacetimes?,''
  {\href{http://dx.doi.org/10.1007/JHEP01(2014)062}
  { {\em JHEP} {\bf 1401}, (2014) 062,}}
 \href {http://arxiv.org/abs/arXiv:1308.5689}
  {\tt [arXiv:1308.5689 [hep-th]]}.

\bibitem{Keeler:2014lia} 
  C.~Keeler, G.~Knodel and J.~T.~Liu,
  ``Hidden horizons in non-relativistic AdS/CFT,''
{ \href{ http://dx.doi.org/10.1007/JHEP08(2014)024}
  { {\em JHEP} {\bf 1408}, 024 (2014),}}
 \href{ http://arxiv.org/abs/arXiv:1404.4877}
  {\tt [arXiv:1404.4877 [hep-th]]}.
  
 \bibitem{Keeler:2015afa}
  C.~Keeler, G.~Knodel, J.~T.~Liu and K.~Sun,
  ``Universal features of Lifshitz GreenÕs functions from holography,''
 { \href{http://dx.doi.org/10.1007/JHEP08(2015)057}
  {{\em JHEP} {\bf 1508} (2015) 057,}}
 \href{http://arxiv.org/abs/arXiv:1505.07830}
  {\tt [arXiv:1505.07830 [hep-th]]}.
 
 \bibitem{Skenderis:2008dh}
  K.~Skenderis and B.~C.~van Rees,
  ``Real-time gauge/gravity duality,''
  {\href{http://dx.doi.org/10.1103/PhysRevLett.101.081601}
   {{\em Phys.\ Rev.\ Lett.\ }  {\bf 101} (2008) 081601,}}
  {\href{http://arxiv.org/abs/arXiv:0805.0150} {\tt [arXiv:0805.0150 [hep-th]]}}.

 \bibitem{Skenderis:2008dg} 
  K.~Skenderis and B.~C.~van Rees,
  ``Real-time gauge/gravity duality: Prescription, Renormalization and Examples,''
  {\href{http://dx.doi.org/10.1088/1126-6708/2009/05/085}
   {{\em JHEP} {\bf 0905}, 085 (2009),}}
  \href{http://arxiv.org/abs/arXiv:0812.2909}
{\tt   [arXiv:0812.2909 [hep-th]]}.
  

\bibitem{Horowitz:2011gh}
  G.~T.~Horowitz and B.~Way,
  ``Lifshitz Singularities,''
\href{  http://dx.doi.org/10.1103/PhysRevD.85.046008}
{  {\em Phys.\ Rev.\ D {\bf 85} (2012) 046008,} }
\href{http://arxiv.org/abs/arXiv:1111.1243}
 {\tt  [arXiv:1111.1243 [hep-th]]}.

\bibitem{Bao:2012yt}
  N.~Bao, X.~Dong, S.~Harrison and E.~Silverstein,
  ``The Benefits of Stress: Resolution of the Lifshitz Singularity,''
\href{http://dx.doi.org/10.1103/PhysRevD.86.106008}
{ {\em   Phys.\ Rev.\ D } {\bf 86} (2012) 106008,}
\href{http://arxiv.org/abs/arXiv:1207.0171}
{\tt   [arXiv:1207.0171 [hep-th]]}.
  
\bibitem{Harrison:2012vy}
  S.~Harrison, S.~Kachru and H.~Wang,
  ``Resolving Lifshitz Horizons,''
\href{  http://dx.doi.org/10.1007/JHEP02(2014)085}
 { {\em JHEP }{\bf 1402} (2014) 085,}
 \href{http://arxiv.org/abs/arXiv:1202.6635}
{\tt  [arXiv:1202.6635 [hep-th]]}.



\bibitem{Knodel:2013fua}
  G.~Knodel and J.~T.~Liu,
  ``Higher derivative corrections to Lifshitz backgrounds,''
 { \href{http://dx.doi.org/10.1007/JHEP10(2013)002}
  {{\em JHEP} {\bf 1310} (2013) 002,}}
 \href{ http://arxiv.org/abs/arXiv:1305.3279}
  {\tt [arXiv:1305.3279 [hep-th]]}.

\bibitem{Azeyanagi:2009pr}
  T.~Azeyanagi, W.~Li and T.~Takayanagi,
  ``On String Theory Duals of Lifshitz-like Fixed Points,''
{ \href{ http://dx.doi.org/10.1088/1126-6708/2009/06/084}
{ {\em JHEP} {\bf 0906} (2009) 084}}
\href{http://arxiv.org/abs/arXiv:0905.0688}
{\tt  [arXiv:0905.0688 [hep-th]]}.

\bibitem{Li:2009pf}
  W.~Li, T.~Nishioka and T.~Takayanagi,
  ``Some No-go Theorems for String Duals of Non-relativistic Lifshitz-like Theories,''
{\href{  http://dx.doi.org/10.1088/1126-6708/2009/10/015}
 { { \em JHEP} {\bf 0910} (2009) 015}}
 \href{http://arxiv.org/abs/arXiv:0908.0363}
{\tt   [arXiv:0908.0363 [hep-th]]}.

\bibitem{Blaback:2010pp}
  J.~Blaback, U.~H.~Danielsson and T.~Van Riet,
  ``Lifshitz Backgrounds from 10d Supergravity,''
{\href{  http://dx.doi.org/10.1088/1751-8113/43/6/065401}
{ {\em  JHEP }{\bf 1002} (2010) 095,}}
\href{  http://arxiv.org/abs/arXiv:0908.2611}
{ {\tt  [arXiv:1001.4945 [hep-th]]}}.


\bibitem{Hartnoll:2009ns}
  S.~A.~Hartnoll, J.~Polchinski, E.~Silverstein and D.~Tong,
  ``Towards strange metallic holography,''
{\href{  http://dx.doi.org/10.1007/JHEP04(2010)120}
{ {\em  JHEP } {\bf 1004} (2010) 120,}}
\href{http://arxiv.org/abs/arXiv:0912.1061} 
{\tt  [arXiv:0912.1061 [hep-th]]}.

\bibitem{Dey:2012tg}
  P.~Dey and S.~Roy,
  ``Lifshitz-like space-time from intersecting branes in string/M theory,''
{\href{  http://dx.doi.org/10.1007/JHEP06(2012)129}
{ {\em  JHEP } {\bf 1206} (2012) 129,}}
\href{ http://arxiv.org/abs/arXiv:1203.5381}
{\tt  [arXiv:1203.5381 [hep-th]]};
  P.~Dey and S.~Roy,
  ``Intersecting D-branes and Lifshitz-like space-time,''
{\href{  http://dx.doi.org/10.1103/PhysRevD.86.066009}
{ {\em  Phys.\ Rev.\ D } {\bf 86} (2012) 066009,}}
\href{ http://arxiv.org/abs/arXiv:1204.4858}
{\tt  [arXiv:1204.4858 [hep-th]]}.
  
\bibitem{Narayan:2012hk}
  K.~Narayan,
  ``On Lifshitz scaling and hyperscaling violation in string theory,''
{\href{  http://dx.doi.org/10.1103/PhysRevD.85.106006}
{ {\em  Phys.\ Rev.\ D } {\bf 85} (2012) 106006,}}
\href{http://arxiv.org/abs/arXiv:1202.5935}  
{\tt  [arXiv:1202.5935 [hep-th]]}.




\bibitem{Donos:2010ax}
  A.~Donos, J.~P.~Gauntlett, N.~Kim and O.~Varela,
  ``Wrapped M5-branes, consistent truncations and AdS/CMT,''
\href{http://dx.doi.org/10.1007/JHEP12(2010)003}{{\em  JHEP} {\bf 1012} (2010) 003},
\href{http://arxiv.org/abs/arXiv:1009.3805}{{\tt  arXiv:1009.3805 [hep-th]}}.

\bibitem{Gregory:2010gx}
R.~Gregory, S.~L. Parameswaran, G.~Tasinato, and I.~Zavala, ``{Lifshitz
  solutions in supergravity and string theory},''
  \href{http://dx.doi.org/10.1007/JHEP12(2010)047}{{\em JHEP} {\bf 1012} (2010)
   047},
\href{http://arxiv.org/abs/1009.3445}{{\tt [arXiv:1009.3445 [hep-th]]}}.

\bibitem{Braviner:2011kz}
H.~Braviner, R.~Gregory, and S.~F. Ross, ``{Flows involving Lifshitz
  solutions},'' \href{http://dx.doi.org/10.1088/0264-9381/28/22/225028}{{\em
  Class.Quant.Grav.} {\bf 28} (2011)  225028},
\href{http://arxiv.org/abs/1108.3067}{{\tt arXiv:1108.3067 [hep-th]}}.

\bibitem{Barclay:2012he}
L.~Barclay, R.~Gregory, S.~Parameswaran, G.~Tasinato, and I.~Zavala,
  ``{Lifshitz black holes in IIA supergravity},''
  \href{http://dx.doi.org/10.1007/JHEP05(2012)122}{{\em JHEP} {\bf 1205} (2012)
   122},
\href{http://arxiv.org/abs/1203.0576}{{\tt [arXiv:1203.0576 [hep-th]]}}.

\bibitem{Burda:2014jca}
  P.~Burda, R.~Gregory and S.~Ross,
  ``Lifshitz flows in IIB and dual field theories,''
\href{ http://dx.doi.org/10.1007/JHEP11(2014)073}{\em JHEP {\bf 1411} (2014) 073},
\href{http://arxiv.org/abs/arXiv:1408.3271}{{\tt [arXiv:1408.3271 [hep-th]]}}.



\bibitem{Balasubramanian:2011ua}
K.~Balasubramanian and J.~McGreevy, ``{String theory duals of
  Lifshitz-Chern-Simons gauge theories},''
  \href{http://dx.doi.org/10.1088/0264-9381/29/19/194007}{{\em
  Class.Quant.Grav.} {\bf 29} (2012)  194007},
\href{http://arxiv.org/abs/1111.0634}{{\tt [arXiv:1111.0634 [hep-th]]}}.

\bibitem{Mulligan:2010wj}
  M.~Mulligan, C.~Nayak and S.~Kachru,
  ``An Isotropic to Anisotropic Transition in a Fractional Quantum Hall State,''
{\href{ http://dx.doi.org/10.1103/PhysRevB.82.085102}
 { {\em Phys.\ Rev.\ B} {\bf 82} (2010) 085102 }}
\href{http://arxiv.org/abs/arXiv:1004.3570}
{\tt   [arXiv:1004.3570 [cond-mat.str-el]]}.


\bibitem{Copsey:2011ek}
  K.~Copsey and R.~B.~Mann,
  ``Hidden Singularities and Closed Timelike Curves in A Proposed Dual for Lifshitz-Chern-Simons Gauge Theories,''
{\href{http://dx.doi.org/10.1103/PhysRevD.85.121902}
 { {\em  Phys.\ Rev.\ D } {\bf 85} (2012) 121902 }}
\href{ http://arxiv.org/abs/arXiv:1112.0578}
{\tt  [arXiv:1112.0578 [hep-th]]}.


\bibitem{Ardonne:2003wa}
  E.~Ardonne, P.~Fendley and E.~Fradkin,
  ``Topological order and conformal quantum critical points,''
{\href{  http://dx.doi.org/10.1016/j.aop.2004.01.004}
 { {\em  Annals Phys.\  } {\bf 310} (2004) 493}}
\href{ http://dx.doi.org/10.1016/j.aop.2004.01.004}
{\tt   [cond-mat/0311466]}.


\bibitem{Witten:1998qj}
  E.~Witten,
  ``Anti-de Sitter space and holography,''
  Adv.\ Theor.\ Math.\ Phys.\  {\bf 2} (1998) 253
  {\href{http://arxiv.org/abs/hep-th/9802150}
  {[hep-th/9802150]}}.



\bibitem{Korovin:2011kw}
  Y.~Korovin,
  ``Holographic Renormalization for Fermions in Real Time,''
  \href{http://arxiv.org/abs/arXiv:1107.0558}
{\tt   arXiv:1107.0558 [hep-th]}.

\bibitem{Alishahiha:2012nm}
  M.~Alishahiha, M.~R.~Mohammadi Mozaffar and A.~Mollabashi,
  ``Fermions on Lifshitz Background,''
{  \href{http://dx.doi.org/10.1103/PhysRevD.86.026002}
{ {\em  Phys.\ Rev.\ D } {\bf 86} (2012) 026002,}}
\href{  http://arxiv.org/abs/arXiv:1201.1764}
{\tt  [arXiv:1201.1764 [hep-th]]}.

\bibitem{Zingg:2013xla}
T.~Zingg, ``{Logarithmic two-Point Correlation Functions from a z = 2 Lifshitz
  Model},''
{\href{http://dx.doi.org/10.1007/JHEP01(2014)108}
{ {\em JHEP} {\bf 1401} (2014) 108 }}
\href{http://arxiv.org/abs/1310.4778}{{\tt [arXiv:1310.4778 [hep-th]]}}.

\bibitem{Sybesma:2015oha}
  W.~Sybesma and S.~Vandoren,
  ``Lifshitz quasinormal modes and relaxation from holography,''
 {\href{ http://dx.doi.org/10.1007/JHEP05(2015)021}
{ {\em  JHEP } {\bf 1505} (2015) 021,}}
\href{  http://arxiv.org/abs/arXiv:1503.07457}
{\tt  [arXiv:1503.07457 [hep-th]]}.


\bibitem{Andrade:2012xy}
T.~Andrade and S.~F. Ross, ``{Boundary conditions for scalars in Lifshitz},''
  \href{http://dx.doi.org/10.1088/0264-9381/30/6/065009}{{\em
  Class.Quant.Grav.} {\bf 30} (2013)  065009},
\href{http://arxiv.org/abs/1212.2572}{{\tt [arXiv:1212.2572 [hep-th]]}}.

\bibitem{Keeler:2012mb}
C.~Keeler, ``{Scalar Boundary Conditions in Lifshitz Spacetimes},''
{\href{http://dx.doi.org/10.1007/JHEP01(2014)067}
{ {\em JHEP} {\bf 1210} (2012) 152,}
}
\href{http://arxiv.org/abs/1212.1728}{{\tt [arXiv:1212.1728 [hep-th]]}}.

\bibitem{Andrade:2013wsa}
T.~Andrade and S.~F. Ross, ``{Boundary conditions for metric fluctuations in
  Lifshitz},''
 { \href{ http://dx.doi.org/10.1088/0264-9381/30/19/195017 }
{  {\em Class.\ Quant.\ Grav.\ } {\bf 30} (2013) 195017,}}
\href{http://arxiv.org/abs/1305.3539}{{\tt [arXiv:1305.3539 [hep-th]]}}.

\bibitem{Henkel}M.~Henkel,
  ``Schrodinger invariance in strongly anisotropic critical systems,''
{\href{  http://dx.doi.org/10.1007/BF02186756}
{ {\em  J.\ Statist.\ Phys.\   } {\bf 75} (1994) 1023,}}
\href{http://arxiv.org/abs/hep-th/9310081}  
{  [hep-th/9310081]}.



\bibitem{Bertoldi:2009vn}
G.~Bertoldi, B.~A. Burrington, and A.~Peet, ``{Black Holes in asymptotically
  Lifshitz spacetimes with arbitrary critical exponent},''
  \href{http://dx.doi.org/10.1103/PhysRevD.80.126003}{{\em Phys.Rev.} {\bf D80}
  (2009)  126003},
\href{http://arxiv.org/abs/0905.3183}{{\tt [arXiv:0905.3183 [hep-th]]}}.

\bibitem{Bertoldi:2009dt}
G.~Bertoldi, B.~A. Burrington, and A.~W. Peet, ``{Thermodynamics of black
  branes in asymptotically Lifshitz spacetimes},''
  \href{http://dx.doi.org/10.1103/PhysRevD.80.126004}{{\em Phys.Rev.} {\bf D80}
  (2009)  126004},
\href{http://arxiv.org/abs/0907.4755}{{\tt [arXiv:0907.4755 [hep-th]]}}.



  
\bibitem{Papadimitriou:2005ii}
  I.~Papadimitriou and K.~Skenderis,
  ``Thermodynamics of asymptotically locally AdS spacetimes,''
 {\href{ http://dx.doi.org/10.1088/1126-6708/2005/08/004}
{ {\em  JHEP } {\bf 0508} (2005) 004}}
{\href{  http://arxiv.org/abs/hep-th/0505190}
{\tt  [hep-th/0505190]}}.

\bibitem{Wald:1993nt}
  R.~M.~Wald,
  ``Black hole entropy is the Noether charge,''
{\href{  http://dx.doi.org/10.1103/PhysRevD.48.R3427}
{ {\em  Phys.\ Rev.\ D } {\bf 48} (1993) 3427}}
 \href{ http://arxiv.org/abs/gr-qc/9307038}
{ \tt  [gr-qc/9307038]}; V.~Iyer and R.~M.~Wald,
  ``Some properties of Noether charge and a proposal for dynamical black hole entropy,''
 {\href{ http://dx.doi.org/10.1103/PhysRevD.50.846}
{ {\em  Phys.\ Rev.\ D } {\bf 50} (1994) 846}}
{\href{  http://arxiv.org/abs/gr-qc/9403028}
{\tt  [gr-qc/9403028]}}.

  


\bibitem{Wang:2008jy}
  M.~T.~Wang and S.~T.~Yau,
  ``Quasilocal mass in general relativity,''
{\href{  http://dx.doi.org/10.1103/PhysRevLett.102.021101}
{ {\em  Phys.\ Rev.\ Lett.\   } {\bf 102} (2009) 021101}}
{\href{http://arxiv.org/abs/arXiv:0804.1174}
{ \tt  [arXiv:0804.1174 [gr-qc]]}}.
  
  
\bibitem{Devecioglu:2011yi}
  D.~O.~Devecioglu and O.~Sarioglu,
  ``On the thermodynamics of Lifshitz black holes,''
{\href{  http://dx.doi.org/10.1103/PhysRevD.83.124041}
{ {\em  Phys.\ Rev.\ D} {\bf 83} (2011) 124041}}
{\href{  http://arxiv.org/abs/arXiv:1103.1993}
{\tt  [arXiv:1103.1993 [hep-th]]}}.
  
\bibitem{Gim:2014nba}
  Y.~Gim, W.~Kim and S.~H.~Yi,
  ``The first law of thermodynamics in Lifshitz black holes revisited,''
{\href{  http://dx.doi.org/10.1007/JHEP07(2014)002}
{ {\em  JHEP } {\bf 1407} (2014) 002}}
{\href{  http://arxiv.org/abs/arXiv:1403.4704}
{ {\tt  [arXiv:1403.4704 [hep-th]]}}}.
  
  
  \bibitem{Brenna:2015pqa}
  W.~G.~Brenna, R.~B.~Mann and M.~Park,
  ``Mass and Thermodynamic Volume in Lifshitz Spacetimes,''
{\href{  http://dx.doi.org/10.1103/PhysRevD.92.044015}
{ {\em  Phys.\ Rev.\ D  }{\bf 92} (2015) 4,  044015}}
{\href{  http://arxiv.org/abs/arXiv:1505.06331}
{\tt  [arXiv:1505.06331 [hep-th]]}}.

  
  




\bibitem{Hoyos:2013eza}
  C.~Hoyos, B.~S.~Kim and Y.~Oz,
  ``Lifshitz Hydrodynamics,''
{\href{http://dx.doi.org/10.1007/JHEP11(2013)145 } 
  { {\em JHEP} {\bf 1311} (2013) 145,}}
\href{http://arxiv.org/abs/arXiv:1304.7481}
{\tt  [arXiv:1304.7481 [hep-th]]}.

\bibitem{Hoyos:2013qna}
  C.~Hoyos, B.~S.~Kim and Y.~Oz,
  ``Lifshitz Field Theories at Non-Zero Temperature, Hydrodynamics and Gravity,''
 {\href{ http://dx.doi.org/10.1007/JHEP11(2013)145}
  { {\em JHEP} {\bf 1403} (2014) 029}}
\href{http://arxiv.org/abs/arXiv:1309.6794}
 {\tt [arXiv:1309.6794 [hep-th]}.


\bibitem{Chapman:2014hja}
  S.~Chapman, C.~Hoyos and Y.~Oz,
  ``Lifshitz Superfluid Hydrodynamics,''
  {\href{http://dx.doi.org/10.1007/JHEP07(2014)027}
{ {\em  JHEP} {\bf 1407} (2014) 027,}}
\href{http://arxiv.org/abs/arXiv:1402.2981}
{\tt  [arXiv:1402.2981 [hep-th]]}.

\bibitem{Arav:2014goa} 
  I.~Arav, S.~Chapman and Y.~Oz,
  ``Lifshitz Scale Anomalies,''
  {\href{http://dx.doi.org/10.1007/JHEP02(2015)078}
  { {\em JHEP} {\bf 1502}, 078 (2015)}}
 \href{ http://arxiv.org/abs/arXiv:1410.5831}
{ {\tt  [arXiv:1410.5831 [hep-th]]}}.

\bibitem{Jensen:2014ama}
  K.~Jensen,
  ``Aspects of hot Galilean field theory,''
{\href{  http://dx.doi.org/10.1007/JHEP04(2015)123}
{ {\em  JHEP} {\bf 1504} (2015) 123,}}
\href{http://arxiv.org/abs/arXiv:1411.7024}
{\tt  [arXiv:1411.7024 [hep-th]]}.

\bibitem{Hoyos:2015lra}
  C.~Hoyos, A.~Meyer and Y.~Oz,
  ``Parity Breaking Transport in Lifshitz Hydrodynamics,''
 \href{ http://arxiv.org/abs/arXiv:1505.03141}
{\tt  arXiv:1505.03141 [hep-th]}.

\bibitem{Banerjee:2015uta} 
  N.~Banerjee, S.~Dutta and A.~Jain,
  ``On equilibrium partition function for non-relativistic fluid,''
  {\href{http://arxiv.org/abs/arXiv:1505.05677}
 {\tt arXiv:1505.05677 [hep-th]}};
N.~Banerjee, S.~Dutta and A.~Jain,
  ``Null Fluids - A New Viewpoint of Galilean Fluids,''
 {\href{ http://arxiv.org/abs/arXiv:1509.04718}
{\tt  arXiv:1509.04718 [hep-th]}}.





\bibitem{Pinzani-Fokeeva:2014cka}
  N.~Pinzani-Fokeeva and M.~Taylor,
  ``Towards a general fluid/gravity correspondence,''
 {\href{ http://dx.doi.org/10.1103/PhysRevD.91.044001}
 { {\em  Phys.\ Rev.\ D } {\bf 91} (2015) 4,  044001}}
 \href{http://arxiv.org/abs/arXiv:1401.5975}
{\tt  [arXiv:1401.5975 [hep-th]]}.

\bibitem{Danielsson:2009gi}
  U.~H.~Danielsson and L.~Thorlacius,
  ``Black holes in asymptotically Lifshitz spacetime,''
{\href{  http://dx.doi.org/10.1088/1126-6708/2009/03/070}
 { {\em JHEP} {\bf 0903} (2009) 070}}
 \href{http://arxiv.org/abs/arXiv:0812.5088}
{\tt  [arXiv:0812.5088 [hep-th]]}.

\bibitem{Mann:2009yx}
  R.~B.~Mann,
  ``Lifshitz Topological Black Holes,''
{\href{http://dx.doi.org/10.1088/1126-6708/2009/06/075}
{ {\em  JHEP} {\bf 0906} (2009) 075}}
\href{http://arxiv.org/abs/arXiv:0905.1136}
{\tt   [arXiv:0905.1136 [hep-th]]}.

\bibitem{Brynjolfsson:2009ct}
  E.~J.~Brynjolfsson, U.~H.~Danielsson, L.~Thorlacius and T.~Zingg,
  ``Holographic Superconductors with Lifshitz Scaling,''
 {\href{ http://dx.doi.org/10.1088/1751-8113/43/6/065401 }
  {{\em   J.\ Phys.\ A  }{\bf 43} (2010) 065401}}
\href{http://arxiv.org/abs/arXiv:0908.2611}
{\tt  [arXiv:0908.2611 [hep-th]]}.

\bibitem{Bertoldi:2011zr}
  G.~Bertoldi, B.~A.~Burrington, A.~W.~Peet and I.~G.~Zadeh,
  ``Lifshitz-like black brane thermodynamics in higher dimensions,''
\href{ http://dx.doi.org/10.1103/PhysRevD.83.126006}{\em Phys.\ Rev.\ D {\bf 83} (2011) 126006},
\href{http://arxiv.org/abs/arXiv:1101.1980}{{\tt [arXiv:1101.1980 [hep-th]]}}.

\bibitem{Amado:2011nd}
I.~Amado and A.~F. Faedo, ``{Lifshitz black holes in string theory},''
  \href{http://dx.doi.org/10.1007/JHEP07(2011)004}{{\em JHEP} {\bf 1107} (2011)
   004},
\href{http://arxiv.org/abs/1105.4862}{{\tt [arXiv:1105.4862 [hep-th]]}}.


\bibitem{Tarrio:2011de}
J.~Tarrio and S.~Vandoren, ``{Black holes and black branes in Lifshitz
  spacetimes},'' \href{http://dx.doi.org/10.1007/JHEP09(2011)017}{{\em JHEP}
  {\bf 1109} (2011)  017},
\href{http://arxiv.org/abs/1105.6335}{{\tt [arXiv:1105.6335 [hep-th]]}}.




\bibitem{Balasubramanian:2009rx}
  K.~Balasubramanian and J.~McGreevy,
  ``An Analytic Lifshitz black hole,''
 {\href{http://dx.doi.org/10.1103/PhysRevD.80.104039}
{ {\em  Phys.\ Rev.\ D }  {\bf 80} (2009) 104039 }}
\href{http://arxiv.org/abs/arXiv:0909.0263}
{\tt   [arXiv:0909.0263 [hep-th]]}.


\bibitem{Korovin:2013nha}
Y.~Korovin, K.~Skenderis, and M.~Taylor, ``{Lifshitz from AdS at finite
  temperature and top down models},''
{\href{  http://dx.doi.org/10.1007/JHEP11(2013)127}
  { {\em JHEP} {\bf 1311} (2013) 127}}
\href{http://arxiv.org/abs/1306.3344}{{\tt [arXiv:1306.3344 [hep-th]]}}.


\bibitem{Pang:2009ad}
  D.~W.~Pang,
  ``A Note on Black Holes in Asymptotically Lifshitz Spacetime,''
  {\href{http://dx.doi.org/10.1088/0253-6102/62/2/14}
   {  {\em Commun.\ Theor.\ Phys.\  }  {\bf 62} (2014) 265}}
   \href{http://arxiv.org/abs/arXiv:0905.2678}
 {\tt  [arXiv:0905.2678 [hep-th]]}.

\bibitem{Pang:2009pd}
  D.~W.~Pang,
  ``On Charged Lifshitz Black Holes,''
{\href{ http://dx.doi.org/10.1007/JHEP01(2010)116}
{  {\em  JHEP }{\bf 1001} (2010) 116}}
\href{http://arxiv.org/abs/arXiv:0911.2777}
{\tt   [arXiv:0911.2777 [hep-th]]}.

\bibitem{Mann:2011bt}
  R.~Mann, L.~Pegoraro and M.~Oltean,
  ``Lifshitz Solitons,''
{\href{  http://dx.doi.org/10.1103/PhysRevD.84.124047}
{ {\em  Phys.\ Rev.\ D } {\bf 84} (2011) 124047}}
{\href{http://arxiv.org/abs/arXiv:1109.5044}  
{\tt  [arXiv:1109.5044 [hep-th]]}}.


\bibitem{Shu:2014eza}
  F.~W.~Shu, K.~Lin, A.~Wang and Q.~Wu,
  ``Lifshitz spacetimes, solitons, and generalized BTZ black holes in quantum gravity at a Lifshitz point,''
{\href{  http://dx.doi.org/10.1007/JHEP04(2014)056}
{ {\em   JHEP } {\bf 1404} (2014) 056}}
{\href{  http://arxiv.org/abs/arXiv:1403.0946}
{\tt  [arXiv:1403.0946 [hep-th]]}}.



\bibitem{Bhattacharyya:2008jc}
  S.~Bhattacharyya, V.~E.~Hubeny, S.~Minwalla and M.~Rangamani,
  ``Nonlinear Fluid Dynamics from Gravity,''
 {\href{ http://dx.doi.org/10.1088/1126-6708/2008/02/045}
  { {\em JHEP} {\bf 0802} (2008) 045,}}
  \href{http://arxiv.org/abs/arXiv:0712.2456}
{\tt  [arXiv:0712.2456 [hep-th]]}.

\bibitem{Pang:2009wa}
  D.~W.~Pang,
  ``Conductivity and Diffusion Constant in Lifshitz Backgrounds,''
{\href{http://dx.doi.org/10.1007/JHEP01(2010)120}  
{ {\em  JHEP} {\bf 1001} (2010) 120,}}
\href{http://arxiv.org/abs/arXiv:0912.2403 } 
{\tt  [arXiv:0912.2403 [hep-th]]}.



\bibitem{Policastro:2002se}
  G.~Policastro, D.~T.~Son and A.~O.~Starinets,
  ``From AdS / CFT correspondence to hydrodynamics,''
{\href{ http://dx.doi.org/10.1088/1126-6708/2002/09/043}
 { {\em JHEP} {\bf 0209} (2002) 043,}}
\href{http://arxiv.org/abs/hep-th/0205052}
{\tt [hep-th/0205052]}; P.~Kovtun, D.~T.~Son and A.~O.~Starinets,
  ``Viscosity in strongly interacting quantum field theories from black hole physics,''
 {\href{http://dx.doi.org/10.1103/PhysRevLett.94.111601}
{ {\em  Phys.\ Rev.\ Lett.\  } {\bf 94} (2005) 111601,}}
\href{http://arxiv.org/abs/hep-th/0405231}
{ \tt  [hep-th/0405231]}.

\bibitem{Adams:2008zk}
  A.~Adams, A.~Maloney, A.~Sinha and S.~E.~Vazquez,
  ``1/N Effects in Non-Relativistic Gauge-Gravity Duality,''
 {\href{ http://dx.doi.org/10.1088/1126-6708/2009/03/097}
 { { \em  JHEP} {\bf 0903} (2009) 097}}
 \href{http://arxiv.org/abs/arXiv:0812.0166}
{\tt  [arXiv:0812.0166 [hep-th]]}.



\bibitem{Hoyos:2013cba}
  C.~Hoyos, B.~S.~Kim and Y.~Oz,
  ``Bulk Viscosity in Holographic Lifshitz Hydrodynamics,''
{\href{  http://dx.doi.org/10.1007/JHEP03(2014)050}
   { {\em JHEP} {\bf 1403} (2014) 050,}}
\href{   http://arxiv.org/abs/arXiv:1312.6380}
{\tt   [arXiv:1312.6380 [hep-th]]}.
  
\bibitem{Gouteraux:2011qh}
  B.~Gouteraux, J.~Smolic, M.~Smolic, K.~Skenderis and M.~Taylor,
  ``Holography for Einstein-Maxwell-dilaton theories from generalized dimensional reduction,''
{\href{  http://dx.doi.org/10.1007/JHEP01(2012)089}
{ {\em  JHEP }{\bf 1201} (2012) 089,}}
\href{  http://arxiv.org/abs/arXiv:1110.2320}
{\tt  [arXiv:1110.2320 [hep-th]]}.
  

\bibitem{Gouteraux:2011ce}
  B.~Gouteraux and E.~Kiritsis,
  ``Generalized Holographic Quantum Criticality at Finite Density,''
 {\href{ http://dx.doi.org/10.1007/JHEP12(2011)036}
{ {\em  JHEP} {\bf 1112} (2011) 036,}}
\href{  http://arxiv.org/abs/arXiv:1107.2116}
{\tt   [arXiv:1107.2116 [hep-th]]}.


\bibitem{Kiritsis:2015doa}
  E.~Kiritsis and Y.~Matsuo,
  ``Hyperscaling-Violating Lifshitz hydrodynamics from black-holes,''
  \href{http://arxiv.org/abs/arXiv:1508.02494}
{ {\tt  arXiv:1508.02494 [hep-th]}}.

\bibitem{Sun:2013wpa}
  J.~R.~Sun, S.~Y.~Wu and H.~Q.~Zhang,
  ``Novel Features of the Transport Coefficients in Lifshitz Black Branes,''
{\href{  http://dx.doi.org/10.1103/PhysRevD.87.086005}
{ {\em  Phys.\ Rev.\ D } {\bf 87} (2013) 086005,}}
\href{  http://arxiv.org/abs/arXiv:1302.5309}
{\tt  [arXiv:1302.5309 [hep-th]]}.


\bibitem{Sun:2013zga}
  J.~R.~Sun, S.~Y.~Wu and H.~Q.~Zhang,
  ``Mimic the optical conductivity in disordered solids via gauge/gravity duality,''
{\href{  http://dx.doi.org/10.1016/j.physletb.2014.01.005}
{ {\em  Phys.\ Lett.\ B } {\bf 729} (2014) 177,}}
 \href{ http://dx.doi.org/10.1016/j.physletb.2014.01.005}
{\tt  [arXiv:1306.1517 [hep-th]]}.

\bibitem{Roychowdhury:2015jha}
  D.~Roychowdhury,
  ``Magnetoconductivity in chiral Lifshitz hydrodynamics,''
{\href{  http://dx.doi.org/10.1007/JHEP09(2015)145}
{ {\em   JHEP}  {\bf 1509} (2015) 145,}}
\href{ http://arxiv.org/abs/arXiv:1508.02002}
{\tt  [arXiv:1508.02002 [hep-th]]}.
  
\bibitem{Roychowdhury:2015cva}
  D.~Roychowdhury,
  ``Lifshitz holography and the phases of the anisotropic plasma,''
 \href{ http://arxiv.org/abs/arXiv:1509.05229}
{\tt  arXiv:1509.05229 [hep-th]}.





\bibitem{Rangamani:2008gi}
  M.~Rangamani, S.~F.~Ross, D.~T.~Son and E.~G.~Thompson,
  ``Conformal non-relativistic hydrodynamics from gravity,''
{\href{http://dx.doi.org/10.1088/1126-6708/2009/01/075} 
{ {\em  JHEP } {\bf 0901} (2009) 075,}}
\href{ http://arxiv.org/abs/arXiv:0811.2049}
{\tt  [arXiv:0811.2049 [hep-th]]}.
  



\bibitem{Griffin:2011xs}
T.~Griffin, P.~Horava, and C.~M. Melby-Thompson, ``{Conformal Lifshitz Gravity
  from Holography},'' \href{http://dx.doi.org/10.1007/JHEP05(2012)010}{{\em
  JHEP} {\bf 1205} (2012)  010},
\href{http://arxiv.org/abs/1112.5660}{{\tt [arXiv:1112.5660 [hep-th]]}}.


\bibitem{Adam:2009gq}
  I.~Adam, I.~V.~Melnikov and S.~Theisen,
  ``A Non-Relativistic Weyl Anomaly,''
{\href{ http://dx.doi.org/10.1088/1126-6708/2009/09/130}
{  { {\em JHEP} {\bf 0909} (2009) 130,}}}
\href{http://arxiv.org/abs/arXiv:0907.2156}
{\tt  [arXiv:0907.2156 [hep-th]]}.

\bibitem{Baggio:2011cp}
M.~Baggio, J.~de~Boer, and K.~Holsheimer, ``{Hamilton-Jacobi Renormalization
  for Lifshitz Spacetime},''
  \href{http://dx.doi.org/10.1007/JHEP01(2012)058}{{\em JHEP} {\bf 1201} (2012)
   058},
\href{http://arxiv.org/abs/1107.5562}{{\tt [arXiv:1107.5562 [hep-th]]}}.

\bibitem{Baggio:2011ha}
M.~Baggio, J.~de~Boer, and K.~Holsheimer, ``{Anomalous Breaking of Anisotropic
  Scaling Symmetry in the Quantum Lifshitz Model},''
  \href{http://dx.doi.org/10.1007/JHEP07(2012)099}{{\em JHEP} {\bf 1207} (2012)
   099},
\href{http://arxiv.org/abs/1112.6416}{{\tt [arXiv:1112.6416 [hep-th]]}}.


\bibitem{Gomes:2011di}
  P.~R.~S.~Gomes and M.~Gomes,
  ``On Ward Identities in Lifshitz-like Field Theories,''
{\href{ http://dx.doi.org/10.1103/PhysRevD.85.085018}
 { {\em Phys.\ Rev.\ D} {\bf 85} (2012) 065010}}
 \href{http://arxiv.org/abs/arXiv:1107.6040}
{\tt   [arXiv:1112.3887 [hep-th]]}.


\bibitem{Jensen:2014hqa}
  K.~Jensen,
 ``Anomalies for Galilean fields,''
\href{ http://arxiv.org/abs/arXiv:1412.7750}
{\tt  arXiv:1412.7750 [hep-th]}.

\bibitem{Auzzi:2015fgg}
  R.~Auzzi, S.~Baiguera and G.~Nardelli,
  ``On Newton-Cartan trace anomalies,''
\href{  http://arxiv.org/abs/arXiv:1511.08150}
{\tt  arXiv:1511.08150 [hep-th]}.


\bibitem{Henningson:1998gx}
  M.~Henningson and K.~Skenderis,
  ``The Holographic Weyl anomaly,''
 { \href{http://dx.doi.org/10.1088/1126-6708/1998/07/023}
  {{ \em JHEP} {\bf 9807} (1998) 023}}
\href{  http://arxiv.org/abs/hep-th/9806087}
  {{ \tt [hep-th/9806087]}}.

\bibitem{deHaro:2000xn}
  S.~de Haro, S.~N.~Solodukhin and K.~Skenderis,
  ``Holographic reconstruction of space-time and renormalization in the AdS / CFT correspondence,''
  {\href{http://dx.doi.org/10.1007/s002200100381}
 { { \em Commun.\ Math.\ Phys.\ }  {\bf 217} (2001) 595}}
\href{ http://arxiv.org/abs/hep-th/0002230}
  {{ \tt [hep-th/0002230]}}.
  
\bibitem{Papadimitriou:2004ap}
  I.~Papadimitriou and K.~Skenderis,
  ``AdS / CFT correspondence and geometry,''
{\href{http://inspirehep.net/record/975894}
 {73rd Meeting between Theoretical Physicists and Mathematicians: (A)ds-CFT Correspondence, Strasbourg, France 2003} }
 \href{ http://arxiv.org/abs/hep-th/0404176}
  {{\tt [hep-th/0404176]}};  I.~Papadimitriou and K.~Skenderis,
  ``Correlation functions in holographic RG flows,''
 { \href{http://dx.doi.org/10.1088/1126-6708/2004/10/075}
  {{\em JHEP} {\bf 0410} (2004) 075}}
\href{http://arxiv.org/abs/hep-th/0407071}
  {{\tt [hep-th/0407071].}}
  
  \bibitem{Korovin:2013bua}
Y.~Korovin, K.~Skenderis, and M.~Taylor, ``{Lifshitz as a deformation of
  Anti-de Sitter},''
{\href{ http://dx.doi.org/10.1007/JHEP08(2013)026}
  { {\em JHEP} {\bf 1308} (2013) 026}}  
\href{http://arxiv.org/abs/1304.7776}{{\tt [arXiv:1304.7776 [hep-th]]}}.

\bibitem{Ross:2009ar}
S.~F. Ross and O.~Saremi, ``{Holographic stress tensor for non-relativistic
  theories},'' \href{http://dx.doi.org/10.1088/1126-6708/2009/09/009}{{\em
  JHEP} {\bf 0909} (2009)  009},
\href{http://arxiv.org/abs/0907.1846}{{\tt [arXiv:0907.1846 [hep-th]}}.

\bibitem{Ross:2011gu}
S.~F. Ross, ``{Holography for asymptotically locally Lifshitz spacetimes},''
  \href{http://dx.doi.org/10.1088/0264-9381/28/21/215019}{{\em
  Class.Quant.Grav.} {\bf 28} (2011)  215019},
\href{http://arxiv.org/abs/1107.4451}{{\tt [arXiv:1107.4451 [hep-th]]}}.

\bibitem{Mann:2011hg}
R.~B. Mann and R.~McNees, ``{Holographic Renormalization for Asymptotically
  Lifshitz Spacetimes},'' \href{http://dx.doi.org/10.1007/JHEP10(2011)129}{{\em
  JHEP} {\bf 1110} (2011)  129},
\href{http://arxiv.org/abs/1107.5792}{{\tt [arXiv:1107.5792 [hep-th]]}}.



\bibitem{Holsheimer:2013ula}
K.~Holsheimer, ``{On the Marginally Relevant Operator in $z=2$ Lifshitz
  Holography},''
{\href{http://dx.doi.org/10.1007/JHEP03(2014)084}
{  {\em JHEP} {\bf 1403} (2014) 084}}
 \href{http://arxiv.org/abs/1311.4539}{{\tt [arXiv:1311.4539 [hep-th]]}}. 

\bibitem{Hohm:2010jc}
  O.~Hohm and E.~Tonni,
  ``A boundary stress tensor for higher-derivative gravity in AdS and Lifshitz backgrounds,''
{\href{ http://dx.doi.org/10.1007/JHEP04(2010)093,}
{ {\em 
  JHEP} {\bf 1004} (2010) 093}}
{\href{http://arxiv.org/abs/arXiv:1001.3598}
{\tt   [arXiv:1001.3598 [hep-th]]}}.

\bibitem{Papadimitriou:2014lia}
  I.~Papadimitriou,
  ``Hyperscaling violating Lifshitz holography,''
 \href{http://arxiv.org/abs/arXiv:1411.0312}
  {{ \tt arXiv:1411.0312 [hep-th]}.}


\bibitem{Kanitscheider:2008kd}
  I.~Kanitscheider, K.~Skenderis and M.~Taylor,
  ``Precision holography for non-conformal branes,''
{\href{  http://dx.doi.org/10.1088/1126-6708/2008/09/094}
  {  {\em JHEP} {\bf 0809} (2008) 094,}}
\href{  http://dx.doi.org/10.1088/1126-6708/2008/09/094}
{\tt  [arXiv:0807.3324 [hep-th]]}.


\bibitem{Jensen:2014aia}
  K.~Jensen,
  ``On the coupling of Galilean-invariant field theories to curved spacetime,''
  \href{http://arxiv.org/abs/arXiv:1408.6855}
 {\tt arXiv:1408.6855 [hep-th]}.

\bibitem{Andringa:2010it}
  R.~Andringa, E.~Bergshoeff, S.~Panda and M.~de Roo,
  ``Newtonian Gravity and the Bargmann Algebra,''
{\href{http://dx.doi.org/10.1088/0264-9381/28/10/105011}
{ {\em  Class.\ Quant.\ Grav.\  } {\bf 28} (2011) 105011,}}
\href{http://arxiv.org/abs/arXiv:1011.1145}
{\tt   [arXiv:1011.1145 [hep-th]].}
  
\bibitem{Christensen:2013rfa}
  M.~H.~Christensen, J.~Hartong, N.~A.~Obers and B.~Rollier,
  ``Boundary Stress-Energy Tensor and Newton-Cartan Geometry in Lifshitz Holography,''
{ \href{ http://dx.doi.org/10.1007/JHEP01(2014)057}
  { {\em JHEP} {\bf 1401} (2014) 057}}
  \href{http://arxiv.org/abs/arXiv:1311.6471}
{\tt  [arXiv:1311.6471 [hep-th]]}.

\bibitem{Hartong:2014oma}
  J.~Hartong, E.~Kiritsis and N.~A.~Obers,
  ``Lifshitz spaceÐtimes for Schršdinger holography,''
{\href{  http://dx.doi.org/10.1016/j.physletb.2015.05.010}
  { {\em  Phys.\ Lett.\ B } {\bf 746} (2015) 318,}}
\href{  http://arxiv.org/abs/arXiv:1409.1519}
{\tt  [arXiv:1409.1519 [hep-th]]}.

\bibitem{Hartong:2014pma}
  J.~Hartong, E.~Kiritsis and N.~A.~Obers,
  ``Schroedinger Invariance from Lifshitz Isometries in Holography and Field Theory,''
 \href{ http://arxiv.org/abs/arXiv:1409.1522}
{\tt  arXiv:1409.1522 [hep-th]}.

\bibitem{Hartong:2015wxa}
  J.~Hartong, E.~Kiritsis and N.~A.~Obers,
  ``Field Theory on Newton-Cartan Backgrounds and Symmetries of the Lifshitz Vacuum,''
 { \href{http://dx.doi.org/10.1007/JHEP08(2015)006}
  { {\em JHEP} {\bf 1508} (2015) 006,}}
\href{  http://arxiv.org/abs/arXiv:1502.00228}
{\tt   [arXiv:1502.00228 [hep-th]]}.

 \bibitem{Hartong:2015zia}
  J.~Hartong and N.~A.~Obers,
  ``Ho\v{r}ava-Lifshitz gravity from dynamical Newton-Cartan geometry,''
 {\href{ http://dx.doi.org/10.1007/JHEP07(2015)155}
  { {\em JHEP} {\bf 1507} (2015) 155,}}
\href{http://arxiv.org/abs/arXiv:1504.07461}
{\tt   [arXiv:1504.07461 [hep-th]]}.
  
  \bibitem{Bergshoeff:2014uea}
  E.~A.~Bergshoeff, J.~Hartong and J.~Rosseel,
  ``Torsional Newton-Cartan geometry and the Schrodinger algebra,''
{\href{ http://dx.doi.org/10.1088/0264-9381/32/13/135017} 
  { {\em Class.\ Quant.\ Grav.\  }  {\bf 32} (2015) 13,  135017,}}
\href{http://arxiv.org/abs/arXiv:1409.5555}
{\tt   [arXiv:1409.5555 [hep-th]]}.

\bibitem{Bekaert:2014bwa}
  X.~Bekaert and K.~Morand,
  ``Connections and dynamical trajectories in generalised Newton-Cartan gravity I. An intrinsic view,''
{\href{  http://arxiv.org/abs/arXiv:1412.8212}
{\tt  arXiv:1412.8212 [hep-th]}}.



\bibitem{Hofman:2014loa}
  D.~M.~Hofman and B.~Rollier,
  ``Warped Conformal Field Theory as Lower Spin Gravity,''
{\href{http://dx.doi.org/10.1016/j.nuclphysb.2015.05.011}  
 { {\em  Nucl.\ Phys.\ B } {\bf 897} (2015) 1,}}
{\href{  http://arxiv.org/abs/arXiv:1411.0672}
{\tt  [arXiv:1411.0672 [hep-th]]}}.
  
\bibitem{Son:2013rqa}
  D.~T.~Son,
  ``Newton-Cartan Geometry and the Quantum Hall Effect,''
{\href{  http://arxiv.org/abs/arXiv:1306.0638}
{\tt  arXiv:1306.0638 [cond-mat.mes-hall]}}.

\bibitem{Geracie:2014nka}
  M.~Geracie, D.~T.~Son, C.~Wu and S.~F.~Wu,
  ``Spacetime Symmetries of the Quantum Hall Effect,''
{\href{http://dx.doi.org/10.1103/PhysRevD.91.045030}  
{ {\em  Phys.\ Rev.\ D } {\bf 91} (2015) 045030,}}
{\href{  http://arxiv.org/abs/arXiv:1407.1252}
{\tt  [arXiv:1407.1252 [cond-mat.mes-hall]]}}.

\bibitem{Chemissany:2011mb}
W.~Chemissany and J.~Hartong, ``{From D3-Branes to Lifshitz Space-Times},''
  \href{http://dx.doi.org/10.1088/0264-9381/28/19/195011}{{\em
  Class.Quant.Grav.} {\bf 28} (2011)  195011},
\href{http://arxiv.org/abs/1105.0612}{{\tt [arXiv:1105.0612 [hep-th]]}}.

\bibitem{Chemissany:2012du}
W.~Chemissany, D.~Geissbuhler, J.~Hartong, and B.~Rollier, ``{Holographic
  Renormalization for z=2 Lifshitz Space-Times from AdS},''
  \href{http://dx.doi.org/10.1088/0264-9381/29/23/235017}{{\em
  Class.Quant.Grav.} {\bf 29} (2012)  235017},
\href{http://arxiv.org/abs/1205.5777}{{\tt [arXiv:1205.5777 [hep-th]]}}.

\bibitem{Papadimitriou:2011qb}
I.~Papadimitriou, ``{Holographic Renormalization of general dilaton-axion
  gravity},'' \href{http://dx.doi.org/10.1007/JHEP08(2011)119}{{\em JHEP} {\bf
  1108} (2011)  119},
\href{http://arxiv.org/abs/1106.4826}{{\tt [arXiv:1106.4826 [hep-th]]}}.


\bibitem{Christensen:2013lma}
M.~H. Christensen, J.~Hartong, N.~A. Obers, and B.~Rollier, ``{Torsional
  Newton-Cartan Geometry and Lifshitz Holography},''
{\href{http://dx.doi.org/10.1103/PhysRevD.89.061901}
{ {\em Phys.\ Rev.\ D} {\bf 89} (2014) 061901}}
\href{http://arxiv.org/abs/1311.4794}{{\tt [arXiv:1311.4794 [hep-th]]}}.


\bibitem{Horava:2009uw}
P.~Horava, {``Quantum Gravity at a Lifshitz Point,"}  
{\href{http://dx.doi.org/10.1103/PhysRevD.79.084008}
{ {\em Phys.Rev.} {\bf
  D79} (2009) 084008, }}
  \href{http://xxx.lanl.gov/abs/0901.3775}{{\tt
  arXiv:0901.3775}}.

\bibitem{Bergshoeff:2015uaa}
  E.~Bergshoeff, J.~Rosseel and T.~Zojer,
  ``Newton-Cartan (super)gravity as a non-relativistic limit,''
{\href{http://arxiv.org/abs/arXiv:1505.02095}
{ \tt  arXiv:1505.02095 [hep-th]}}.

\bibitem{Bergshoeff:2015ija}
  E.~Bergshoeff, J.~Rosseel and T.~Zojer,
  ``Newton-Cartan supergravity with torsion and Schroedinger supergravity,''
{\href{  http://arxiv.org/abs/arXiv:1509.04527}
{ \tt  arXiv:1509.04527 [hep-th]}}.


\bibitem{Bagchi:2009my}
  A.~Bagchi and R.~Gopakumar,
  ``Galilean Conformal Algebras and AdS/CFT,''
{\href{  http://dx.doi.org/10.1088/1126-6708/2009/07/037}
{ {\em  JHEP } {\bf 0907} (2009) 037,}}
\href{  http://arxiv.org/abs/arXiv:0902.1385}
{\tt  [arXiv:0902.1385 [hep-th]]}.

\bibitem{Bagchi:2009pe}
  A.~Bagchi, R.~Gopakumar, I.~Mandal and A.~Miwa,
  ``GCA in 2d,''
{\href{  http://dx.doi.org/10.1007/JHEP08(2010)004}
{ {\em  JHEP}  {\bf 1008} (2010) 004,}}
 \href{ http://arxiv.org/abs/arXiv:0912.1090}
{\tt  [arXiv:0912.1090 [hep-th]]}.

\bibitem{Janiszewski:2012nb}
S.~Janiszewski and A.~Karch,
  ``Non-relativistic holography from Horava gravity,''
{\href{  http://dx.doi.org/10.1007/JHEP02(2013)123}
{ {\em  JHEP} {\bf 1302} (2013) 123,}}
\href{ http://arxiv.org/abs/arXiv:1211.0005} 
{\tt  [arXiv:1211.0005 [hep-th]]}. 

\bibitem{Janiszewski:2012nf}
S.~Janiszewski and A.~Karch,
  ``String Theory Embeddings of Nonrelativistic Field Theories and Their Holographic Horava Gravity Duals,''
{\href{ http://dx.doi.org/10.1103/PhysRevLett.110.081601} 
{ {\em  Phys.\ Rev.\ Lett.\ } {\bf 110} (2013) 8,  081601,}}
\href{ http://arxiv.org/abs/arXiv:1211.0010}
{\tt   [arXiv:1211.0010 [hep-th]]}.

\bibitem{Jensen:2014wha}
  K.~Jensen and A.~Karch,
  ``Revisiting non-relativistic limits,''
{\href{ http://dx.doi.org/10.1007/JHEP04(2015)155 }
{ {\em  JHEP} {\bf 1504} (2015) 155,}}
\href{http://arxiv.org/abs/arXiv:1412.2738}
{\tt  [arXiv:1412.2738 [hep-th]]}.




\bibitem{Griffin:2012qx}
  T.~Griffin, P.~Ho?ava and C.~M.~Melby-Thompson,
  ``Lifshitz Gravity for Lifshitz Holography,''
{\href{ http://dx.doi.org/10.1103/PhysRevLett.110.081602 }
{ {\em  Phys.\ Rev.\ Lett.\  } {\bf 110} (2013) 8,  081602,}}
\href{http://arxiv.org/abs/arXiv:1211.4872} 
{\tt   [arXiv:1211.4872 [hep-th]]}.

\bibitem{Bekaert:2015xua}
  X.~Bekaert and K.~Morand,
  ``Connections and dynamical trajectories in generalised Newton-Cartan gravity II. An ambient perspective,''
{\href{  http://arxiv.org/abs/arXiv:1505.03739}
{\tt  arXiv:1505.03739 [hep-th]}}.

\bibitem{Hartong:2015xda}
  J.~Hartong,
  ``Gauging the Carroll Algebra and Ultra-Relativistic Gravity,''
{\href{  http://dx.doi.org/10.1007/JHEP08(2015)069}
  { {\em JHEP} {\bf 1508} (2015) 069,}}
 \href{ http://arxiv.org/abs/arXiv:1505.05011}
{\tt  [arXiv:1505.05011 [hep-th]]}.





  
\bibitem{Ryu:2006bv}
  S.~Ryu and T.~Takayanagi,
  ``Holographic derivation of entanglement entropy from AdS/CFT,''
{\href{  http://dx.doi.org/10.1103/PhysRevLett.96.181602}
{ {\em   Phys.\ Rev.\ Lett.\ }  {\bf 96} (2006) 181602,}}
{\href{  http://arxiv.org/abs/hep-th/0603001}
{\tt  [hep-th/0603001]}}.
 
 \bibitem{Alishahiha:2014cwa}
  M.~Alishahiha, A.~F.~Astaneh and M.~R.~M.~Mozaffar,
  ``Thermalization in backgrounds with hyperscaling violating factor,''
{\href{  http://dx.doi.org/10.1103/PhysRevD.90.046004}
{ {\em  Phys.\ Rev.\ D } {\bf 90} (2014) 4,  046004,}}
{\href{  http://arxiv.org/abs/arXiv:1401.2807}
{  [arXiv:1401.2807 [hep-th]]}}.
 
\bibitem{Chakraborty:2014lfa}
  S.~Chakraborty, P.~Dey, S.~Karar and S.~Roy,
  ``Entanglement thermodynamics for an excited state of Lifshitz system,''
{\href{  http://dx.doi.org/10.1007/JHEP04(2015)133}
{ {\em  JHEP } {\bf 1504} (2015) 133,}}
\href{ http://arxiv.org/abs/arXiv:1412.1276}
{\tt  [arXiv:1412.1276 [hep-th]]}.
  
 
\bibitem{Fischetti:2014zja}
  S.~Fischetti and D.~Marolf,
  ``Complex Entangling Surfaces for AdS and Lifshitz Black Holes?,''
{\href{  http://dx.doi.org/10.1088/0264-9381/31/21/214005}
{ {\em  Class.\ Quant.\ Grav.\  } {\bf 31} (2014) 21,  214005,}}
\href{  http://arxiv.org/abs/arXiv:1407.2900}
{\tt  [arXiv:1407.2900 [hep-th]]}.
  
   \bibitem{Hosseini:2015gua}
  S.~M.~Hosseini and A.~Veliz-Osorio,
  ``Entanglement and mutual information in 2d nonrelativistic field theories,''
 \href{ http://arxiv.org/abs/arXiv:1510.03876}
{\tt  arXiv:1510.03876 [hep-th]}.

 
  
\bibitem{Sachdev:2011wg}
  S.~Sachdev,
  ``What can gauge-gravity duality teach us about condensed matter physics?,''
{\href{  http://dx.doi.org/10.1146/annurev-conmatphys-020911-125141}
{ {\em  Ann.\ Rev.\ Condensed Matter Phys.\  } {\bf 3} (2012) 9,}}
\href{  http://arxiv.org/abs/arXiv:1108.1197}
{\tt  [arXiv:1108.1197 [cond-mat.str-el]]}.
 
 
\bibitem{Adams:2012th}
  A.~Adams, L.~D.~Carr, T.~SchŠfer, P.~Steinberg and J.~E.~Thomas,
  ``Strongly Correlated Quantum Fluids: Ultracold Quantum Gases, Quantum Chromodynamic Plasmas, and Holographic Duality,''
{\href{  http://dx.doi.org/10.1088/1367-2630/14/11/115009}
{ {\em  New J.\ Phys.\ }  {\bf 14} (2012) 115009,}}
 \href{ http://arxiv.org/abs/arXiv:1205.5180}
{\tt  [arXiv:1205.5180 [hep-th]]}.

 
\bibitem{Charmousis:2010zz}
  C.~Charmousis, B.~Gouteraux, B.~S.~Kim, E.~Kiritsis and R.~Meyer,
  ``Effective Holographic Theories for low-temperature condensed matter systems,''
 {\href{ http://dx.doi.org/10.1007/JHEP11(2010)151}
{ {\em  JHEP } {\bf 1011} (2010) 15,}}
\href{  http://arxiv.org/abs/arXiv:1005.4690}
{\tt  [arXiv:1005.4690 [hep-th]]}.
  
\bibitem{Dong:2012se}
  X.~Dong, S.~Harrison, S.~Kachru, G.~Torroba and H.~Wang,
  ``Aspects of holography for theories with hyperscaling violation,''
{\href{ http://dx.doi.org/10.1007/JHEP06(2012)041} 
{ {\em  JHEP} {\bf 1206} (2012) 041,}}
\href{ http://arxiv.org/abs/arXiv:1201.1905}
{\tt [arXiv:1201.1905 [hep-th]]}.
 

  
\bibitem{Gouteraux:2013oca}
  B.~GoutŽraux,
  ``Universal scaling properties of extremal cohesive holographic phases,''
{\href{  http://dx.doi.org/10.1007/JHEP01(2014)080}
{ {\em  JHEP}  {\bf 1401} (2014) 080,}}
\href{  http://arxiv.org/abs/arXiv:1308.2084}
{\tt   [arXiv:1308.2084 [hep-th]]};
 B.~GoutŽraux,
  ``Charge transport in holography with momentum dissipation,''
{\href{  http://dx.doi.org/10.1007/JHEP04(2014)181}
{ {\em  JHEP}  {\bf 1404} (2014) 181,}}
\href{  http://arxiv.org/abs/arXiv:1401.5436}
{\tt  [arXiv:1401.5436 [hep-th]]}.
 
\bibitem{Karch:2014mba}
  A.~Karch,
  ``Conductivities for Hyperscaling Violating Geometries,''
{\href{  http://dx.doi.org/10.1007/JHEP06(2014)140}
{ {\em  JHEP} {\bf 1406} (2014) 140,}}
\href{  http://arxiv.org/abs/arXiv:1405.2926}
{\tt  [arXiv:1405.2926 [hep-th]]}.
 
 
\bibitem{Hartnoll:2015sea}
  S.~A.~Hartnoll and A.~Karch,
  ``Scaling theory of the cuprate strange metals,''
{\href{http://dx.doi.org/10.1103/PhysRevB.91.155126}
 { { \em  Phys.\ Rev.\ B } {\bf 91} (2015) 15,  155126,}}
\href{ http://arxiv.org/abs/arXiv:1501.03165}
{\tt  [arXiv:1501.03165 [cond-mat.str-el]]}.
  
  

\bibitem{Skenderis:2009nt}
  K.~Skenderis, M.~Taylor and B.~C.~van Rees,
  ``Topologically Massive Gravity and the AdS/CFT Correspondence,''
{\href{  http://dx.doi.org/10.1088/1126-6708/2009/09/045}
{ {\em  JHEP} {\bf 0909} (2009) 045,}}
\href{ http://arxiv.org/abs/arXiv:0906.4926} 
{\tt  [arXiv:0906.4926 [hep-th]]}.




  

  
    
  
  
 
 
  

 
  
  
  
  
\end{thebibliography}
\end{document}